\documentclass[12pt]{article}

\usepackage{putex}
\usepackage{bm}

\usepackage{tikz-feynman}
\usetikzlibrary{external}             
\immediate\write18{mkdir -p pgf-img}  
\usepackage{graphicx}
\usepackage{caption}
\usepackage{amsmath}
\usepackage{array}
\usepackage{subcaption}
\usepackage{epstopdf}
\usepackage{enumerate}
\usepackage{mathrsfs}
\usepackage{cite}
\usepackage{youngtab}
\usepackage{tensor}
\usepackage{slashed}
\usepackage[aligntableaux=center]{ytableau}
\usepackage[utf8]{inputenc}
\usepackage{rotating}
\usepackage[
      colorlinks=true,
      linkcolor=blue,
      urlcolor=blue,
      filecolor=black,
      citecolor=red,
      ]{hyperref}
\usepackage{yfonts}

\newcommand{\es}[2] {\begin{equation} \label{#1} \begin{split} #2 \end{split} \end{equation}}

\newcommand{\com}[1]{}

\newcommand{\R}{\mathbb{R}}

\newcommand{\cN}{{\mathcal N}}
\newcommand{\cO}{{\mathcal O}}

\newcommand{\beq}{\begin{equation}}
\newcommand{\eeq}{\end{equation}}

\newcommand {\be} {\begin {equation}}
\newcommand {\ee} {\end {equation}}

\newcommand {\bes} {\begin {equation*}}
\newcommand {\ees} {\end {equation*}}

\def\ie{\begin{equation}\begin{aligned}}
\def\fe{\end{aligned}\end{equation}}

\newcommand{\m}{\mu}
\newcommand{\n}{\nu}
\newcommand{\pa}{\nabla}

\newcommand{\da}{{\dot\alpha}}
\newcommand{\db}{{\dot\beta}}

\numberwithin{equation}{section}

\def\<{\langle}
\def\>{\rangle}

\newcommand{\fcy}[1]{\mathcal{#1}}

\newcommand{\ak}{\alpha}        
\newcommand{\bk}{\beta}         
\newcommand{\gk}{\gamma}        \newcommand{\Gk}{\Gamma}
\newcommand{\dk}{\delta}        \newcommand{\Dk}{\Delta}
   
\newcommand{\zk}{\zeta}         
          
\newcommand{\qk}{\theta}        
         
\newcommand{\kk}{\kappa}        
\newcommand{\lk}{\lambda}       \newcommand{\Lk}{\Lambda}
          
\newcommand{\sk}{\sigma}        \newcommand{\Sk}{\Sigma}
\newcommand{\tk}{\tau}          
\newcommand{\fk}{\phi}          \newcommand{\Fk}{\Phi}
\newcommand{\ck}{\chi}          
\newcommand{\yk}{\psi}          \newcommand{\Yk}{\Psi}
\newcommand{\wk}{\omega}        \newcommand{\Wk}{\Omega}

\newcommand{\nb}{\partial}      \newcommand{\Nb}{\nabla}

\def\da{{\dot\alpha}}
\def\db{{\dot\beta}}
\def\dg{{\dot\gamma}}

\def\bT{{\bar T}}
\def\pa{\partial}

\newcommand{\co}{{\mathcal O}}

\begin{document}

\preprint{PUPT-2617 \\
MIT-CTP/5190}

\date{March 2020}

\institution{PU}{Joseph Henry Laboratories, Princeton University, Princeton, NJ 08544, USA}
\institution{Stanford}{Stanford Institute for Theoretical Physics, Department of Physics, \cr Stanford University,  Stanford, CA 94305, USA}
\institution{MIT}{Center for Theoretical Physics and Department of Mathematics,  \cr Massachusetts Institute of Technology, Cambridge, MA 02139, USA}

\title{
A Bispinor Formalism for Spinning Witten Diagrams
}

\authors{Damon J.~Binder,\worksat{\PU} Daniel Z.~Freedman,\worksat{\Stanford, \MIT} and Silviu S.~Pufu\worksat{\PU}}

\abstract{
We develop a new embedding-space formalism for AdS$_4$ and CFT$_3$ that is useful for evaluating Witten diagrams for operators with spin.  The basic variables are Killing spinors for the bulk AdS$_4$ and conformal Killing spinors for the boundary CFT$_3$.  The more conventional embedding space coordinates $X^I$ for the bulk and $P^I$ for the boundary are bilinears in these new variables.  We write a simple compact form for the general bulk-boundary propagator, and, for boundary operators of spin $\ell \geq 1$, we determine its conservation properties at the unitarity bound.    In our CFT$_3$ formalism, we identify an $\mathfrak{so}(5,5)$ Lie algebra of differential operators that includes the basic weight-shifting operators.  These operators, together with a set of differential operators in AdS$_4$, can be used to relate Witten diagrams with spinning external legs to Witten diagrams with only scalar external legs.   We provide several applications that include Compton scattering and the evaluation of an $R^4$ contact interaction in AdS$_4$. Finally, we derive bispinor formulas for the bulk-to-bulk propagators of massive spinor and vector gauge fields and evaluate a diagram with spinor exchange.
}

\maketitle

\tableofcontents

\section{Introduction}

There has recently been renewed interest in computing correlation functions of local operators in top-down models of holography in various dimensions \cite{Rastelli:2017ymc,Rastelli:2017udc,Rastelli:2016nze,Zhou:2017zaw,Chester:2018aca,Chester:2018dga,Binder:2018yvd,Binder:2019jwn,Binder:2019mpb,Alday:2018pdi,Alday:2018kkw,Chester:2019pvm}.  Interest has been spurred by the observation that Witten diagrams can be efficiently bootstrapped in Mellin space  \cite{Mack:2009gy,Mack:2009mi},\footnote{See~\cite{Penedones:2019tng} for recent work on defining Mellin amplitudes for general CFT correlators that do not necessarily have a holographic interpretation.} where they obey simple analytic properties that are reminiscent of the analytic properties of scattering amplitudes in flat space \cite{Penedones:2010ue,Fitzpatrick:2011ia,Fitzpatrick:2011hu,Fitzpatrick:2011dm}. In addition to these analytic properties, the bootstrap conditions include crossing symmetry and constraints required by supersymmetry.  Remarkably, this program has led to the successful evaluation of contact Witten diagrams corresponding to higher derivative corrections to 10d or 11d supergravity, or even to certain one-loop diagrams with higher derivative vertices, even though the complete forms of these higher-derivative interaction vertices remain unknown \cite{Chester:2018aca,Chester:2018dga,Binder:2018yvd,Binder:2019jwn,Binder:2019mpb,Alday:2018pdi,Alday:2018kkw,Chester:2019pvm}.

The work referenced above involves correlation functions of scalar operators only.  (See, however, \cite{Goncalves:2014rfa,Costa:2014kfa,Nishida:2018opl} for some work on Witten diagrams for spinning correlators.)   The restriction to scalar operators occurs for several reasons.  First, in the cases studied thus far, maximal or near-maximal supersymmetry relates  correlators of operators with spin to scalar correlators \cite{Binder:2018yvd,Binder:2019jwn,Binder:2019mpb}.  With less supersymmetry, spinning correlators require separate study. Furthermore,
 the CFT structures needed for spinning operators are cumbersome to work with and become more and more complicated with increasing spin.  Lastly,  the Mellin representation of spinning correlators has not yet been developed, and consequently, the analytic properties of spinning correlators are not yet fully understood.

The goal of this paper is to initiate a systematic study of holographic correlators of spinning operators by developing a new embedding-space formalism that makes it easier to evaluate Witten diagrams for external operators of any spin.  We will restrict our work to the case AdS$_4$/CFT$_3$.  Within the formalism we develop, we define various differential operators that can be used to ``spin up'' scalar correlators. This idea of changing the spin of the operators in a correlation function is not new, and has been used in the past to study both Witten diagrams \cite{Paulos:2011ie,Goncalves:2014rfa,Costa:2014kfa} and conformal blocks \cite{Costa:2011dw, Costa:2011mg, SimmonsDuffin:2012uy, Iliesiu:2015qra, Iliesiu:2015akf, Hijano:2015zsa,Costa:2016hju,Karateev:2017jgd,Dyer:2017zef}.

Our formalism is a variant of the embedding space formalisms for AdS/CFT which includes some features of the spinor-helicity formalism for scattering amplitudes.  The standard embedding space formalism linearizes conformal transformations by embedding both AdS$_{d+1}$ and its boundary $\R^{d-1, 1}$ within $\R^{d,2}$ \cite{Dirac:1936fq,Ferrara:1973yt,Mack:1969rr,Boulware:1970ty,Ferrara:1973eg}.   In particular, the manifold AdS$_{d+1}$ is realized as the subset of $\mathbb R^{d,2}$ (with signature $-++ \cdots ++-$) for which\footnote{Here we took the radius of AdS, $L$, to be $L=1$.  If $L$ were not set to one, then $-L^2$ would appear on the RHS of \eqref{Hyp} instead of $-1$.}
 \es{Hyp}{
X \cdot X \equiv -(X^0)^2+(X^1)^2+\dots+(X^d)^2-(X^{d+1})^2 = -1\,.
 }
The boundary $\R^{d-1,1}$ of this manifold can be identified with the null light cone $P\cdot P = 0$, up to rescalings $P^I\sim\lk P^I$, with $I = 0, \ldots, d+1$. To describe spinning fields/operators in this formalism, one must introduce fields/operators with $SO(d,2)$ indices possessing gauge redundancies  and/or obeying constraints \cite{Weinberg:2012mz,Costa:2011mg,Costa:2014kfa,Iliesiu:2015qra,Elkhidir:2014woa}.  For example, in CFT$_d$, symmetric tensor operators ${\cal O}^{\mu \nu}(x)$, with $\mu, \nu = 0, \ldots, d-1$, lift to symmetric tensors ${\cal O}^{IJ}(P)$ in embedding space obeying the transversality constraint $P_I {\cal O}^{IJ}(P) = 0$ as well as the gauge redundancy ${\cal O}^{IJ}(P) \sim {\cal O}^{IJ}(P) + P^{(I} \Lambda^{J)}(P)  $, with $\Lambda^I(P)$ transverse, $P_I \Lambda^I(P) = 0$, but otherwise arbitrary.  Dealing with such gauge redundancies and constraints can be cumbersome.

Similar difficulties are familiar from the study of scattering amplitudes. Indeed, traditionally, massless amplitudes are described using polarization vectors and gauge conditions. In the last two decades however, much progress has been made using spinor helicity variables---variables which transform in spinor representations of both the Lorentz group and also the little group of the particle. These variables have not only led to many technical improvements in the computation of perturbative scattering amplitudes, but they have also allowed a deeper understand of scattering amplitudes, and in particular they have illuminated otherwise hidden connections between theories with quite different matter content. For a good textbook introduction to these methods with many references, see \cite{Elvang:2015rqa}.

Our aim is to develop analogous methods for describing correlators in AdS$_4$/CFT$_3$.  In AdS$_4$, the analogues of the spinor-helicity variables are the Killing spinors $T^A_\alpha$ and their conjugates $\bar T^A_{\dot \alpha}$, which transform in the spinor representation of the $SO(3, 2)$ isometry group as exhibited by the index $A = 1, \ldots, 4$, and in the spinor representations of the $SO(3, 1)$ local Lorentz group, as exhibited by the indices $\alpha$ and $\dot \alpha$.  On the boundary CFT$_3$, the analogues of the spinor-helicity variables are the conformal Killing spinors $S^A_a$, which also transform in the spinor representation of $SO(3, 2)$ as well as the spinor representation of the Lorentz group $SO(2, 1)$.  They are the boundary limits of the $T^A_\alpha$ and $\bar T^A_{\dot \alpha}$ defined in the bulk.

For scattering amplitudes, the components of the momentum of each particle can be written as quadratic expressions in the spinor-helicity variables. Similarly in our approach the embedding space coordinates $X^I$ (for AdS$_4$) and $P^I$ (for CFT$_3$) can be written as quadratic expressions in the Killing spinors and conformal Killing spinors, respectively.  Because of this fact, the fields in the bulk of AdS$_4$ can be thought of as functions of the $T^A_{\alpha}$ and $\bar T^A_{\dot \alpha}$ and the operators in the boundary CFT$_3$ can be thought of as functions of the $S^A_a$.  The $T^A_\alpha$ and $\bar T^A_{\dot \alpha}$ contain 16  real components and  constraints are needed to reduce this number to  $4$  independent ones, as appropriate to parametrize the four-dimensional space AdS$_4$. Likewise the real $S^A_a$ obey constraints such that their $8$ components are reduced to $3$, which is the number of dimensions of $\R^{2,1}$.  We discuss these constraints in detail in the next section.  When defining a spinning field (or operator) in this formalism, we do not need transversality constraints or gauge redundancies on the fields (or operators).  Instead we simply specify the appropriate transformation properties under $SO(3, 1)$ (or under $SO(2, 1)$).  In this way the difficulty of dealing with tensor operators and gauge redundancies is avoided.  This formalism is particularly suitable for Witten diagram computations because bulk-boundary propagator take a particularly simple form.

After establishing this formalism, we are guided by conformal symmetry to define various differential operators with respect to our new coordinates $T^A_{\alpha}$, $\bar T^A_{\dot \alpha}$, and $S^A_a$.  These differential operators become particularly useful when evaluating contact and exchange Witten diagrams. As we will see, all diagrams can be reduced to diagrams where the external legs are scalar propagators.   On the boundary, the differential operators we use to spin up the external legs are more unified versions of the weight-shifting operators for the spinor representation of $SO(3, 2)$, which were defined in \cite{Karateev:2017jgd} in the embedding-space formalism of \cite{Iliesiu:2015qra}.  

The body of this paper is organized as follows.  Section~\ref{BISPINOR} introduces the bispinors needed to describe the bulk and boundary in AdS$_4$/CFT$_3$ and outlines their relation to Killing spinors.  The general form of all bulk-boundary propagator is then derived from conformal invariance. In Section~\ref{DIFFOPS}, we define the full set of conformal covariant differential operators needed to relate Witten diagrams with spinning external lines to diagrams with external scalars.  Many examples of contact and tree level exchange diagrams are presented in Section~\ref{WITDIAG}.  In Section~\ref{BULKBULK}, we derive the bulk-bulk propagators for massive spinor and gauge vector fields and show that their boundary limits agree with previous bulk-boundary propagators.  We evaluate a 4-point Witten diagram with spinor exchange.  Finally we determine the bulk-bulk propagators of the purely chiral and anti-chiral fields obtained by covariant differentiation of their parent Lagrangian fields.  We end with a discussion of our results in Section~\ref{DISCUSSION}. There are six appendices which present our conventions and provide more detail on ideas from the the main text.

\section{Bispinors for the bulk and boundary}
\label{BISPINOR}

The chief innovation of our study is the use of bispinor variables\footnote{Similar variables are used in \cite{Nishida:2018opl} for AdS$_5$/CFT$_4$, but the approach is developed differently.}  to describe both CFT operators and their dual AdS fields in the bulk.  We describe the key features of our work in this section, with applications left for later sections.  

First, let us establish some conventions.  The symmetry group of embedding space is viewed as $SO(3,2)$ or its double cover $Sp(4)$. Indices of the $SO(3,2)$ spinor representation are denoted by capitals $A,B,\dots$. From the $Sp(4)$ perspective this representation is the fundamental, and spinor indices can be lowered and raised with the symplectic form ${\epsilon_{AB} = -\epsilon_{BA}}$, as in $\psi_A = \epsilon_{AB} \psi^B$ and $\psi^A = \epsilon^{AB} \psi_B$.  To work with $SO(3, 2)$ spinors we define 5 real $\Gk^I$ matrices,\footnote{A specific set of matrices is presented in Appendix~\ref{sec:NORMSCONS}\@.}  and 10 group generators  $\Gk^{IJ} = [\Gk^I, \Gk^J]/2$.  When these matrices act as linear transformations, we use up/down indices, e.g. $(\Gk^I)^A{}_B$.  When they are bilinear forms, we lower or raise indices with the symplectic matrix, e.g $\Gk^I_{AB} = \epsilon_{AC} (\Gk^I)^C{}_B$.  As bilinear forms, the $\Gk^I_{AB}$ are anti-symmetric, and the $\Gk^{IJ}_{AB}$ are symmetric. We will find it convenient to introduce angle brackets to suppress $Sp(4)$ indices, defining for spinors $S^A, T^A$ and vectors $P^I_i$ the quantity:
\begin{equation}
\langle S P_1.... P_n T\rangle \equiv S^A\epsilon_{AB}{(\slashed P_1)^B}_C{(\slashed P_2)^C}_D\dots{(\slashed P_n)^E}_FT^F\,, \quad \text{ where } {(\slashed P_i)^B}_C = P^I{(\Gk_I)^B}_C\,.
\end{equation}

Having established these conventions, let us now turn to the definition of spinning fields in the bulk AdS$_4$ and the boundary $\R^{2, 1}$.

\subsection{Bulk}

To describe spinning fields in AdS$_4$ we use frame fields. These are necessary to describe spinor fields on curved manifolds, but as we will see are also convenient for bosonic fields. Recall that in the frame field formalism for general relativity, we rewrite the metric $g_{ij}(x)$ in terms of frame fields $e_i^\mu(x)$:
\begin{equation}
g_{ij}(x) = e_i^\mu(x) e_j^\nu(x) \eta_{\mu\nu}
\end{equation}
where $\eta_{\mu\nu}$ is the Minkowski metric, $\mu,\nu,\ldots = 0, \ldots, 3$ are the local Lorentz spacetime indices, and $x$ is any arbitrary point on our manifold. The metric is invariant under local $SO(3,1)$ transformations
\begin{equation}
e_i^\mu(x) \longrightarrow {\Lk^\mu}_\nu(x) e_i^\nu(x)\,,
\end{equation}
signifying that at each point on the manifold we are free to choose any orthonormal basis for the tangent space we like. 

Spinning fields living on a general 4d manifold transform covariantly under local $SO(3,1)$ transformations, or more correctly as representations of its double cover $SL(2;\mathbb C)$. We shall call this group the ``little group,'' because it plays the same role in our formalism as the little group plays in spinor helicity methods. To describe bulk fields we will use Weyl spinor indices $\ak, \da, \bk,\db,\ldots$. For example, $V_{\ak\da}(x) \equiv \sk_{\ak\da}^\mu V_\mu(x)$ is a bulk vector,\footnote{Our conventions for the sigma matrices $\sk^\mu_{\ak\da}$ are given in appendix \ref{sec:NORMSCONS}. In particular, with our conventions ${V_\mu = -\frac 12 \sigma_\mu^{\ak\da}V_{\ak\da}}$ and ${V_\mu V^\mu = -\frac12V_{\ak\da}V^{\ak\da}}$.} while $\psi_\ak(x)$ is a left-handed Weyl spinor and its conjugate $\bar\psi_{\dot\ak}(x)$ is a right-handed spinor. A Dirac fermion consists of two independent Weyl spinors. 

Now let us consider how to describe spinning fields living on AdS$_4$. We would like to think of fields, such as a vector field $V_{\ak\dot\ak}(X)$, as functions of embedding space vector $X^I$. But there is an obvious difficulty; $X^I$ carries an $SO(3,2)$ index but the vector field $V_{\ak\dot\ak}(X)$ carries an $SO(3,1)$ index. As mentioned in the Introduction, we resolve this mismatch by describing AdS$_4$ not with an embedding space vector $X^I$, but instead by a bispinor $T^A_\ak$ and its conjugate $\bar T^A_{\dot\ak}$ transforming covariantly under both $SO(3,2)$ and the little group $SO(3,1)$.   As in the spinor helicity formalism where one writes the momentum vector as a product of two spinor-helicity variables, we construct the vector $X^I$ as a product of bispinors
\begin{equation}\label{Xdef}
  X^I = - \frac 14 \langle T_\alpha \Gamma^I T^\alpha \rangle =  - \frac 14 \langle \bar T_{\da}  \Gamma^I \bar T^{\da} \rangle \,.
\end{equation}
This means that the bulk spacetime coordinates $X^I$ are determined by the $T$'s or $\bar T$'s; the converse is only true up to $SO(3,1)$ gauge transformations.  We therefore take the pair $T^A_\ak, \bT^A_\da$ as the basic variables in the bulk. For example, we can express the bulk vector field as $V_{\ak\da}(T,\bT)$. A general bulk field can be viewed as a symmetric multi-spinor $\Phi^{\ak_1 \ldots\ak_{m}}_{\da_1\ldots\da_{n}}(T,\bT)$. It transforms in the $(m/2,n/2)$ irreducible representation of $SO(3,1)$. 

The bispinor $T^A_\ak$ and its conjugate $\bT^A_{\da}$ contain sixteen real degrees of freedom, while a point in AdS$_4$ is described by only four. We eliminate the additional degrees of freedom in the following way:
\begin{enumerate}
	\item The $SO(3,1)$ little group is $6$ dimensional, and so this redundancy removes $6$ degrees of freedom.
	\item Four real degrees of freedom are removed by requiring that
	\begin{equation}\label{eq:TbTcond2}
	\langle T_\ak \bar T_{\dot\ak}\rangle = T^A_\ak\bar T^B_{\dot\ak}\epsilon_{AB} = 0\,.
	\end{equation}
	\item Finally, two degrees of freedom are removed by enforcing
	 \es{Proj3}{
	  \langle T_\ak T_\bk\rangle = T^A_\ak T_{\bk A} = 2 i\epsilon_{\ak\bk} \,, \qquad
	    \langle \bar T_\da \bar T_\db\rangle = \bar T^A_\da \bar T_{\db A} = -2 i\epsilon_{\da\db}
	   \,.
	   }
	The choice of normalization is arbitrary; our choice is such that using \eqref{Proj3} together with \eqref{Xdef} implies $X \cdot X = -1$, as appropriate for describing AdS$_4$ of unit curvature scale.
\end{enumerate}
From \eqref{Proj3} and \eqref{Xdef}, we can also derive that
 \es{GotX}{
  \slashed{X}^{AB}  = - T^A_\ak T^B_\bk\epsilon^{\ak\bk} + i \epsilon^{AB}
   = - \bar T^A_\da \bar T^B_\db \epsilon^{\da\db} - i \epsilon^{AB} \,.
 } 
It is furthermore straightforward to verify that
\begin{equation}\label{eq:XTEq}
\slashed X^A_{\ B}T^B_\ak = iT^A_\ak\,,\qquad \slashed X^A_{\ B}\bar T^B_{\dot\ak} = -i\bar T^A_{\dot\ak}\,.
\end{equation}

To find an explicit form for the $T^A_\alpha$ and $\bar T^A_{\dot \alpha}$, let us parametrize AdS$_4$ in Poincar\'e coordinates by writing 
\begin{equation}
X^I  = \frac 1 {z}\left(\vec x,\frac {1-\vec x^2-z^2}2,\frac {1+\vec x^2+z^2}2\right) \,,
\end{equation}
where $\vec x$ is a 3-vector and $z>0$ the radial coordinate which vanishes at the boundary of AdS$_4$.  Note that in this limit we can identify the $P^I$ as
\begin{equation}\label{eq:PForm}
P^I = \lim_{z\to 0}\, z X^I =  \left(\vec x,\frac{1-\vec x^2}2,\frac{1+\vec x^2}2\right)\,,
\end{equation}
up to the rescalings $P^I \sim \lk P^I$.  From \eqref{eq:XTEq}, one useful choice of explicit parametrization is
 \es{TTbarExplicit}{
T^A_\ak(\vec x,z) = \frac 1 {\sqrt{z}}\begin{pmatrix} 1 & 0\\ 0 & 1 \\ -x^0+x^1 & -i z-x^2 \\ i z-x^2 & -x^0-x^1 \\ \end{pmatrix}  \,, \qquad
 \bar T^A_\da(\vec x,z) = \frac 1 {\sqrt{z}}\begin{pmatrix} 1 & 0\\ 0 & 1 \\ -x^0+x^1 & i z-x^2 \\ -i z-x^2 & -x^0-x^1 \\ \end{pmatrix} \,.
 }

One can observe that the expressions \eqref{TTbarExplicit} represent the Killing spinors on AdS$_4$.  The Killing spinor equations obeyed by \eqref{TTbarExplicit} are\footnote{We show in Appendix~\ref{sec:KILLSPIN} that the equations \eqref{2kill} are equivalent to the more familiar Killing spinor equations in AdS$_4$.} 
 \begin{equation} \label{2kill}
\nabla_{\bk\db} T^C_\gk = i \epsilon_{\bk\gk} \bT^C_\db \,, \qquad \nabla_{\bk\db} \bT^C_\dg = -i \epsilon_{\db\dg} T^C_\bk \,.
\end{equation}
Killing spinors are widely used in supersymmetry and supergravity, but they are independently useful, and this is part of the reason why they appear here.

When acting on Killing spinors, the covariant derivative appearing in \eqref{2kill} can be written as
 \begin{equation}\label{nabada}
\nabla_{\ak\da} \equiv i\left(T^A_\ak\frac{\pa}{\pa \bT^{A\da}}-\bT^A_\da\frac{\pa}{\pa T^{A\ak}}\right)\,.
\end{equation}
It is not hard to check that this is the unique (up to normalization) first order differential operator preserving conditions \eqref{eq:TbTcond2} and \eqref{Proj3}. We have normalized $\Nb_{\ak\da}$ such that bulk Laplacian is $\Nb^2 \equiv -\frac12 \Nb^{\ak\da}\Nb_{\ak\da} = \Nb^\mu\Nb_\mu$. Since we can write any AdS$_4$ field as a function of $T$ and $\bar T$, we can use this expression to compute any covariant derivative.  In Appendix \ref{CONNECTION} we show how the conventional AdS$_4$ metric, frame field and spin connection can be computed in any coordinate system using our formalism; in particular the Poincar\'e patch results can be derived using the parametrization \eqref{TTbarExplicit} of $T$ and $\bT$. It is a remarkable and simplifying feature of our approach that when $\nabla_{\alpha \dot \alpha}$, as given in \eqref{nabada}, acts on any spinning field, all effects of the conventional spin connection are included.

Killing vectors are $\Gk$-matrix bilinears matrix of these spinors, as we now discuss.  The adjoint representation of $SO(3,2)$ can be identified with anti-symmetric $5\times 5$ matrices $M^{[IJ]}$. The
vectors $V_{\ak\da}^{[IJ]} = T^A_\ak \Gk_{AB}^{[IJ]}\bT^B_\da$ are Killing vectors.  To show this is trivial. One simply applies \eqref{2kill} which gives
\begin{equation}\label{symkill}
\nabla_{\ak\da}V_{\bk\db}^{[IJ]}+\nabla_{\bk\db}V_{\ak\da}^{[IJ]} =0 \,.
\end{equation}
It is even simpler from the viewpoint of $Sp(4)$, whose adjoint representation consists of the $4\times 4$ symmetric $N^{(AB)}$.   The vector $U^{(AB)}_{\ak\da} = T^{(A}_{\ak} \bT^{B)}_\da$ is a Killing vector;  its symmetric covariant derivative also vanishes.

In this section, we have derived the bispinor formalism by analogy to the spinor helicity formalism. We defined the embedding space vector $X^I$ in terms of bispinors $T^A_\ak$ and $\bT^A_\da$ which were little group covariant, and found that this required conditions \eqref{eq:TbTcond2} and \eqref{Proj3} in order remove spurious degrees of freedom and enforce $X\cdot X = -1$. In Appendix \ref{COSET} we present an alternative derivation of the bispinor formalism which uses the coset construct $AdS_4\approx Sp(4)/SL(2,\mathbb C)$. While this approach is more abstract, it has the advantage of generalizing more easily to other spacetime dimensions.

\subsection{Boundary}
The discussion above of bulk physics has an analogue for the boundary. The boundary theory is a conformal field theory, and local operators transform covariantly under representations of $SO(2,1)\times \mathbb R_+$ of Lorentz and scale transformations, which are the manifest symmetries in radial quantization. The group $SO(2,1)\times \mathbb R_+$ plays the role that the little group $SO(3,1)$ played in the previous section.\footnote{Using frame fields, we can define a conformal structure on a $3d$ manifold via fields $e_i^\mu(x)$, where $i$ is a tangent vector index and $\mu$ is a $SO(2,1)\times \mathbb R_+$ index. In this case the metric $g_{ij} = e_i^\mu e_j^\nu \eta_{\mu\nu}$ is now defined only up to local rescalings $g_{ij}(x) \rightarrow \lk(x)^2g_{ij}(x)$. Explicit formulas for the metric and frame field in our formalism can be found in Appendix \ref{sec:boundaryMet}.}

The basic spinor representation of $SO(2,1)$ is 2-dimensional and real, and we use indices $a,b,\ldots$ for these spinors. The analogue of the bulk $T^A_\ak, \bT^A_\da$ bispinors is a single, real, bispinor $S^A_a$ which satisfies the constraint
\begin{equation}\label{eq:SCons}
\langle S_a S_b\rangle = S^A_aS^B_b\epsilon_{AB} = 0\,,
\end{equation}
and transforms under a local scale transformation with parameter $\lk(S) \in \R_+$ as ${S^A_a \rightarrow \sqrt{\lk(S)} S^A_a}$. While $S^A_a$ has 8 degrees of freedom, the gauge symmetry $SO(2,1)\times \mathbb R_+$ removes four real degrees of freedom and \eqref{eq:SCons} removes one more. This leaves three degrees of freedom, precisely the right number we need to describe a boundary point.

Using $S^A_a$ we can define a null vector $P^I$ via
\begin{equation}\label{bdycons}
S^A_aS^B_b\epsilon^{ab} = -P^I\Gamma_I^{AB}\,,\qquad P^I = -\frac 14\Gamma^I_{AB}S^A_aS^B_b\epsilon^{ab}\,.
\end{equation}
In the parametrization \eqref{eq:PForm} we can compute $S^A_a(\bar x)$ as the boundary limit\footnote{We will discuss the boundary limit of bulk points in greater detail in Section \ref{BOUNDLIM}. In \eqref{eq:CFTSpinor} we are choosing a particular identification of the bulk and boundary frame fields, which will not be preserved under frame field rotations. In spite of this issue, \eqref{eq:CFTSpinor} provides a convenient parametrization of the boundary.} of $T^A_\ak(\bar x,z)$:
\begin{equation}\label{eq:CFTSpinor}
S^A_a(\bar x) = \lim_{z\to 0}\,\frac{\sqrt z}2\left(T^A_a(\bar x,z) + \bar T^A_a(\bar x,z)\right) = \begin{pmatrix} 1 & 0\\ 0 & 1 \\ -x^0+x^1 & -x^2 \\ -x^2 & -x^0-x^1 \\ \end{pmatrix}\,.
\end{equation}
For each fixed $A$, the bispinor $S^A_a$ is a conformal Killing spinor.  In general, a conformal Killing spinor $\Sk$ obeys the equation $\Nb_\mu \Sk = \gamma_\mu \hat\Sk$, where $\hat\Sk = \frac 13 \slashed{\Nb} \Sk$ is another spinor.  We see that in flat space with the standard frame (with vanishing spin connection), the first two rows of \eqref{eq:CFTSpinor} obey the conformal Killing spinor equation with vanishing $\hat\Sk$, while the last two rows obey it with constant $\hat\Sk$. In Appendix \ref{sec:boundaryMet} we show that the $S^A_a$ are conformal Killing spinors for any conformally flat boundary metric.

The relations \eqref{bdycons} tell us that we can regard the bi-spinor $S^A_a$ as the basic descriptor for boundary operators in embedding space, rather than $P^I$. A general CFT$_3$ operator of spin $\ell$ and conformal dimension $\Dk$ transforms as a symmetric rank $2\ell$ spinor $\co_{a_1\dots a_{2\ell}}(S)$ obeying
 \es{LamScaling}{
\co_{a_1\dots a_{2\ell}}(\sqrt{\lk} S) = \lk^{\Dk} \co_{a_1\dots a_{2\ell}}(S)\,.
 }

\subsection{Polarized operators}

As is common in the literature, it is very convenient to polarize multi-rank expressions in order to avoid proliferation of indices. On the boundary we use the constant $s^a$ transforming in the $SL(2;\mathbb R)$ fundamental, and we can then define
 \es{OsDef}{
\fcy O(s,S) \equiv s^{a_1}\dots s^{a_{2\ell}}\co_{a_1\dots a_{2\ell}}(S)\,.
 }
To recover the indices we simply differentiate with respect to $s^a$. Since $\fcy O(s,S)$ is an $SL(2;\mathbb R)$ singlet it is not hard to see that it depends on $s^a$ and $S^A_a$ only through the $SL(2;\mathbb R)$ invariants $P^I$ and $\sk^A \equiv s^a S^A_a$, which automatically satisfies
\begin{equation}{\slashed P^A}_B\sk^B = 0 \,.  \end{equation}
Thus, we may write $\co(\sk,P)$ in place of $\co(s,S)$ if we wish.  Combining \eqref{LamScaling} and \eqref{OsDef}, little group invariance now reduces to the condition
\begin{equation}\label{eq:boundscal}
\cO(\mu_1 \sk,\mu_2 P) = \mu_1^{2\ell}\mu_2^{-\Dk-\ell}\cO(\sk,P)\,,
\end{equation}
so that it is now straightforward to construct all possible conformally invariant structures. For instance, the unique (up to our choice of normalization) two-point function for a spin-$\ell$ operator is
\begin{equation}\label{eq:scalarNorm}
\langle\co(\sk_1,P_1)\co(\sk_2,P_2)\rangle = 2^{3-2\Dk}\pi^2\Gk(2\Dk-1)\frac{\langle\sk_1\sk_2\rangle^{2\ell}}{(-2P_1\cdot P_2)^{\Dk+\ell}}
\end{equation}

In the bulk we frequently polarize little group indices using $t^\ak$ or $\bar t^\da$, defining
\begin{equation} 
\varphi(t,\bar t,T,\bT) = t^{\ak_1}\dots t^{\ak_m} \bar t^{\da_1}\dots \bar t^{\da_n}\varphi_{\ak_1\ldots\ak_m\da_1\ldots\da_n}(T,\bT).
\end{equation}
Defining the $SO(3,1)$ invariants
\begin{equation}
\tk^A= t^\ak T^A_\ak\,,\qquad \bar \tk^A = \bar t^\da \bT^A_\da\,,
\end{equation}
we then find that the field $\varphi(t,\bar t,T,\bT) \equiv \varphi(\tk,\bar\tk,X)$ depends only on $\tk^A$, $\bar\tk^A$, and $X^I$, and satisfies
\begin{equation}\label{eq:bulkscal}
\varphi(\mu\tk,\nu\bar\tk,X) = \mu^m \nu^n \varphi(\tk,\bar\tk,X)\,.
\end{equation}

\subsection{Bulk-boundary propagators}
\label{sec:bulkbound}

The main applications of our formalism are Witten diagram computations in AdS/CFT\@.  In top-down constructions of AdS$_4$/CFT$_3$, one always has a weakly-coupled gravitational theory (or higher spin theory) in AdS$_4$, and boundary correlation functions are computed via Witten diagrams, perturbatively in the Newton constant.  An important quantity in Witten diagram computations is the bulk-boundary propagator $G_{B \partial}(P; X)$, which, for a boundary operator $\cO(P)$ dual to a bulk field $\phi(X)$ quantifies how a delta function insertion of $\cO(P)$ in the boundary CFT sources the dual bulk field $\phi(X)$.    As will become clear, the bulk-boundary propagator in our formalism is fixed by conformal symmetry up to an overall normalization constant.  For example, for a scalar boundary operator of scaling dimension $\Delta$ dual to a bulk scalar field of mass $m^2 = \Delta(\Delta - 3)$, it takes the form
 \es{BulkBdryScalar}{
   G_{B \partial}(P; X) \equiv \langle \cO(P) \phi(X) \rangle   =  \frac{\cN}{(-2 P \cdot X)^\Delta}    \,,
 }
where $\cN$ is a normalization constant.   In \eqref{BulkBdryScalar}, we denoted the bulk-boundary propagator as a two-point function $\langle \cO(P) \phi(X) \rangle$, which is the notation we will use from now on.  It can be justified because one way to compute CFT correlators in AdS/CFT is to first compute bulk correlation functions in the effective theory in AdS and then taking the bulk points to the boundary.  (See, for example, \cite{Penedones:2016voo} and Section~\ref{BOUNDLIM} below.)  In this framework, the bulk-boundary propagator in \eqref{BulkBdryScalar} can be viewed as the limit of the two-point function of two bulk operators when one of these two operators is taken to the boundary.\footnote{One should note that, in AdS/CFT, the bulk theory always contains gravity, so bulk operators (and consequently bulk correlators and the bulk-boundary propagator) are not gauge-invariant under bulk diffeomorphisms.  However, the boundary limits of the bulk operators are gauge-invariant, so one can compute the CFT correlators by first computing the bulk correlators or the bulk-boundary propagator in a specific diffeomorphism gauge, and then taking their boundary limits.  Going beyond AdS/CFT, the same framework applies to a QFT in AdS for which one can define boundary observables by simply taking the bulk operators to the boundary.  The difference between this case and the usual AdS/CFT set-up where the bulk theory contains gravity is that for a QFT in AdS, the bulk operators are now well-defined observables.\label{DiffeoFootnote}}

For operators with spin, let us consider the bulk-boundary propagator $\<\co(\sk,P)\varphi(\tk,\bar\tk,X)\>$ between a bulk field $\varphi$  and its CFT dual operator $\co$.  The simplest type is the propagator between the bosonic field $\varphi_{\ak_1\dots\ak_\ell;\,\da_1\dots\da_\ell}(X)$ and its dual operator $\co_{a_1\dots a_{2\ell}}$ of integer spin $\ell$ and weight $\Dk$.  The scaling relations \eqref{eq:boundscal} and \eqref{eq:bulkscal} fix this bulk-boundary propagator (up to normalization) to take the form
\begin{equation}\label{eq:bbProp1}
\<\cO(\sk,P)\varphi(\tk,\bar\tk,X)\> = \fcy N_{\cO \varphi} \frac{\<\sk\tk\>^\ell\<\sk\bar\tk\>^\ell}{(-2 P\cdot X)^{\Dk+\ell}}\,.
\end{equation}
It is not hard (and a good exercise!) to check that its bulk divergence vanishes
\begin{equation}\label{eq:bbDiv}
\nabla^{\ak\da}\frac{\pa}{\pa t^\da}\frac{\pa}{\pa \bar t^\da} \<\cO(\sk,P)\varphi(\tk,\bar\tk,X)\> = 0\,.
\end{equation} 
One can understand \eqref{eq:bbDiv} as a consequence of the bulk field equations for a massive field, but since all we have used to derive it is conformal invariance, it in fact holds more generally in any quantum field theory in AdS$_4$.  

For massless fields with $\ell \geq 1$, the vanishing of the divergence instead corresponds to a gauge choice, and, while other gauge choices are possible, these gauge choices are not $SO(3,2)$ invariant.  A massless field is dual to a spin-$\ell$ conserved current (of conformal dimension $\Dk=\ell+1$) on the boundary.  We will discuss in Section~\ref{sec:FurtherDOs} how the conservation condition is implemented in our formalism. It turns out that the boundary divergence of  $\<\cO(\sigma,P)\varphi(\tk,\bar\tk,X)\>$ does not vanish. Instead, it is compensated by a bulk gauge transformation, and this leads to Ward identities for Witten diagrams, as will be discussed in Appendix~\ref{sec:wards}.

It is easy to generalize this discussion to any bulk-boundary propagator between a bulk field $\varphi$ with spin $(m,n)$, and boundary field $\fcy O$ of spin $\ell$:
\begin{equation}\label{eq:bulkBound}
\langle \fcy O(\sk,P) \varphi(\tk,\overline \tk,X)\rangle = \fcy N_{\cO \varphi} \frac{\langle \tk\sk\rangle^{\ell+m-n}\langle \overline \tk\sk\rangle^{\ell-m+n}\langle \tk P\overline \tk\rangle^{m+n-\ell}}{(-2X\cdot P)^{\Delta + m+n}}\,.
\end{equation}
The right-hand expression only exists if each of the three angle brackets appears with non-negative integer power. This implies that $\langle\fcy O\varphi\rangle$ can be non-zero only if both 
\begin{equation}\label{eq:bboundCond}
{m+n\geq \ell \geq |m-n|}\quad \text{and}\quad 2\ell \equiv 2(m+n) \mod 2\,.\end{equation}
We should emphasise that because \eqref{eq:bulkBound} is fixed purely by conformal invariance, it holds for all operators in any quantum field theory in AdS$_4$---see Footnote~\ref{DiffeoFootnote}. In particular, it holds when $\varphi$ can be written as the derivative of another bulk field.

To gain a more intuitive understanding of \eqref{eq:bulkBound}, let us define $L = m+n$ to be the total spin of a bulk field and $H = m-n$ to be the ``handedness'' of the operator. When we differentiate a bulk field, we can shift either $L$ or $H$ by $\pm 1$, depending on how we choose to contract the derivative indices. So for example, a bulk vector field $A_{\ak\da}$ has $L = 1$ and $H = 0$. We can decompose the derivative of the field, $\Nb_{\ak\da}A_{\bk\db}$, into four irreducible $SO(3,1)$ representations
\begin{equation}
\Nb^{(\ak}{}_{(\da}A^{\bk)}{}_{\db)}\,,\qquad \Nb_{\ak\dot\ak}A^{\ak\dot\ak}\,,\qquad \Nb_{(\ak}{}^\da A_{\bk)\da}\,,\qquad \Nb_{\ak(\da} A^\alpha{}_{\db)}\,.
\end{equation}
For the first and second operators $H$ remains zero but $L$ shift to $2$ or $0$ respectively, while for the last two operators $L = 1$ but $H = \pm1$. Shifting $L$ changes the power of $\<\tk P\bar\tk\>$ appearing in the bulk-boundary propagator, while shifting $H$ changes the relative powers of $\<\sk\tk\>$ and $\<\sk\bar\tk\>$. In particular taking the divergence of a bulk field reduces $L$ by one, and so reduces the power of $\<\tk P\bar\tk\>$ by one. If our initial field had $L = \ell$, the bulk divergence has total spin $L - 1$, which would imply a negative power of $\<\tk P\bar\tk\>$ in the bulk-boundary propagator. Since this is impossible, the bulk divergence of a bulk-boundary propagator with $L = \ell$ must vanish. The equation \eqref{eq:bbDiv} is a special case of this more general result.

Since the bulk (connection) Laplacian $\Nb^2 \equiv \Nb^\mu\Nb_\mu = -\frac 12 \Nb^{\ak\da}\Nb_{\ak\da}$ is a Lorentz scalar, it follows from \eqref{eq:bulkBound} that $\langle\fcy O\varphi\rangle$ and $\langle\fcy O\Nb^2\varphi\rangle$ must be proportional to each other. Indeed, using \eqref{nabada} it is not hard to check that
\begin{equation}\label{eq:BBEoM}\begin{split}
\langle \fcy O(\sk,P) (\Nb^2 - \lk) \varphi(\tk,\overline \tk,X)\rangle & = 0\,,\\
 \text{where}\quad \lk &= \Dk(\Dk-3)+\ell(\ell+1)-2m(m+1)-2n(n+1)\,.
\end{split}\end{equation}
For bulk fermion field $\yk^{\ak\bk_1\dots\bk_n\db_1\dots\db_n}(T,\bT)$ and its conjugate $\bar\yk^{\bk_1\dots\bk_n\dot\ak\db_1\dots\db_n}(T,\bT)$ coupled to a real fermionic operator on the boundary $\Yk^{a_1\dots a_{2\ell}}(S)$, a similar computation allows us to deduce that
\begin{equation}\label{eq:fermEoM}\begin{split}
\Bigg\<\Yk(\sk,P)&\left(i{\Nb_\ak}^{(\da|}\yk^{\ak\bk_1\dots\bk_n|\db_1\dots\db_n)}(T,\bT)-\mu\bar\yk^{\bk_1\dots\bk_n\dot\ak\db_1\dots\db_n}(T,\bT)\right)\Bigg\> = 0\\
 \text{ where }&\quad \mu = e^{i\theta}\frac{(\Dk+\ell-2)(2\ell+1)}{2(n+1)}\,, \quad e^{i\qk} = \frac{\fcy N_{\Yk\psi}}{\fcy N_{\Yk\psi}^*}\,.
\end{split}\end{equation}
By performing a field redefinition $\yk\rightarrow e^{i\qk/2}\yk$ we can set the phase $\qk = 1$ if we wish.

We again emphasise that these equations are satisfied by the bulk-boundary propagator for arbitrary theories in AdS$_4$. For the special case that these field are free fields, however, we can use these equations to relate the mass of the bulk field to the conformal dimension of the boundary field. So for instance, a free scalar field $\phi$ satisfies the equation of motion $(\Nb^2-m^2)\phi = 0$, and so from \eqref{eq:BBEoM} we see that $m^2 = \Dk(\Dk-3)$.

\section{Differential operators}
\label{DIFFOPS}

In this section we study differential operators which may act on either bulk fields or boundary operators. We have already met one such operator, the bulk covariant derivative $\Nb_{\ak\da}$, which as we shall see can be related to the bulk conformal and little group generators. On the boundary, differential operators can be used to shift both the spin and conformal dimension of operators, allowing us to construct ``weight-shifting'' operators for Witten diagrams and conformal blocks.

\subsection{Bulk symmetry generators}
The adjoint representation of $SO(3,2)$ can be described as symmetric matrices $M^{AB}$ with spinor indices, or alternatively, as antisymmetric matrices $M^{IJ}$ with vector indices. An infinitesimal conformal transformation $\lk^{AB}$ acts on spinors $T^A_\ak$ and $\bar T^A_\da$ as:
\begin{equation}
\delta_{\lambda} T^A_\ak = \lambda^{AB}T_{B\ak} = -i\lambda^{BC}D_{BC}T^A_\ak\,,\qquad \dk_\lk\bar T^A_\da = \lambda^{AB}\bT_{B\da}  = -i\lambda^{BC}D_{BC}\bT^A_\da\,,
\end{equation}
where we define the differential operator
\begin{equation}\label{eq:bulkconf}
D_{BC} \equiv i\left(T_{(B|\ak}\frac{\nb}{\nb T^{|C)}_\ak} +\bT_{(B|\da}\frac{\nb}{\nb \bT^{|C)}_\da}\right)\,.
\end{equation}
The $D_{AB}$ satisfy the $Sp(4)$ commutation relations
 \es{DCommutators}{
[D_{AB},D_{CD}] = -\frac i 2\left(\epsilon_{AC}D_{BD}+\epsilon_{AD}D_{BC}+\epsilon_{BC}D_{AD}+\epsilon_{BD}D_{AC}\right)\,.
 }

We can likewise implement infinitesimal $\mathfrak{so}(3,1)\approx\mathfrak{sl}(2)\times\mathfrak{sl}(2)$ little group transformations by the differential operators:
\begin{equation}\label{eq:LDef}
L_{\ak\bk}\equiv -i T^A_{(\ak|}\frac{\nb}{\nb T^{A|\bk)}}\,,\qquad \bar L_{\dot\ak\dot\bk}\equiv -i\bT^A_{(\da|}\frac{\nb}{\nb \bT^{A|\db)}}  \,.
\end{equation}
When they act on a bulk field $\varphi_{\ak_1\dots\ak_{m}\dot\ak_1\dots\dot\ak_{n}}$ they simply rotate the little group indices:
\begin{equation}\begin{split}\label{eq:littleAct2}
L_{\bk\gk}\varphi_{\ak_1\dots\ak_{m}\dot\ak_1\dots\dot\ak_{n}}(T,\bar T) &= -i\sum_{k = 1}^{m} \epsilon_{\ak_k(\bk|}\varphi_{\ak_1\dots |\gk)\dots \ak_{m}\dot\ak_1\dots\dot\ak_{n}}(T,\bT)\,, \\
\bar L_{\db\dot\gk}\varphi_{\ak_1\dots\ak_{m}\dot\ak_1\dots\dot\ak_{n}}(T,\bar T) &= -i\sum_{k = 1}^{n} \epsilon_{\da_k(\db|}\varphi_{\ak_1\dots\ak_{m}\dot\ak_1\dots|\dot\gk)\dots\dot\ak_{n}}(T,\bT)\,.
\end{split}\end{equation}

Given our definitions for the covariant derivative \eqref{nabada}, conformal generator \eqref{eq:bulkconf}, and little group generators \eqref{eq:LDef}, it is straightforward to check that
\begin{equation}\label{eq:confgensnab}
D_{AB} = \frac i2 T_{(A}^\ak\bar T_{B)}^{\dot\ak}\Nb_{\ak\dot\ak} - \frac i2\left(T_A^\ak T_B^\bk L_{\ak\bk} - \bar T_A^{\dot\ak}\bar T_B^{\dot\bk}L_{\dot\ak\dot\bk}\right)\,.
\end{equation}
This equation states that the conformal variation of some operator $\varphi(T,\bT)$ can be realized as the sum of a translation and a little group rotation. Contracting both sides with $T^A_\ak$ and/or $\bT^A_\da$, we then find that
\begin{equation}\label{eq:nabconfgen}
\Nb_{\ak\dot\ak} = -i T_\ak^{A}\bar T_{\dot\ak}^{B}D_{AB}\,,\qquad L_{\ak\bk} = -\frac i2 T^A_\ak T^B_\bk D_{AB}\,,\qquad \bar L_{\da\db} = \frac i2 \bT^A_\da \bT^B_\db D_{AB}\,.
\end{equation}

From  \eqref{DCommutators}, we can determine the commutators of the conformal generators $D_{AB}$ with the operators in \eqref{eq:nabconfgen},
\begin{equation}\label{DcomV}
[D_{AB},\Nb_{\ak\da}] = [D_{AB},L_{\ak\bk}] = [D_{AB},\bar L_{\da\db}] = 0\,,
\end{equation}
as well as the commutator of two covariant derivatives: 
\begin{equation}
[\Nb_{\ak\da},\Nb_{\bk\db}] = -i[\Nb_{\ak\da},T_\bk^{A}\bar T_{\dot\bk}^{B}D_{AB}] = -i\left([\Nb_{\ak\da},T_\bk^{A}]\bT_\db^B+T_\bk^{A}[\Nb_{\ak\da},\bT_\db^B]\right)D_{AB}\,. 
\end{equation}
The commutator of $\Nb_{\ak\da}$ with $T^A_\ak$ and $\bT^A_\da$ can be found from \eqref{2kill}, allowing us to deduce that
\begin{equation}\label{Nbcom}
[\Nb_{\ak\da},\Nb_{\bk\db}] = -i\left(i\epsilon_{\ak\bk}\bT_\da^A\bT_\db^B-i\epsilon_{\da\db}T_\bk^{A}T_\ak^B\right)D_{AB} = -2i\left(\epsilon_{\da\db}L_{\ak\bk} + \epsilon_{\ak\bk}\bar L_{\da\db}\right) \,.
\end{equation}
Applying this identity to spinor fields $\yk_\ak$ and $\bar\yk_\da$, we can compute
\begin{equation}
[\Nb_{\ak\da},\Nb_{\bk\db}]\yk_\gk = -2\epsilon_{\da\db}\epsilon_{\gk(\ak}\yk_{\bk)}\,, \qquad [\Nb_{\ak\da},\Nb_{\bk\db}]\bar\yk_{\dot\gamma} = -2\epsilon_{\ak\bk}\epsilon_{\dot\gk(\da}\yk_{\db)}\,.
\end{equation}
Rewriting the spinor indices as vector indices, we find that
\begin{equation}
[\Nb_\mu,\Nb_\nu]\yk_\gk = \frac 12 (\sk_{[\mu})_{\gk\da}(\sk_{\nu]})^{\bk\da}\yk_\bk\,,\qquad [\Nb_\mu,\Nb_\nu]\bar\yk_\dg = \frac 12 (\sk_{[\mu})_{\ak\dot\gk}(\sk_{\nu]})^{\ak\db}\bar\yk_\db\,,
\end{equation}
and hence conclude that the Riemann tensor on AdS$_4$ is\footnote{Using the conventions for Dirac fermions introduced in Appendix \ref{sec:KILLSPIN}, one can further check that $[\Nb_\mu,\Nb_\nu]\Yk = -\frac12[\gk_\mu,\gk_\nu]\Yk$, where $\Yk$ is a Dirac spinor and $\gk_\mu$ are the 4d gamma matrices.}
\begin{equation}
(R_{\mu\nu})_\ak{}^\bk = \frac 12 (\sk_{[\mu})_{\ak\da}(\sk_{\nu]})^{\bk\da}\,, \qquad (\bar R_{\mu\nu})_\da{}^\db = \frac 12 (\sk_{[\mu})_{\ak\da}(\sk_{\nu]})^{\ak\db}\,.
\end{equation}

\subsection{Boundary symmetry generators}
Symmetries on the boundary act analogously. An infinitesimal conformal transformation $\lk^{AB}$ acts on a spinor $S^A_a$ as:
\begin{equation}\begin{split}
\delta_{\lambda} S^A_a = \lambda^{AB}S_{Ba} = -i\lk^{BC}\fcy D_{BC}S^A_a\,,\quad 
\text{ where }\quad \fcy D_{AB} = iS_{(A|a}\frac{\nb}{\nb S^{|B)}_a}\,.
\end{split}\end{equation}

We can also write the generators of the $GL(2;\mathbb R)\approx SL(2;\mathbb R)\times \mathbb R$ little group:
\begin{equation}\label{eq:lilGens}\begin{aligned}
\fcy L_{ab} &= -i S^A_{(a|}\frac{\nb}{\nb S^{A|b)}}\,,\qquad &\fcy D &= -\frac12 S^A_a\frac{\nb}{\nb S^A_a} \,,\\
\end{aligned}\end{equation}
where $\fcy L_{ab}$ generates $SL(2;\mathbb R)$ and $\fcy D$ generates $\mathbb R$. On an operator $\fcy O^{a_1\dots a_{2\ell}}(S)$ with conformal dimension $\Delta$, these act as:
\begin{equation}
\fcy L_{bc}\fcy O_{a_1\dots a_{2\ell}}(S) = -i\sum_{k = 1}^{2\ell} \epsilon_{a_k(b|}\fcy O_{a_1\dots |c)\dots a_{2\ell}}(S)\,,\qquad \fcy D\fcy O_{a_1\dots a_{2\ell}}(S)  = \Delta\fcy O_{a_1\dots a_{2\ell}}(S) \,.
\end{equation}
For a polarized operator,
\begin{equation}\label{eq:littleAct}
\fcy L_{ab}s^b\fcy O(\sk,P) = -i\ell s_a\fcy O(\sk,P) \,,\qquad \fcy D\fcy O(\sk,P) = \Delta \fcy O(\sk,P)\,.
\end{equation}

With this new technology, we can rederive \eqref{eq:BBEoM} in a more abstract fashion, using the conformal Casimir. Because the bulk-boundary propagator is conformally invariant, it follows that
\begin{equation}\label{eq:ConfInv}
(\fcy D_{AB}+D_{AB})\langle\fcy O(S)\varphi(T,\bar T)\rangle = 0
\end{equation}
and hence that
\begin{equation}\label{eq:casEquation}
(\fcy D_{AB}\fcy D^{AB}-D_{AB}D^{AB})\langle\fcy O(S)\varphi(T,\bT)\rangle = 0\,.
\end{equation}
Using \eqref{eq:confgensnab}, we find that for the bulk Casimir operator $D_{AB}D^{AB}$ can be rewritten as
\begin{equation}
D_{AB}D^{AB} = \Nb^2-L^{\ak\bk}L_{\ak\bk} - \bar L^{\da\db}\bar L_{\da\db}\,.
\end{equation}
From this we can deduce that
\begin{equation}\label{eq:casbulk}
D_{AB}D^{AB}\varphi(T,\bT) = \left[\Nb^2+2m(m+1)+2n(n+1)\right]\varphi(T,\bT)\,,
\end{equation}
where $(m,n)$ is the $SO(3, 1)$ spin of $\varphi(T,\bT)$. On the boundary after some work we find that the Casimir is given by
\begin{equation}
\fcy D_{AB}\fcy D^{AB} = \fcy D\left(\fcy D + 3\right) + \frac12\fcy L^{ab}\fcy L_{ab}
\end{equation}
so that when acting on $\fcy O(S)$ with spin $\ell$ and conformal dimension $\Dk$,
\begin{equation}\label{eq:casOp}
\fcy D_{AB}\fcy D^{AB} \fcy O(S) = \left[\Dk(\Dk-3)+\ell(\ell+1)\right]\fcy O(S)\,.
\end{equation}
Substituting \eqref{eq:casbulk} and \eqref{eq:casOp} into \eqref{eq:casEquation} gives us \eqref{eq:BBEoM}.

\subsection{Other differential operators}
\label{sec:FurtherDOs}

The boundary symmetry generators $\fcy D_{AB}$, $\fcy D$, and $\fcy L$ do not exhaust the full pool of differential operators available to us. Other differential operators, and in particular the weight-shifting operators introduced in \cite{Karateev:2017jgd}, have proven useful when studying spinning operators in CFTs, and so in this section we shall find expressions for such differential operators in the bispinor formalism.

Before discussing other operators, however, we should first discuss why we are not free to simply differentiate with respect to $S^A_a$. This is not allowed because conformal correlators are only defined for bispinors $S^A_a$ satisfying the condition $\langle S_aS^a\rangle = 0$. As a consequence, we should treat any two functions $F$ and $G$ satisfying
\begin{equation}\label{eq:gaugeCon}
F(S) - G(S) = \langle S_aS^a\rangle f(S) \,,
\end{equation}
for an arbitrary function $f(S)$, as physically equivalent. But if we differentiate \eqref{eq:gaugeCon} we find that
\begin{equation}
\frac\nb{\nb S^A_a}\left(\langle S_bS^b\rangle f(S)\right) \bigg|_{\langle S_aS^a\rangle = 0} = 2S_A^af(S) \neq 0\,,
\end{equation}
and so $\frac\nb{\nb S^A_a}$ is not a well-defined operator. To resolve this problem, we should consider only differential operators $\fcy G$ satisfying
\begin{equation}\label{eq:diffidCond}
\fcy G\left(\langle S_aS^a\rangle f(S)\right)\Big|_{\langle S_aS^a\rangle = 0} = 0
\end{equation}
for arbitrary $f(S)$. We will call such an operator a conformally-covariant differential operator.\footnote{Similar considerations imply that bulk differential operators should preserve the conditions ${\langle T_\ak T_{\bk}\rangle = 2i\epsilon_{\ak\bk}}$ and $\langle T_\ak \bar T_{\da}\rangle = 0$. It is not hard to check that this is indeed the case for all of the bulk operators we have considered so far, $\Nb_{\ak\da}$, $D_{AB}$, $L_{\ak\bk}$, and $\bar L_{\da\db}$.}

The first such differential operator we shall consider is
\begin{equation}\label{divIab}
\nb_{ab}^I \equiv -S^A_{(a|}(\Gk^I)_{AB}\frac{\nb}{\nb S^{B|b)}} \,,
\end{equation}
which transforms as both an $SO(3,2)$ and $SO(2,1)$ vector. It is straightforward to check that
\begin{equation}
\nb_{ab}^I \langle S_cS^c\rangle = 0\,,
\end{equation}
and from this it then follows that $\nb_{ab}^I$ satisfies \eqref{eq:diffidCond}. It is also easy to see that
\begin{equation}
P_I\nb^I_{ab} = 0\,.
\end{equation}
In appendix \ref{sec:boundaryMet} we show that $\nb_{ab}^I$ is closely related to the boundary covariant derivative. In particular, in embedding space the conservation condition becomes
\begin{equation}\label{eq:conscon}
\nb_{a_1a_2}^IJ^{a_1a_2a_3\dots a_{2\ell}}(S) = 0\,,
\end{equation}
which holds for operators satisfying $\Dk = \ell+1$ and $\ell\geq 1$. We show in Appendix \ref{sec:boundaryMet} that this is equivalent to imposing the conservation condition
\begin{equation}
\Nb_{i_1}J^{i_1\dots i_\ell}(x) = 0
\end{equation}
on every conformally flat metric, where the $i_k$ indices are tangent and cotangent indices.

Another differential operator we will find very useful is
\begin{equation}\label{eq:fcyE}\begin{split}
\fcy E^A_a = \frac 12 S^A_a \frac{\nb}{\nb S^B_b}\frac{\nb}{\nb S_B^b} +\left(\fcy D-\frac32\right)\frac{\nb}{\nb S_A^a}\,.
\end{split}\end{equation}
The first term in \eqref{eq:fcyE} annihilates any arbitrary function $f(P)$ which depends on $S^A_a$ only through $P^I$, and so:
\begin{equation}\label{eq:EonScal}
\fcy E^A_af(P) = (\Dk-1)\frac{\nb}{\nb S^a_A}f(P)
\quad \text{ if } f(P) \text{ satisfies }f(\lk P) = \lk^{-\Delta}f(P)\,.\end{equation}
We can thus think of $\fcy E^A_a$ as a conformally covariant version of $\frac{\nb}{\nb S^A_a}$.

Although our introduction of the operator $\fcy E^A_a$ may seem a little \emph{ad hoc}, its importance is underlined by the following two results:
\begin{enumerate}
\item The space of linear operators spanned by $S^A_a$, $\fcy E^A_a$, $\fcy D^{AB}$, $\fcy D-\frac 32$, $\fcy L_{ab}$ and $\nb_{ab}^I$, when considered as a Lie algebra, is isomorphic to $\mathfrak{so}(5,5)$. The subspace generated by $\fcy D^{AB}$, $\fcy L_{ab}$ and $\nb_{ab}^I$ is isomorphic to $\mathfrak{so}(4,4)$.
\item Any conformally-covariant differential operator can be constructed as sums and products of $\fcy E^A_a$ and $S^A_a$.
\end{enumerate}
We will prove both of these facts in Appendix \ref{sec:diffOpAp}. Here we will focus on commutation relations, which will prove useful both for understanding the algebraic structure of the differential operator, and also for computing Witten diagrams in the next section. The symmetry generators and $\nb_{ab}^I$ appear in the commutator of $\fcy E^A_a$ with $S^A_a$:
\begin{equation}\label{opComs0}
[\fcy E^A_a,S^B_b] = -\frac i2\epsilon_{ab}\fcy D^{AB} + \epsilon_{ab}\epsilon^{AB}\left(\fcy D-\frac32\right) + \frac i4\epsilon^{AB}\fcy L_{ab} + \frac14 (\Gk_I)^{AB}\nb_{ab}^I\,.
\end{equation}
By contracting both sides of this identity with $Sp(4)$ and $SL(2;\mathbb R)$ invariants we can express each of the differential operators on the right-hand side purely in terms of $\fcy E^A_a$ and $S^A_a$. It is also straightforward to check that:
\begin{equation}\label{opComs}\begin{aligned}
&[S^A_a,S^B_b] = 0\,, \qquad &[\nb_{ab}^I,S^A_c] = (\Gk^I)^A{}_B\epsilon_{c(a}S^B_{b)}\,,\\
&[\fcy E^A_a,\fcy E^B_b] = 0\,,\qquad &[\nb_{ab}^I,\fcy E^A_c] = (\Gk^I)^A{}_B\epsilon_{c(a}\fcy E^B_{b)}\,,\\
\end{aligned}\end{equation}
$$
[\nb^I_{ab},\nb^J_{cd}] = -i \delta^{IJ}\left(\epsilon_{a(c}\fcy L_{d)b}+\epsilon_{b(c}\fcy L_{d)a}\right) -i\epsilon_{a(c}\epsilon_{d)b}(\Gk^I\Gk^J)^{AB}\fcy D_{AB}\,.
$$

To finish, let us relate the operators $S^A_a$ and $\fcy E^A_a$ to the fundamental weight-shifting operators given in \cite{Karateev:2017jgd}. Consider some primary operator $\fcy O_{a_1\dots a_{2\ell}}$ with spin $\ell$ and conformal dimensional $\Dk$. By acting with $S^A_a$ and then symmetrizing or antisymmetrizing the little group indices, we can construct operators of spin $\ell\mp\frac12$ and dimension $\Dk-\frac12$. In polarized notation, we can express this construction as the weight-shifting operators:
\begin{equation}\begin{split}
\fcy W^{--}_A[\fcy O(\sk,P)] &= \frac{1}{2\ell}S_A^a\frac{\nb}{\nb s^a}\fcy O(\sk,P) = \frac{-1}{2\ell}\slashed P_A{}^B\frac{\nb}{\nb\sk^B}\fcy O(\sk,P)\,,\\
\fcy W^{-+}_A[\fcy O(\sk,P)]  &= s_a S_A^a \fcy O(\sk,P) = -\sk_A\fcy O(\sk,P)\,.
\end{split}\end{equation}
Up to an overall normalization, these are precisely the first two fundamental weight-shifting operators in (2.71) of \cite{Karateev:2017jgd}. We can similarly act with $\fcy E^A_a$ to construct the other two fundamental weight-shifting operators that increase the scaling dimension by $1/2$:
\begin{equation}\begin{split}
\fcy W^{+-}_A[\fcy O(\sk,P)] &= \frac{1}{2\ell}\fcy E^A_a\frac{\nb}{\nb s_a}\fcy O(\sk,P)\\
&= \frac{-1}{8\ell}\Bigg[-4(\Dk-1)(1+\ell-\Dk)\frac{\nb}{\nb\sk^A} + 2(1+\ell-\Dk)\slashed P_{AB}(\Gk^I)^B{}_C\frac{\nb}{\nb\sk_C} \\
&\ \ \ \ \ \ \ \ \ \  \ -  \sk_A\slashed P_{BC}(\Gk^I)^C{}_D\frac{\nb}{\nb\sk_B}\frac{\nb}{\nb\sk_D} \Bigg]\fcy O(\sk,P)
\,,\\
\fcy W^{++}_A[\fcy O(\sk,P)]  &= s_a \fcy E_A^a\fcy O(\sk,P) \\
&= \frac12\left[2(\Dk-1)\Gk^I_{AB}\sk^B\frac{\nb}{\nb P^I} + \sk_A\sk^B\Gk^I_{BC}\frac{\nb}{\nb P^I}\frac{\nb}{\nb \sk_C}\right]\fcy O(\sk,P)\,.
\end{split}\end{equation}
We hence conclude that $S^A_a$ and $\fcy E^A_a$ are bispinor analogues of the fundamental weight-shifting operators of \cite{Karateev:2017jgd}.  Indeed, $S^A_a$ packages together the weight-shifting operators $\fcy W^{--}_A$ and $\fcy W^{-+}_A$ in one expression, while $\fcy E^A_a$ packages together $\fcy W^{+-}_A$ and $\fcy W^{++}_A$.

\section{Witten diagrams}
\label{WITDIAG}
We will now use the differential operators introduced in the previous section to evaluate Witten diagrams. The basic idea is to use these differential operators to rewrite spinning external legs as derivatives of scalar external legs. We begin by illustrating our methods in the simplest possible setting, that of contact diagrams.

\subsection{Contact diagrams}
\label{CONTDIAG}
Consider a theory of bulk scalar fields $\phi_i(T,\bar T)$ dual to boundary scalar operators $\Fk_i(S)$ with conformal dimension $\Dk_i$. We normalize the bulk-boundary propagator such that
\begin{equation}\label{sbbp}
\<\Fk_i(S)\phi_j(T,\bT)\> = \dk_{ij} \frac{\Gk(\Dk_i)}{(-2P\cdot X)^{\Dk_i}}\,.
\end{equation}
An interaction $g_0\phi_1\phi_2\phi_3\phi_4$ gives, at leading order in $g_0$, the contact Witten diagram:
\begin{equation}\begin{split}
\feynmandiagram [inline=(b.base), horizontal = f to b] {
a [particle=$\Phi_1$] --  b --  c [particle=$\Phi_2$],
d [particle=$\Phi_3$] --  b -- [opacity=0] e ,
f -- [opacity=0] b --  g [particle=$\Phi_4$],
h   -- [opacity=0]  b -- [opacity=0] i
};
\equiv g_0\Pi_0(P_i) &= ig_0\int dX \prod_{i = 1}^4\<\Fk_i(S_i)\phi_i(T,\bT)\> \\
&= ig_0\int dX \prod_{i = 1}^4\frac{\Gamma(\Dk_i)}{(-2P_i\cdot X)^{\Dk_i}} \\
&= ig_0\left(\prod_{i = 1}^4\Gk(\Dk_i)\right)D_{\Dk_1\Dk_2\Dk_3\Dk_4}(P_i)\,,
\end{split}\end{equation}
where $D_{\Dk_1\Dk_2\Dk_3\Dk_4}(P_i)$ is a $D$-function (see Appendix~\ref{sec:SCLWITDIA} for more details). 

Now let us consider the analogous diagram for the derivative interaction $g_1(\Nb^{\ak\da}\fk_1)(\Nb_{\ak\da}\fk_2)\fk_3\fk_4$, which we can evaluate using the conformal generators. Applying first \eqref{eq:ConfInv} and then \eqref{eq:confgensnab}, it is easy to see that
\begin{equation}
\fcy D_{AB} \<\Fk_i(S)\phi_j(T,\bT)\> = -D_{AB}\<\Fk_i(S)\phi_j(T,\bT)\> = \frac i2T^\ak_{(A}\bT^\da_{B)}\<\Fk_i(S)\Nb_{\ak\da}\fk_j(T,\bT)\>\,.
\end{equation}
Defining $(\fcy D_i)_{AB}$ to be the conformal generator acting on the $i^{\text{th}}$ spinor $(S_i)^A_a$, we then find that:
\begin{equation}\begin{split}
(\fcy D_1\cdot\fcy D_2)&\<\Fk_1(S_1)\phi_1(T,\bT)\>\<\Fk_2(S_2)\phi_2(T,\bT)\> \\
&= -\frac12\<\Fk_1(S_1)\Nb^{\ak\da}\phi_1(T,\bT)\>\<\Fk_2(S_2)\Nb_{\ak\da}\phi_2(T,\bT)\> \,,
\end{split}\end{equation}
where $\fcy D_i\cdot\fcy D_j \equiv (\fcy D_i)^{AB}(\fcy D_j)_{AB}$. We now find that our derivative interaction gives rise to the Witten diagram $g_1 \Pi_1(P_i)$, with
\begin{equation}\label{eq:addDer}\begin{split}
\Pi_1(P_i) = -2\fcy D_1\cdot\fcy D_2\,\Pi_0(P_i) = -2i\left(\prod_{i = 1}^4\Gk(\Dk_i)\right)(\fcy D_1\cdot\fcy D_2)D_{\Dk_1\Dk_2\Dk_3\Dk_4}(P_i)\,.
\end{split}\end{equation}

As a second example consider the interaction $\frac14gB^{\ak\dot\ak}B_{\ak\dot\ak}\phi^2$ coupling a massive scalar $\phi$ to a massive vector $B^{\ak\dot\ak}$. We will denote the boundary operator dual to these fields by $\Fk$ and $J_{ab}$ respectively, normalizing these operators such that
\begin{equation}
\langle\Fk(P)\fk(X)\rangle = \frac{\Gk(\Dk_\Fk)}{(-2P\cdot X)^{\Dk_\Fk}}\,,\qquad \langle J(\sk,P)B(\tk,\bar\tk,X) \rangle = \frac{\Gk(\Dk_J+1)}{\Dk_J-1}\frac{\langle\sk\tk\rangle\langle\sk\bar\tk\rangle}{(-2P\cdot X)^{\Dk_J+1}}\,.
\end{equation}
Our aim will be to evaluate
\begin{equation}
g\Pi_V(S_i) \equiv \feynmandiagram [inline=(b.base), horizontal = f to b] {
a [particle=$J(S_1)$] -- [boson] b -- c [particle=$\Phi(S_3)$],
d [particle=$\Phi(S_4)$] -- b -- [opacity=0] e ,
f -- [opacity=0] b -- [boson] g [particle=$J(S_2)$],
h   -- [opacity=0]  b -- [opacity=0] i
};  \,,
\end{equation}
which gives the $O(g)$ contribution to the correlator $\langle \Fk\Fk JJ\rangle$. To evaluate this diagram we first note that
\begin{equation}\begin{split}
\nb_{ab}^I\frac{\Gk(\Dk_J)}{(-2P\cdot X)^{\Dk_J}} &= \frac{\Gk(\Dk_J+1)\langle S_a\Gk^IXS_b\rangle}{(-2P\cdot X)^{\Dk_J+1}} \\
&= \frac {-i(\Dk_J-1)}2\langle T_\ak\Gk^I\bar T_{\dot\ak}\rangle\langle J_{ab}(S)B^{\ak\dot\ak}(T,\bar T) \rangle\,.
\end{split}\end{equation}
Using the Fierz identity
\begin{equation}
\langle R_1\Gk^IR_2\rangle\langle R_3\Gk_IR_4\rangle = \langle R_1R_3\rangle\langle R_2R_4\rangle - \langle R_1R_4\rangle\langle R_2R_3\rangle\,,
\end{equation}
we can simplify
\begin{equation}
\langle T_\ak\Gk^I\bar T_{\dot\ak}\rangle\langle T_\bk\Gk_I\bar T_{\dot\bk}\rangle = 4\epsilon_{\ak\bk}\epsilon_{\dot\ak\dot\bk}\,,
\end{equation}
and hence find that
\begin{equation}\label{eq:massV1}\begin{split}
\langle J_{ab}(P_1)B^{\ak\dot\ak}&(T,\bar T)\rangle\langle J_{cd}(P_2)B_{\ak\dot\ak}(T,\bar T) \rangle\\
&= -(\Dk_J-1)^2 (\nb_1)^I_{ab}(\nb_2)_{Icd} \frac{\Gk(\Dk_J)^2}{(-2P_1\cdot X)^{\Dk_J}(-2P_2\cdot X)^{\Dk_J}}\,,
\end{split}\end{equation}
where $(\nb_i)^I_{ab}$ acts on the $i^{\text{th}}$ position. We thus have shown that
\begin{equation}\label{eq:massV2}
\Pi_V(S_i) = -i(\Dk_J-1)^2\Gk(\Dk_\Fk)^2\Gk(\Dk_J)^2(\nb_1)^I_{ab}(\nb_2)_{Icd}D_{\Dk_J\Dk_J\Dk_\Fk\Dk_\Fk}(P_i)\,.
\end{equation}

Extending these computations to higher spin particles is straightforward. Consider an interaction $\frac14g\varphi^{\ak_1\dots\ak_\ell\dot\ak_1\dots\dot\ak_{\ell}}\varphi_{\ak_1\dots\ak_\ell\dot\ak_1\dots\dot\ak_{\ell}}\phi^2$, where $\fk$ is a scalar and $\varphi^{\ak_1\dots\ak_\ell\dot\ak_1\dots\dot\ak_{\ell}}$ is a massive field of $SO(3, 1)$ spin $(\ell,\ell)$. We will use $\fcy O(S)$ to denote the boundary operator dual to $\varphi$. By convention we use the bulk-boundary propagator
\begin{equation}
\langle \fcy O(\sk,P)\varphi(\tau,\bar\tau,X)\rangle = \frac{\Gk(\Dk_{\fcy O}+\ell)}{(\Dk_{\fcy O}-1)_{\ell}}\frac{\langle\sk\tk\rangle^\ell\langle\sk\bar\tk\rangle^\ell}{(-2P\cdot X)^{\Dk_{\fcy O}+\ell}}\,,
\end{equation}
where $(a)_\ell = \frac{\Gamma(a+\ell)}{\Gamma(a)}$ is the Pochhammer symbol. 

To simplify our expressions we will use polarized operators, and will find it convenient to define $\nb_{\bm{ii}}^I \equiv (s_i)^a(s_i)^b(\nb_i)^I_{ab}$. We can then compute
\begin{equation}\begin{split}
\left(\nb_{\bm{ii}}^{I_1}\dots\nb_{\bm{ii}}^{I_\ell}\right)\frac{\Gk(\Dk_{\fcy O})}{(-2P_i\cdot X)^{\Dk_{\fcy O}}} &= (-i)^\ell\Gk(\Dk_{\fcy O}+\ell)\frac{\langle\sk_i\Gk^{I_1}X\sk_i\rangle\dots\langle\sk_i\Gk^{I_\ell}X\sk_i\rangle}{(-2P_i\cdot X)^{\Dk_{\fcy O}+\ell}} \\
&= \frac{(-i)^\ell(\Dk_{\fcy O}-1)_\ell}{2^\ell}\langle \fcy O(\sk_i,P_i)\varphi^{\ak_1\dots\ak_\ell\dot\ak_1\dots\dot\ak_{\ell}}(T,\bT)\rangle\prod_{j = 1}^\ell \langle T_{\ak_j}\Gk^{I_j}\bT_{\da_j}\rangle\,.
\end{split}\end{equation}
We can now to generalise \eqref{eq:massV1} and \eqref{eq:massV2} to the spin $\ell$ case, and so find that leading contribution $\Pi_H(S_i)$ to the correlator $\langle \fcy O\fcy O\fk\fk\rangle$ is
\begin{equation}\begin{split}
\Pi_H(S_i) &=  \kk(\nb_{\bm{11}}^I\nb_{I\bm{22}})^\ell D_{\Dk_{\fcy O}\Dk_{\fcy O}\Dk_\Fk\Dk_\Fk}(P_i)\,, \\
\text{ where }\quad \kk &= i(-1)^\ell(\Dk_{\fcy O}-1)^2\Gk(\Dk_{\fcy O}+\ell-1)^2\Gk(\Dk_\Fk)^2 \,.
\end{split}\end{equation}
Thus, we have seen that various differential operators can be used to spin up the external legs.

\subsection{Compton scattering}
\label{sec:COMPTON}
We will now apply our tools to compute something new: Compton scattering in scalar QED in AdS$_4$. We will use $A^{\ak\dot\ak} = \sk^{\ak\da}_\mu A^\mu$ for the bulk gauge field and $\phi$ for the complex scalar.  The Lagrangian is:
\begin{equation}\begin{split}\label{eq:sQEDLag}
\fcy L &= -\frac 14 F^{\mu\nu}F_{\mu\nu}-(D^\mu\phi)^*D_\mu\phi-m^2\phi^*\phi \\
&= -\frac 14 F^{\mu\nu}F_{\mu\nu}+\frac12\left(\Nb^{\ak\da}+ieA^{\ak\da}\right)\phi^*\left(\Nb_{\ak\da}-ieA_{\ak\da}\right)\phi -m^2\phi^*\phi \,.
\end{split}\end{equation}

On the boundary these fields are dual to a conserved current $J^{ab}$, and a complex scalar $\Fk$, respectively. We will normalize the bulk-boundary propagator $\langle JA\rangle$ such that
\begin{equation}\label{eq:JA2pt}
\langle J(\sk,P)A^{\ak\dot\ak}(\tau,\bar \tau, X)\rangle = \frac{2\langle\sk \tau\rangle\langle\sk\bar\tau\rangle}{(-2P\cdot X)^{3}}\,.
\end{equation}

To study the correlator $\langle\Phi(S_1)\Phi^\dagger(S_2)J(S_3)J(S_4)\rangle$ at tree level we must consider three diagrams:
\begin{equation}\label{eq:comptWit}
\Pi_{\text{comp}}(S_i) = \feynmandiagram [inline=(b.base), horizontal = b to e] {
a [particle=$\Phi(S_1)$] -- [fermion] b -- [boson] c [particle=$J(S_3)$],
d [particle=$\Phi^\dagger(S_2)$] -- [anti fermion]  e -- [boson] f [particle=$J(S_4)$],
b -- [fermion] e
}; + \feynmandiagram [inline=(b.base), horizontal = b to e] {
a [particle=$\Phi(S_1)$] -- [fermion] b -- [boson] c [particle=$J(S_4)$],
d [particle=$\Phi^\dagger(S_2)$] -- [anti fermion]  e -- [boson] f [particle=$J(S_3)$],
b -- [fermion] e
}; +\feynmandiagram [inline=(b.base), horizontal = f to b] {
a [particle=$J(S_3)$] -- [boson] b -- [anti fermion]  c [particle=$\Phi(S_1)$],
d [particle=$\Phi^\dagger(S_2)$] -- [anti fermion]  b -- [opacity=0] e ,
f -- [opacity=0] b -- [boson] g [particle=$J(S_4)$],
h   -- [opacity=0]  b -- [opacity=0] i
};
\end{equation}
Let us begin with the last diagram. This is a special case of 2-vector 2-scalar diagram which we evaluated in \eqref{eq:massV2}, and so we find that
\begin{equation}
\feynmandiagram [inline=(b.base), horizontal = f to b] {
a [particle=$J(S_3)$] -- [boson] b -- [anti fermion]  c [particle=$\Phi(S_1)$],
d [particle=$\Phi^\dagger(S_2)$] -- [anti fermion]  b -- [opacity=0] e ,
f -- [opacity=0] b -- [boson] g [particle=$J(S_4)$],
h   -- [opacity=0]  b -- [opacity=0] i
}; = -ie^2\Gk(\Dk)^2 (\nb_3)^I_{ab}\nb_4)_{Icd} D_{\Dk\Dk22}(P_i)\,.
\end{equation}

Now let us compute the first diagram in \eqref{eq:comptWit}. This $t$-channel diagram is given by\footnote{We can always use integration by parts to guarantee that the derivative acts on the boundary leg of the diagram. Note that as a consequence of \eqref{eq:bulkBound}, $\langle J_{ab}\Nb^{\ak\dot\ak}A_{\ak\dot\ak}\rangle$ automatically vanishes.}
\begin{equation}\begin{split}
\Pi_{t\text{-exch}} &= -e^2\int dX_1 dX_2\Big( \langle\Fk(P_1)\Nb_{\ak\dot\ak}\fk^\dagger(X_1)\rangle\langle J(\sk_3,P_3)A^{\ak\dot\ak}(T_1,\bar T_1)\rangle\\
&\times\langle\fk(X_1)\fk^\dagger(X_2)\rangle\langle\Fk^\dagger(P_2)\Nb_{\bk\dot\bk}\fk(X_2)\rangle\langle J(\sk_4,P_4)A^{\bk\dot\bk}(T_2,\bar T_2)\rangle\Big)\,.
\end{split}\end{equation}
In order to evaluate this let us first simplify:
\begin{equation}\begin{split}\label{eq:comptVert}
\langle\Fk(P_1)\Nb_{\ak\dot\ak}\fk^\dagger(X_1)\rangle\langle J(\sk_3,P_3)A^{\ak\dot\ak}(T_1,\bar T_1)\rangle &= \frac{-2i\Gk(\Dk+1)\langle T_{1\ak}P_1\bar T_{1\dot\ak}\rangle\langle\sk_3T_1^\ak\rangle\langle\sk_3\bar T_1^{\dot\ak}\rangle}{(-2P_1\cdot X)^{\Dk+1}(-2P_3\cdot X)^3} \\
&=\frac{2\Gk(\Dk+1)\langle\sigma_3[P_1,X]\sigma_3\rangle}{(-2P_1\cdot X)^{\Dk+1}(-2P_3\cdot X)^3}\,.
\end{split}\end{equation}
Using the identity
\begin{equation}
\fcy D_{AB}\left(\frac{\Gk(\Dk)}{(-2X\cdot P)^\Dk}\right) = \frac{-i\Gk(\Dk+1)[\slashed P,\slashed X]_{AB}}{2(-2X\cdot P)^{\Dk+1}} \,,
\end{equation}
we can rewrite \eqref{eq:comptVert} in terms of a differential operator acting on scalar bulk-boundary propagators:
\begin{equation}\begin{split}\label{eq:comptVert2}
\langle\Fk(P_1)\Nb_{\ak\dot\ak}\fk^\dagger(X_1)\rangle\langle J(\sk_3,P_3)A^{\ak\dot\ak}(T_1,\bar T_1)\rangle 
&= \langle\sigma_3\fcy D_1\sigma_3\rangle\frac{4i\Gk(\Dk)}{(-2P_1\cdot X_1)^\Dk(-2P_3\cdot X_1)^3}\,.
\end{split}\end{equation}
This allows us to express $\Pi_{t\text{-exch}}$ in terms of the scalar Witten diagram
\begin{equation}\label{eq:compT}\begin{split}
\Pi_{t\text{-exch}}(S_i) &= -4e^2\langle\sigma_3\fcy D_1\sigma_3\rangle\langle\sigma_4\fcy D_2\sigma_4\rangle\Pi^{\text{scalar}}_{t\text{-exch}}(S_i) \\
\Pi^{\text{scalar}}_{t\text{-exch}}(S_i) &= \feynmandiagram [inline=(b.base), horizontal = b to e] {
a [particle=$\Dk_\Phi$] -- b -- c [particle=$3$],
d [particle=$\Dk_\Phi$] -- e -- f [particle=$3$],
b -- [edge label=$\Dk_\Fk$] e
};\,.
\end{split}\end{equation}
General scalar exchange diagrams have been computed in Mellin space \cite{Penedones:2010ue,Fitzpatrick:2011ia,Paulos:2011ie}; see Appendix~\ref{sec:SCLWITDIA} for more details. For the specific diagram appearing in \eqref{eq:compT}, we find that
\begin{equation}\begin{split}
M_{t\text{-exch}}(\gk_{ij}) &= \frac{-i\pi^{5/2}\Gk(\Dk)^2}{4}\sum_{m = 0}^\infty \frac {a_m}{2\gk_{13}+2m-3}\,, \\
\text{ where }\quad a_m &= \frac{1}{m!\Gk\left(\frac32-m\right)^2\Gk\left(\Dk+m-\frac12\right)}\,,
\end{split}\end{equation}
and so
\begin{equation}
\Pi_{t\text{-exch}}(S_i) = -4e^2\langle\sigma_3\fcy D_1\sigma_3\rangle\langle\sigma_4\fcy D_2\sigma_4\rangle\int [d\gk]M_{t\text{-exch}}(\gk_{ij})\prod_{i<j}\frac{\Gk(\gk_{ij})}{(-2P_i\cdot P_j)^{\gk_{ij}}} \,.
\end{equation}

Finally, let us consider the $u$-channel diagram, which is the second diagram in \eqref{eq:comptWit}. There is nothing new for us to calculate here, as the $t$ and $u$ channels can be related by interchanging $P_3\leftrightarrow P_4$. We can hence state the leading contribution to Compton scattering in scalar QED:
\begin{equation}\begin{split}
\Pi_{\text{comp}}(S_i)  = & -4e^2\langle\sigma_3\fcy D_1\sigma_3\rangle\langle\sigma_4\fcy D_2\sigma_4\rangle\Pi^{\text{scalar}}_{t\text{-exch}}(P_i) + (3\leftrightarrow 4)\\
& +i e^2\Gk(\Dk)^2 \left(\nb_{\bm{33}}\cdot \nb_{\bm{44}}\right)D_{\Dk\Dk22}(P_i)\,.
\end{split}\end{equation} 

Now consider the more general scalar QED diagram
\begin{equation}\label{genScQED}
\Pi(S_1,S_2,\dots) = {\scriptsize\feynmandiagram [inline=(b.base), horizontal = b to e] {
a [particle=$\Fk(S_1)$] -- [fermion] b -- [photon] c [particle=$J(S_2)$],
d -- e [blob]-- f,
g -- e -- h,
b -- [fermion,edge] e
};}\,.
\end{equation}
The blob represents the rest of the Witten diagram, whose specific details are not important. Using \eqref{eq:comptVert2}, we find that
\begin{equation}\label{sQEDvr}
\Pi(S_1,S_2,\dots) = 2e\<\sk_2\fcy D_1\sk_2\> {\scriptsize\feynmandiagram [inline=(b.base), horizontal = b to e] {
a [particle=$\Fk(S_1)$] -- [fermion] b --  c [particle=$3$],
d -- e [blob]-- f,
g -- e -- h,
b -- [fermion,edge] e
};}\,,
\end{equation}
where in the right-hand diagram the photon has been replaced by a dimension $3$ scalar, but has otherwise been left unchanged. This allows us to generalize our Compton scattering calculation to a large class of scalar QED diagrams.

\subsection{Fermion scattering}
\label{FERMSCT}

In this section we study Witten diagrams involving external fermions. A free Dirac fermion in AdS$_4$ can be written as a pair of Weyl spinors $\psi^\ak$ and $\chi^\ak$, with Lagrangian
\begin{equation}\label{eq:DircLag}
\fcy L = i\bar\psi^\da\Nb_{\ak\da}\psi^\ak + i\bar\chi^\da\Nb_{\ak\da}\ck^\ak - m\left(\psi^\ak\chi_\ak + \bar\psi_\da\bar\chi^\da\right)\,.
\end{equation}
The mass term is uniquely fixed, up to field redefinitions, by the condition that the two Weyl fermions have equal mass. Our choice of mass term makes the global $U(1)$ symmetry of the Dirac Lagrangian manifest, under which $\psi_\ak$ has charge $+1$ and $\chi_\ak$ has charge $-1$. This will prove particularly convenient later on when we discuss QED.\footnote{Another useful choice is to define $\xi_1 = \frac 1{\sqrt2}(\psi+i\chi)$ and $\xi_2 = \frac 1{\sqrt 2}(\chi+i\psi)$. This choice diagonalizes the mass term so that the two Weyl fermions decouple.}

On the boundary of AdS$_4$ the Dirac fermion is dual to a complex spinor $\Yk_a(S)$ which has charge $+1$ and conformal dimension $\Dk_\Yk$. We will normalize our bulk-boundary propagators such that:
\begin{equation}\begin{split}
\langle \Yk^\dagger(\sk,P)\yk(\tau,X)\rangle = \langle \Yk(\sk,P)\ck(\tau,X)\rangle &= +\Gk\left(\Dk_\Yk+\frac12\right)\frac{\<\sk\tk\>}{(-2P\cdot X)^{\Dk_\Yk+\frac12}}\,,\\
\langle \Yk(\sk,P)\bar\yk(\bar\tau,X)\rangle = \langle \Yk^\dagger(\sk,P)\bar\ck(\bar\tau,X)\rangle &= -\Gk\left(\Dk_\Yk+\frac12\right)\frac{\<\sk\bar\tk\>}{(-2P\cdot X)^{\Dk_\Yk+\frac12}}\,.
\end{split}\end{equation}
We can use \eqref{eq:fermEoM} to verify that these bulk-boundary propagators satisfy the equations of motion implied by the Dirac Lagrangian \eqref{eq:DircLag}, with $\Dk_\Yk = m + \frac 32$.

Now consider coupling our Dirac fermion to a real scalar field $\sk$ via the Yukawa interaction
\begin{equation}\label{eq:yukLag}
\fcy L_{\text{Yukawa}} = g\sk\left(\yk^\ak\ck_\ak + \bar\yk_\da\bar\ck^\da\right)\,.
\end{equation}
We can then study Witten diagrams of the form
\begin{equation}\label{sclExchFerm}
\Pi(S_1,S_2,\dots) = {\scriptsize\feynmandiagram [inline=(b.base), horizontal = b to e] {
a [particle=$\Psi(S_1)$] -- [fermion] b -- [fermion] c [particle=$\Psi^\dagger(S_2)$],
d -- e [blob]-- f,
g -- e -- h,
b -- [scalar,edge label=$\sk$] e
};}\,.
\end{equation}
The blob represents the rest of the Witten diagram, whose specific details are not important. Our goal will be to relate \eqref{sclExchFerm} to diagrams where the fermions have been replaced with scalars. 

To this end, consider complex scalar field $\fk(X)$ in AdS$_4$, coupled to $\sigma$ via the Lagrangian
\begin{equation}
\fcy L = -(\nb^\mu\fk^*)(\nb_\mu\fk) - m_\fk^2\fk^*\fk - g\sk\fk^*\fk\,.
\end{equation} 
Let $\Fk(S)$ be the boundary operator dual to $\fk(X)$, with bulk-boundary propagator
\begin{equation}\<\Fk(P)\fk^\dagger(X)\> = \<\Fk^\dagger(P)\fk(X)\> = \frac{\Gk(\Dk_\Fk)}{(-2P\cdot X)^{\Dk_\Fk}}\,.\end{equation}
We will now consider the Witten diagram
\begin{equation}\label{sclExchScl}
\Pi^{\text{scalar}}(S_i;\Dk_\Fk) = {\scriptsize\feynmandiagram [inline=(b.base), horizontal = b to e] {
a [particle=$\Fk(S_1)$] -- [fermion] b -- [fermion] c [particle=$\Fk^\dagger(S_2)$],
d -- e [blob]-- f,
g -- e -- h,
b -- [scalar,edge label=$\sk$] e
};}\,.
\end{equation}
The blob here represents the same Witten diagram as \eqref{sclExchFerm}. We now show that the two diagrams, which differ only in the spin of the external operators, are related by the equation:
\begin{equation}\begin{split}\label{sctoferm}
\Pi(S_i) = -2i\< S_{1a}S_{2b} \> &\Pi^{\text{scalar}}\left(S_i;\Dk_\Yk+\frac12\right) \\
= \frac{8i\<\fcy E_{1a}\fcy E_{2b} \>}{(2\Dk_\Yk-3)^2} &\Pi^{\text{scalar}}\left(S_i;\Dk_\Yk-\frac12\right)
\end{split}\end{equation}
Therefore, once we evaluate $\Pi^{\text{scalar}}(S_i;\Dk_\Fk)$ it is easy to compute $\Pi(S_i)$.

To derive \eqref{sctoferm}, let us introduce the notation $\fk_\pm(X)$ and $\Fk_\pm(P)$ to specifically describe a complex scalar field and its boundary dual with conformal dimension $\Dk_\pm = \Dk_\Yk \pm \frac 12$. Acting with $S^A_a$ on the bulk-boundary propagator ${\langle\Fk_+\fk_-\rangle}$, we find that
\begin{equation}\label{eq:fermS}\begin{split}
S^A_a\<\Fk_+^\dagger(P)\fk_+(X)\> &= -\frac{i}2\left\langle\Yk_a^\dagger\left(\yk_\ak T^{A\ak}-\bar\chi_{\dot\ak}\bar T^{A\dot\ak}\right)\right\rangle \,,\\
S^A_a\<\Fk_+(P)\fk_+^\dagger(X)\> &= -\frac i2\left\langle\Yk_a\left(\bar\yk_{\da}\bar T^{A\da}-\chi_\ak T^{A\ak}\right)\right\rangle \,.\\ 
\end{split}\end{equation}
We can then compute
\begin{equation}\begin{split}
&\langle S_{1a}S_{2b}\rangle\<\Fk_+(P_1)\fk_+^\dagger(X)\> \<\Fk_+^\dagger(P_2)\fk_+(X)\>\\
&= \frac i 2\<\Yk_a(S_1)\ck_\ak\>\<\Yk_b^\dagger(S_2)\yk^\ak\> -\frac i 2\<\Yk_a(S_1)\bar\yk_\da\>\<\Yk_b^\dagger(S_2)\bar\ck^\da\> \,,
\end{split}\end{equation}
from which the first line of \eqref{sctoferm} then follows.

Likewise, when we apply $\fcy E^A_a$ to the bulk-boundary propagator ${\<\Fk_-\fk_-\>}$, we find that
\begin{equation}\label{eq:fermE}\begin{split}
{\fcy E}^A_a\<\Fk_-^\dagger(P)\fk_-(X)\> &= -\frac 12\left(\Dk_\Yk-\frac32\right)\left\langle\Yk_a^\dagger\left(\yk_\ak T^{A\ak}+\bar\chi_{\dot\ak}\bar T^{A\dot\ak}\right)\right\rangle \,, \\
{\fcy E}^A_a\<\Fk_-(P)\fk_-^\dagger(X)\> &=\frac 12\left(\Dk_\Yk-\frac32\right)\left\langle\Yk_a\left(\bar\yk_\da \bT^{A\da}+\chi_\ak T^{A\ak}\right)\right\rangle \,.
\end{split}\end{equation}
We can then compute
\begin{equation}\begin{split}
\frac{4\langle \fcy E_{1a}\fcy E_{2b}\rangle}{\left(2\Dk_\Yk-3\right)^2}&\<\Fk_-(P_1)\fk_-^\dagger(X)\> \<\Fk_-^\dagger(P_2)\fk_-(X)\>\\
&= -\frac i 2\<\Yk_a(S_1)\ck_\ak\>\<\Yk_b^\dagger(S_2)\yk^\ak\> + \frac i 2\<\Yk_a(S_1)\bar\yk_\da\>\<\Yk_b^\dagger(S_2)\bar\ck^\da\> \\
&=-\langle S_{1a}S_{2b}\rangle\<\Fk_+(P_1)\fk_+^\dagger(X)\> \<\Fk_+^\dagger(P_2)\fk_+(X)\>
\end{split}\end{equation}
and thus the second line of \eqref{sctoferm} follows.

As a concrete example, consider the four fermion diagram
\begin{equation}\label{eq:4fermd}
\Pi(S_i) = {\scriptsize\feynmandiagram [inline=(b.base), horizontal = b to e] {
a [particle=$\Yk(S_1)$] -- [fermion] b -- [fermion] c [particle=$\Yk^\dagger(S_2)$],
d [particle=$\Yk(S_3)$] -- [fermion] e -- [fermion] f [particle=$\Yk^\dagger(S_4)$],
b -- [scalar,edge label=$\sigma$] e
};}\,.
\end{equation}
Using \eqref{sctoferm} we can reduce this task to that of computing a scalar diagram:
\begin{equation}\label{eq:4ferms}\begin{split}
\Pi(S_i) &=  -4\<S_{1a}S_{2b}\>\<S_{3c}S_{4d}\> \Pi^\text{scalar}(S_i) \,,\\
\text{ with } \Pi^\text{scalar}(S_i) &= {\scriptsize\feynmandiagram [inline=(b.base), horizontal = b to e] {
a [particle=$\Fk_+(S_1)$] -- [fermion] b -- [fermion] c [particle=$\Fk_+^\dagger(S_2)$],
d [particle=$\Fk_+(S_3)$] -- [fermion] e -- [fermion] f [particle=$\Fk_+^\dagger(S_4)$],
b -- [scalar,edge label=$\sigma$] e
};}
\end{split}\end{equation}
General four-scalar exchange diagrams have been computed in by \cite{Penedones:2010ue,Fitzpatrick:2011ia,Paulos:2011ie}, (see Appendix~\ref{sec:SCLWITDIA} for general expressions). For the specific diagram in \eqref{eq:4ferms}, we find that
\begin{equation}
\Pi^{\text{scalar}}(S_i) = \frac{\pi^{3/2}}2\int [d\gk]\,\fcy M(\gk_{ij})\prod_{i<j}\frac{\Gk(\gk_{ij})}{(-2P_i\cdot P_j)^{\gk_{ij}}}
\end{equation}
where the Mellin amplitude is given by the infinite series
\begin{equation}
\fcy M(s,t) = \sum_{m = 0}^\infty \frac{1}{s-\Dk_\sk-2m}\frac{\left(\frac 12 \Dk_\sk-\Dk_\Yk+\frac12\right)_m^2}{m!\left(\Dk_\sk-\frac12\right)_m}\,.\\
\end{equation}
We have thus been able to evaluate \eqref{eq:4fermd} in terms of spinning conformal structures multiplying a Mellin amplitude.  

So far we have considered diagrams of the form \eqref{sclExchFerm}, where the two fermions interact with a scalar field $\sk$. Now let us consider QED
\begin{equation}\label{QEDlag}
\fcy L = \bar\psi^\da\left(i\Nb_{\ak\da}+eA_{\ak\da}\right)\psi^\ak + \bar\chi^\da\left(i\Nb_{\ak\da}-eA_{\ak\da}\right)\ck^\ak - m\left(\psi^\ak\chi_\ak + \bar\psi_\da\bar\chi^\da\right) - \frac 14 F^{\m\n}F_{\m\n}\,,
\end{equation}
and Feynman diagrams of the form
\begin{equation}\label{AExchFerm}
\Pi(S_i) = {\scriptsize\feynmandiagram [inline=(b.base), horizontal = b to e] {
a [particle=$\Yk(S_1)$] -- [fermion] b -- [fermion] c [particle=$\Yk^\dagger(S_2)$],
d -- e [blob]-- f,
g -- e -- h,
b -- [photon,edge label=$A_{\ak\da}$] e
};}\,.
\end{equation}
As done previously, we evaluate this diagram by relating it to the scalar diagram
\begin{equation}\label{AExchScl}
\Pi^{\text{scalar}}(S_i;\Dk_\Fk) = {\scriptsize\feynmandiagram [inline=(b.base), horizontal = b to e] {
a [particle=$\Fk(S_1)$] -- [fermion] b -- [fermion] c [particle=$\Fk^\dagger(S_2)$],
d -- e [blob]-- f,
g -- e -- h,
b -- [photon,edge label=$A_{\ak\da}$] e
};}\,,
\end{equation}
where the complex scalar interacts with the gauge field via the scalar QED Lagrangian \eqref{eq:sQEDLag}. We now show that the diagrams \eqref{AExchFerm} and \eqref{AExchScl} are related by the equation
\begin{equation}\label{eq:qedRel}
\Pi(S_i) = -2i\<S_{1a}S_{2b}\>\Pi^{\text{scalar}}\left(S_i;\Dk_\Yk+\frac12\right)+\frac{8i\<\fcy E_{1a}\fcy E_{2b}\>}{(2\Dk_\Yk-3)^2}\Pi^{\text{scalar}}\left(S_i;\Dk_\Yk-\frac12\right) \,.
\end{equation}

To derive this result, first note that the bulk covariant derivative commutes with boundary differential operators. For example, using \eqref{eq:fermS} we find that
\begin{equation}\begin{split}
S^A_a\<\Fk_+^\dagger(P)&\Nb_{\ak\da}\fk_+(X)\> = \Nb_{\ak\da}S^A_a\<\Fk_+^\dagger(P)\fk_+(X)\> \\
&= -\frac{i}2\left\langle\Yk_a^\dagger\left(T^{A\bk}\Nb_{\ak\da}\yk_\bk-\bar T^{A\dot\bk}\Nb_{\ak\da}\bar\chi_{\dot\bk}\right)\right\rangle +\frac 12 \left\langle\Yk_a^\dagger\left(\bT^A_\da\yk_\ak+T^A_\ak\bar\chi_{\da}\right)\right\rangle \,.
\end{split}\end{equation}
Using this expression, we obtain
\begin{equation}\label{qedDev1}\begin{split}
\<S_{1a}S_{2b}\>&\<\Fk_+(P_1)\fk_+^\dagger(X)\>\<\Fk_+^\dagger(P_2)\Nb_{\ak\da}\fk_+(X)\>\\
&=  \frac i2\left[\<\Yk_a(S_1)\ck_\bk\>\<\Yk_b^\dagger(S_2)\Nb_{\ak\da}\yk^\bk\> -\<\Yk_a(S_1)\bar\yk_\db\>\<\Yk_b^\dagger(S_2)\Nb_{\ak\da}\bar\ck^\db\>\right] \\
&+ \frac12\left[\<\Yk_a(S_1)\bar\yk_\da\>\<\Yk_b^\dagger(S_2)\yk_\ak\> + \<\Yk_a(S_1)\ck_\ak\>\<\Yk_b^\dagger(S_2)\bar\ck_\da\>\right]\,.
\end{split}\end{equation}
Analogous calculations with the $\fcy E$ operators reveal that
\begin{equation}\label{qedDev2}\begin{split}
\frac{4\<\fcy E_{1a}\fcy E_{2b}\>}{(2\Dk_\Yk-3)^2}&\<\Fk_-(P_1)\fk_-^\dagger(X)\>\<\Fk_-^\dagger(P_2)\Nb_{\ak\da}\fk_-(X)\> \\ 
&=  \frac i 2\left[\<\Yk_a(S_1)\ck_\bk\>\<\Yk_b^\dagger(S_2)\Nb_{\ak\da}\yk^\bk\> -\<\Yk_a(S_1)\bar\yk_\db\>\<\Yk_b^\dagger(S_2)\Nb_{\ak\da}\bar\ck^\db\>\right] \\
&-\frac12\left[\<\Yk_a(S_1)\bar\yk_\da\>\<\Yk_b^\dagger(S_2)\yk_\ak\> + \<\Yk_a(S_1)\ck_\ak\>\<\Yk_b^\dagger(S_2)\bar\ck_\da\>\right]\,.
\end{split}\end{equation}
By subtracting \eqref{qedDev2} from \eqref{qedDev1} we can eliminate the derivative terms, and \eqref{eq:qedRel} then follows.

As a simple example, we can apply \eqref{eq:qedRel} to the Witten diagram
\begin{equation}
\Pi(S_i) = {\scriptsize\feynmandiagram [inline=(b.base), horizontal = b to e] {
a [particle=$\Yk(S_1)$] -- [fermion] b -- [fermion] c [particle=$\Yk^\dagger(S_2)$],
d [particle=$\Wk(S_3)$] -- [fermion] e -- [fermion] f [particle=$\Wk^\dagger(S_4)$],
b -- [photon,edge label=$A$] e
};}
\end{equation}
where $\Wk(P)$ is the boundary dual of a charge $+e$ bulk scalar field. To calculate $\Pi(S_i)$ we simply need to evaluate
\begin{equation}\label{scalQEDexch}
\Pi^{\text{scalar}}(S_i;\Dk_\Fk) = {\scriptsize\feynmandiagram [inline=(b.base), horizontal = b to e] {
a [particle=$\Fk(S_1)$] -- [fermion] b -- [fermion] c [particle=$\Fk^\dagger(S_2)$],
d [particle=$\Wk(S_3)$] -- [fermion] e -- [fermion] f [particle=$\Wk^\dagger(S_4)$],
b -- [photon,edge label=$A$] e
};}
\end{equation}
for scalar operators $\Fk(P)$ with conformal dimension $\Dk_\Fk = \Dk_\Yk \pm \frac 12$. This diagram has been computed in Mellin space by \cite{Paulos:2011ie}:
\begin{equation}
\Pi^{\text{scalar}}(S_i;\Dk_\Fk) = \frac{\pi^{3/2}}2\int [d\gk]\,\fcy M(\gk_{ij})\prod_{i<j}\frac{\Gk(\gk_{ij})}{(-2P_i\cdot P_j)^{\gk_{ij}}}
\end{equation}
where
\begin{equation}\\
\fcy M(s,t) = \Gk\left(\Dk_\Wk\right)\Gk\left(\Dk_\Fk\right)(t-u)\sum_{m = 0}^\infty \frac{\left(\frac32-\Dk_\Wk\right)_m\left(\frac32-\Dk_\Fk\right)_m}{(s-2m-1)m!\Gk\left(m+\frac32\right)}\,. \\
\end{equation}

It is straightforward to generalize \eqref{sctoferm} and \eqref{eq:qedRel} to arbitrary fermion-fermion-spin $\ell$ vertices:
\begin{equation}\begin{split}
 {\scriptsize\feynmandiagram [inline=(b.base), horizontal = b to e] {
a [particle=$\Yk(S_1)$] -- [fermion] b -- [fermion] c [particle=$\Yk^\dagger(S_2)$],
d -- e [blob]-- f,
g -- e -- h,
b -- [gluon,edge label=$\varphi$] e
};} &=  m_1\<S_{1a}S_{2b}\> {\scriptsize\feynmandiagram [inline=(b.base), horizontal = b to e] {
a [particle=$\Dk_\Fk+\frac 12$] -- [fermion] b -- [fermion] c [particle=$\Dk_\Fk+\frac 12$],
d -- e [blob]-- f,
g -- e -- h,
b -- [gluon,edge label=$\varphi$] e
};} +  m_2\<\fcy E_{1a}\fcy E_{2b}\> {\scriptsize\feynmandiagram [inline=(b.base), horizontal = b to e] {
a [particle=$\Dk_\Fk-\frac 12$] -- [fermion] b -- [fermion] c [particle=$\Dk_\Fk-\frac 12$],
d -- e [blob]-- f,
g -- e -- h,
b -- [gluon,edge label=$\varphi$] e
};} \\
&+ m_3\<S_{1a}\fcy E_{2b}\> {\scriptsize\feynmandiagram [inline=(b.base), horizontal = b to e] {
a [particle=$\Dk_\Fk+\frac12$] -- [fermion] b -- [fermion] c [particle=$\Dk_\Fk-\frac12$],
d -- e [blob]-- f,
g -- e -- h,
b -- [gluon,edge label=$\varphi$] e
};} + m_4\<\fcy E_{1a}S_{2b}\> {\scriptsize\feynmandiagram [inline=(b.base), horizontal = b to e] {
a [particle=$\Dk_\Fk-\frac12$] -- [fermion] b -- [fermion] c [particle=$\Dk_\Fk+\frac12$],
d -- e [blob]-- f,
g -- e -- h,
b -- [gluon,edge label=$\varphi$] e
};}\,,
\end{split}\end{equation}
where the $m_i$'s are constants which depend on the specific fermion-fermion-boson vertex in the left-hand Witten diagram. For parity preserving vertices $m_3 = m_4 = 0$ while for parity violating vertices $m_1 = m_2 = 0$.

\subsection{Higher-derivative corrections for massless particles}
\label{DEVCOR}

As a final example, we shall discuss higher-derivative corrections to massless spinning correlators. For simplicity we begin by studying a $U(1)$ gauge field $A^{\ak\da}(T,\bT)$, which is dual to a conserved current $J_{ab}(S)$ on the boundary. In an effective theory in the bulk where such a $U(1)$ gauge field is present, such as in the low-energy expansion of the M-theory action around AdS$_4\times \fcy M_7$ for some compact Sasaki-Einstein manifold $\fcy M_7$ with a $U(1)$ isometry, we expect that the Maxwell Lagrangian will receive an infinite number of higher-derivative corrections. If the theory is parity preserving, the first few such interactions are
\begin{equation}\begin{split}
\fcy L = -\frac 1 {4}F_{\m\n}F^{\m\n} &+ \frac{g_1\ell_s^4}{8}\left(F^{\ak\bk}F_{\ak\bk}+\bar F^{\dot\ak\dot\bk}\bar F_{\dot\ak\dot\bk}\right)^2 + \frac{g_2\ell_s^4}{8}\left(F^{\ak\bk}F_{\ak\bk}-\bar F^{\dot\ak\dot\bk}\bar F_{\dot\ak\dot\bk}\right)^2\\
&+\frac{g_3\ell_s^4}{8}\left(F_{\ak\bk}F^{\bk\gk}F_{\gk\dk}F^{\dk\ak}+\bar F_{\dot\ak\dot\bk}\bar F^{\dot\bk\dot\gk}\bar F_{\dot\gk\dot\dk}\bar F^{\dot\dk\dot\ak}\right) + O(\ell_s^6)\,,
\end{split}\end{equation}
where the $\ell_s$ is some interaction length scale, the coefficients $g_i$ are dimensionless, and we define
\begin{equation}\label{FabDef}
F^{\ak\bk} \equiv {\Nb^{(\ak}}_{\dot\ak} A^{\bk)\dot\ak} = \frac12(\sk^\m)^\ak{}_\da(\sk^\n)^{\bk\da}F_{\m\n}\,,\quad \bar F^{\da\db} \equiv {\Nb_{(\ak}}^{(\da|} A^{\bk|\db)} = \frac12(\sk^\m)_\ak{}^\da(\sk^\n)^{\ak\db}F_{\m\n} \,.
\end{equation}
Our aim is to use the technology developed so far to derive the leading order (in $g_i$) correction to $\<JJJJ\>$ induced by these higher derivative terms:
\begin{equation}
\Pi(S_i) = \feynmandiagram [inline=(b.base), horizontal = f to b] {
a [particle=$J(S_3)$] -- [boson] b -- [boson]  c [particle=$J(S_1)$],
d [particle=$J(S_2)$] -- [boson]  b -- [opacity=0] e ,
f -- [opacity=0] b -- [boson] g [particle=$J(S_4)$],
h   -- [opacity=0]  b -- [opacity=0] i
};\,.
\end{equation}

As a first step, we can use \eqref{eq:JA2pt} and \eqref{FabDef} to compute the $\<JF\>$ and $\<J\bar F\>$ bulk-boundary propagators,
\begin{equation}
\langle J(\sk,P) F(\tk,X) \rangle = \frac{-2i\langle\sk\tk\rangle^2}{(-2P\cdot X)^3}\,,\quad \langle J(\sk,P) F(\bar\tk,X) \rangle = \frac{+2i\langle\sk\bar\tk\rangle^2}{(-2P\cdot X)^3} \,.
\end{equation}
Contact Witten diagrams are built from products of these bulk-boundary propagators, with all bulk $SO(3,1)$ indices contracted. Our strategy will be to relate the $\<JF\>$ and $\<J\bar F\>$ propagators to those of conformally-coupled scalar fields using the differential operators $S^A_a$ and $\fcy E^A_a$.

Conformally-coupled scalar fields are dual to operators with conformal dimension $\Dk = 1$ or $\Dk = 2$, depending on the choice of boundary conditions. There is however a slight issue we must deal with. As is apparent from \eqref{eq:EonScal}, the differential operator $\fcy E^A_a$ automatically annihilates any scalar function with conformal dimension $\Dk = 1$. To circumvent this issue, we introduce a rescaled version of the $\fcy E^A_a$ operator
 \es{hatE}{
\hat{\fcy E}^A_a = \left(\fcy D-\frac32\right)^{-1}\fcy E^A_a = \frac 1 {(2\fcy D-3)}S^A_a\frac{\nb}{\nb S^B_b}\frac{\nb}{\nb S^b_B} + \frac{\nb}{\nb S^a_A}\,,
 }
which does not suffer from this problem. We can now compute
\begin{equation}\label{jf12}\begin{split}
\hat{\fcy E}^A_{(a}\hat{\fcy E}^B_{b)}\frac{1}{(-2P\cdot X)} &=  \frac i2\left(\langle J_{ab}(S)F_{\ak\bk}(T,\bT)\rangle T^{A\ak}T^{B\bk}-\langle J_{ab}(S)\bar F_{\da\db}(T,\bT)\rangle \bT^{A\da}\bT^{B\db}\right) \,, \\
\hat{\fcy E}^{(A}_{(a}S^{B)}_{b)}\frac{1}{(-2P\cdot X)^2}    &=  -\frac16\left(\langle J_{ab}(S)F_{\ak\bk}(T,\bT)\rangle T^{A\ak}T^{B\bk}+\langle J_{ab}(S)\bar F_{\da\db}(T,\bT)\rangle \bT^{A\da}\bT^{B\db}\right) \,.
\end{split}\end{equation}
Introducing the concise notation ${\hat{\fcy E}^A_{\bm i}} = s_i^a \fcy (\hat{\fcy E}_i)^A_a$, where $(\hat{\fcy E}_i)^A_a$ acts on the $i^{\text{th}}$ operator, we can then immediately write down the leading corrections to $\Pi(S_i)$ in $\kk = L / \ell_s $:
\begin{equation}\begin{split}
\Pi(S_i) = -\frac{i}{\kk^4}\bigg[\Big(&g_1\langle\hat{\fcy E}_{\bm 1}\hat{\fcy E}_{\bm 2}\rangle^2\langle\hat{\fcy E}_{\bm 3}\hat{\fcy E}_{\bm 4}\rangle^2+g_3\langle\hat{\fcy E}_{\bm 1}\hat{\fcy E}_{\bm 2}\rangle\langle\hat{\fcy E}_{\bm 2}\hat{\fcy E}_{\bm 3}\rangle\langle\hat{\fcy E}_{\bm 3}\hat{\fcy E}_{\bm 4}\rangle\langle\hat{\fcy E}_{\bm 4}\hat{\fcy E}_{\bm 1}\rangle\Big)D_{1111}(P_i)\\ 
+9&g_2\langle\hat{\fcy E}_{\bm 1}\hat{\fcy E}_{\bm 2}\rangle\langle\hat{\fcy E}_{\bm 1}{\sk_2}\rangle\langle\hat{\fcy E}_{\bm 3}\hat{\fcy E}_{\bm 4}\rangle\langle\hat{\fcy E}_{\bm 3}{\sk_4}\rangle D_{1212}(P_i)\bigg]+\text{cyclic perms of 234} \,.
\end{split}\end{equation}

The generalization to gravity is straightforward. In this case, the metric perturbation $h^{\ak\bk\da\db}(T,\bT)$ is dual to the stress-tensor $T_{abcd}(S)$. The field strength is the spinor form of the Weyl tensor, given by
\begin{equation}\label{eqRDef}
R^{\ak\bk\gk\dk}(T,\bar T) = {\Nb^{(\gk}}_{\dot\ak}{\Nb^{\dk}}_{\dot\bk}h^{\ak\bk)\dot\ak\dot\bk}(T,\bar T)\,,\qquad \bar R^{\dot\ak\dot\bk\dot\gk\dot\dk}(T,\bar T) = {\Nb_{\ak}}^{(\dot \gk}{\Nb_{\bk}}^{\dot\dk|}h^{\ak\bk|\dot\ak\dot \bk)}(T,\bar T)\,.
\end{equation}
It is invariant under diffeomorphisms, and local corrections to the Einstein-Hilbert action can always be constructed from these tensors.  For simplicity we focus on the $R^4$ in supergravity, which is the square of the Bel-Robinson term \cite{Deser:1977nt}:
\begin{equation}
\fcy L_{R^4} = \frac{\lk\ell_s^8}{2}R^2\bar R^2 = \frac{\lk\ell_s^8}{8}\left[(R^2+\bar R^2)^2 - (R^2-\bar R^2)^2\right]\,,
\end{equation}
and study the contact diagram
\begin{equation}
\Pi_{R^4}(S_i) = \feynmandiagram [inline=(b.base), horizontal = f to b] {
a [particle=$T(S_3)$] -- [gluon] b -- [gluon]  c [particle=$T(S_1)$],
d [particle=$T(S_2)$] -- [gluon]  b -- [opacity=0] e ,
f -- [opacity=0] b -- [gluon] g [particle=$T(S_4)$],
h   -- [opacity=0]  b -- [opacity=0] i
};\,.
\end{equation}

To compute this diagram we should first begin with the $\<TR\>$ and $\<T\bar R\>$ bulk-boundary propagator, which we normalize such that
\begin{equation}
\<T(\sk,P)R(\tk,X)\> = \frac{-24\<\sk\tk\>^4}{(-2P\cdot X)^5}\,,\qquad \<T(\sk,P)\bar R(\bar\tk,X)\> = \frac{-24\<\sk\bar\tk\>^4}{(-2P\cdot X)^5}\,.
\end{equation}
We can then derive analogues of \eqref{jf12} for the graviton:
\begin{equation}\begin{split}
\hat{\fcy E}^A_{(a}\hat{\fcy E}^B_{b}\hat{\fcy E}^C_{c}&\hat{\fcy E}^D_{d)}\frac{1}{(-2P\cdot X)} \\
= -\frac 12&\left(\langle T_{abcd}(S)R_{\ak\bk\gk\dk}\rangle T^{A\ak}T^{B\bk}T^{C\gk}T^{D\dk}+\langle T_{abcd}(S)\bar R_{\da\db\dot\gk\dot\dk}\rangle T^{A\ak}T^{B\bk}T^{C\gk}T^{D\dk}\right) \\
\hat{\fcy E}^A_{(a}\hat{\fcy E}^B_{b}&\hat{\fcy E}^C_{c}S^D_{d)}\frac{1}{(-2P\cdot X)^2} \\
= -\frac i{20}&\left(\langle T_{abcd}(S)R_{\ak\bk\gk\dk}\rangle T^{A\ak}T^{B\bk}T^{C\gk}T^{D\dk}-\langle T_{abcd}(S)\bar R_{\da\db\dot\gk\dot\dk}\rangle \bT^{A\da}\bT^{B\db}\bT^{C\dot\gk}\bT^{D\dot\dk}\right)\,,
\end{split}\end{equation}
and using these relations it is straightforward to show that
\begin{equation}\begin{split}
\Pi_{R^4} =\, &\frac{-i\lk}{16\kk^8}\langle\hat{\fcy E}_{\bm 1}\hat{\fcy E}_{\bm 2}\rangle^3\langle\hat{\fcy E}_{\bm 3}\hat{\fcy E}_{\bm 4}\rangle^3\left(\langle\hat{\fcy E}_{\bm 1}\hat{\fcy E}_{\bm 2}\rangle\langle\hat{\fcy E}_{\bm 3}\hat{\fcy E}_{\bm 4}\rangle D_{1111}(P_i)-
100\langle\hat{\fcy E}_{\bm 1}{\sk_2}\rangle\langle\hat{\fcy E}_{\bm 3}{\sk_4}\rangle D_{1212}(P_i)\right)\\
&+\text{cyclic perms of 234}\,. \\
\end{split}\end{equation}

\section{Spinning bulk-to-bulk propagators}
\label{BULKBULK}

In this section we use the bispinor formalism to describe spinning bulk-to-bulk propagators and their applications.

\subsection{Bulk fermion propagator}
\label{BulkFermion}
In this section we derive the fermion bulk-to-bulk propagator in the bispinor formalism. Using differential operators the fermion bulk-to-bulk propagator can be related to that of scalars, as derived using $\fcy N = 1$ supersymmetry in \cite{Burges:1985qq}, and this was used to compute fermion exchange diagrams in \cite{Kawano:1999au}. As we shall see, the bispinor formalism allows us to rederive these results in a concise fashion.  See also \cite{Faller:2017hyt} for an evaluation of a spinor exchange diagram using the embedding space formalism introduced in \cite{Iliesiu:2015qra}. 

Let us begin by deriving fermion bulk-to-bulk propagator. For simplicity we study a free Majorana fermion
\begin{equation}\label{eq:majlag}
\fcy L = i\bar\psi^\da\Nb_{\ak\da}\yk^\ak - \frac 12 m\yk^\ak\yk_\ak-\frac 12 m^*\bar\yk_\da\bar\yk^\da\,.
\end{equation}
From this Lagrangian we can derive the equations of motion
\begin{equation}\label{eq:majlagmot}
i\Nb_{\ak\da}\yk^\ak - m^*\yk_\da = 0\,,\qquad i\Nb_{\ak\da}\bar\yk^\da+m\yk_\ak = 0\,.
\end{equation}

The phase of $m$ is arbitrary, as under a field redefinition $\yk\rightarrow e^{i\qk}\yk$ and $\bar\yk\rightarrow e^{-i\qk}\bar\yk$ we find that $m\rightarrow m e^{-2i\qk}$. We will fix our conventions for $m$ using the bulk-boundary propagator. The fields $\psi$ and $\bar\psi$ are dual on the boundary to a real spinor $\Yk_a(S)$. We will use the bulk-boundary propagators 
\begin{equation}\label{weylbubo}\begin{split}
\< \Yk^a(S)\yk^\ak(T,\bT)\> &= \frac{\Gamma\left(\Dk_\Yk+\frac12\right)\<S^aT^\ak\>}{(-2P\cdot X)^{\Dk_\Yk+\frac12}}\,,\\ 
\< \Yk^a(S)\bar\yk^\da(T,\bT)\> &= \frac{-\Gamma\left(\Dk_\Yk+\frac12\right)\<S^a\bT^\da\>}{(-2P\cdot X)^{\Dk_\Yk+\frac12}} \,.
\end{split}\end{equation}
By comparing the equation of motion \eqref{eq:majlagmot} to that satisfied by the bulk-boundary propagator \eqref{eq:fermEoM}, we find that 
\begin{equation}
m^* = m  = \Dk_\Yk-\frac32\,.
\end{equation}
We will therefore take $m$ to be real and positive if $\Dk_\Yk>\frac32$ and real and negative if $1\leq\Dk_\Yk<\frac32$. The unitarity bound requires that $\Dk_\Yk\geq1$.

Let us now turn to the bulk-to-bulk propagators
\begin{equation}\begin{split}
\<\yk^\ak(T_1,\bT_1)\bar\yk^\db(T_2,\bT_2)\> &= \<T_1^\ak\bar T_2^\db\>F_1(u)\,, \\
\<\yk^\ak(T_1,\bT_1)\yk^\bk(T_2,\bT_2)\> &= \<T_1^\ak T_2^\bk\>F_2(u)\,,
\end{split}\end{equation}
where $u \equiv \frac 12 (X_1-X_2)^2$ is the chordal distance. Using the equations of motion, we can derive a set of coupled differential equations for $F_1(u)$ and $F_2(u)$. We find that the equations of motion\footnote{We ignore the delta functions which may appear on the right-hand side of \eqref{fermeom1}. These merely fix the $u\rightarrow0$ behaviour of the propagator, which we achieve using \eqref{eq:shortlim}.}
\begin{equation}\label{fermeom1}
\left\<\yk_\ak\left[i(\Nb_2)_{\bk\db}\yk^\bk-m\bar\yk_\db\right]\right\> = 0 \,,\qquad  \left\<\yk_\ak\left[i(\Nb_2)_{\bk\db}\bar\yk^\db+m\yk_\bk\right]\right\> = 0 \,, 
\end{equation}
give the equations
\begin{equation}\label{fermdifop}\begin{split}
uF_1'(u) + 2F_1(u) - mF_2(u) &= 0 \,, \\
(u+2)F_2'(u)+2F_2(u) - m F_1(u) &= 0\,. \\
\end{split}\end{equation}
We can solve these equations in terms of the scalar bulk-to-bulk propagator\footnote{An explicit expression for the propagator can be found Appendix~\ref{sec:SCLWITDIA}.} $G_\Dk(u)$, with
\begin{equation}\begin{split}
F_1(u) &= -\frac i2\left(\left(\Dk_\Yk-\frac12\right)G_{\Dk_\Yk-\frac12}(u)+(u+2) G_{\Dk_\Yk-\frac12}'(u) \right)\,,\\
F_2(u) &= -\frac i2\left(\left(\Dk_\Yk-\frac12\right)G_{\Dk_\Yk-\frac12}(u)+u G_{\Dk_\Yk-\frac12}'(u)\right)\,.\\
\end{split}\end{equation}
The overall normalization is fixed by requiring that
\begin{equation}\label{eq:shortlim}
\lim_{X_1\rightarrow X_2} \<\yk_\ak(T_1,\bT_1)\bar\yk_\db(T_2,\bT_2)\> = \frac{i\<T_{1\ak}\bT_{2\db}\>}{8\pi^2 u^2} + O(u^{-1})\,.
\end{equation}

Having computed the bulk-to-bulk propagator for a Majorana fermion, let us rewrite is in a more computationally convenient form. Using the identities
\begin{equation}\begin{split}
T^A_{1\ak}T^B_{2\bk}(D_2)_{AB} f(u) &= \frac i2\<T_{1\ak}T_{2\bk}\> u f'(u)\,,\\
T^A_{1\ak}\bT^B_{2\db} (D_2)_{AB}f(u) &= \frac i2\<T_{1\ak}\bT_{2\db}\> (u+2)f'(u)\,,
\end{split}\end{equation}
we can rewrite the bulk-to-bulk propagator as:
\begin{equation}\label{eq:fermbulkbulk}\begin{split}
\<\yk_\ak(T_1,\bT_1)\yk_\bk(T_2,\bT_2)\> &= T^A_{1\ak}T^B_{2\bk}F_{AB}\,,\qquad \<\yk_\ak(T_1,\bT_1)\bar\yk_\db(T_1,\bT_1)\> = T^A_{1\ak}\bar T^B_{2\db}F_{AB}\,,
\end{split}\end{equation}
where we define
\begin{equation}\begin{split}
F_{AB} &= \left[D_{AB} - \frac i2\left(\Dk_\Yk-\frac12\right) \epsilon_{AB}\right]G_{\Dk_\Yk-\frac12}(u) \,.
\end{split}\end{equation}
Thus, we can derive the fermion bulk-to-bulk propagator by acting with differential operators on the scalar bulk-to-bulk propagator.

Having derived \eqref{eq:DircLag}, let us now consider the Lagrangian
\begin{equation}
\fcy L = i\bar\psi^\da\Nb_{\ak\da}\yk^\ak - \frac m2 \left(\yk^\ak\yk_\ak+\bar\yk_\da\bar\yk^\da \right)  - \frac12(\Nb\fk)^2 - M^2\fk^2 - \frac 12g\fk\left(\yk^\ak\yk_\ak+\bar\yk_\da\bar\yk^\da \right)\,,
\end{equation}
coupling a Majorana fermion $\yk$ to a real scalar field $\fk$. As usual, $\Fk(S)$ is the boundary dual of $\fk$ with bulk-boundary propagator
\begin{equation}
\< \Fk(S)\fk(T,\bT)\> = \frac{\Gk(\Dk_\Fk)}{(-2 P\cdot X)^{\Dk_\Fk}}\,.
\end{equation}
We will first study the fermion-exchange diagram (see also \cite{Faller:2017hyt}):
\begin{equation}\label{eq:fermExchD}
\Pi(S_i) = \feynmandiagram [inline=(b.base), horizontal = b to e] {
a [particle=$\Fk(S_1)$] -- [scalar] b --  c [particle=$\Yk(S_3)$],
d [particle=$\Fk(S_2)$] -- [scalar]  e --  f [particle=$\Yk(S_4)$],
b --  e
};\,.
\end{equation}
To evaluate this diagram, we relate it to the scalar diagram
\begin{equation}
\Pi^{\text{scalar}}(S_i) = \feynmandiagram [inline=(b.base), horizontal = b to e] {
a [particle=$\Fk(S_1)$] -- [scalar] b  -- c [particle=$\Fk_+(S_3)$],
d [particle=$\Fk(S_2)$] -- [scalar]  e -- f [particle=$\Fk_+(S_4)$],
b -- [edge label=$\fk_-$]  e
};\,,
\end{equation}
where, as in Section \ref{FERMSCT}, $\Fk_\pm(P)$ and $\fk_\pm(X)$ are fictitious real scalars of dimension $\Delta_\pm = \Delta_\Yk\pm\frac12$. This requires us to use the Majorana analogues of \eqref{eq:fermS}
\begin{equation}\label{eq:fermSM}\begin{split}
S^A_a\<\Fk_+(P)\fk_+(X)\> &= -\frac{i}2\left\langle\Yk_a\left(\yk_\ak T^{A\ak}-\bar\yk_{\dot\ak}\bar T^{A\dot\ak}\right)\right\rangle \,,\\
\end{split}\end{equation}
Combining \eqref{eq:fermSM} with \eqref{eq:fermbulkbulk}, it is straightforward to check that
\begin{equation}
\Pi(S_i) = \left[-4\<S_{4b}(\fcy D_2+\fcy D_4)S_{3a}\> + 2i\left(\Dk_\Yk-\frac12\right) \<S_{3a}S_{4b}\> \right] \Pi^{\text{scalar}}(S_i)\,.
\end{equation}
We have thus seen that the fermion exchange diagram can be computed by applying differential operators to the scalar exchange diagram.

\subsection{The boundary limit}
\label{BOUNDLIM}

We will now turn to the task of taking the boundary limit of a bulk point in our formalism. This will allow us to verify that the boundary limit of the bulk-to-bulk propagator gives the correct bulk-boundary propagator. For simplicity, let us start with the case of a scalar field $\fk(X)$. Consider a path $X^I(s)$ in AdS$_4$ which approaches a boundary point $P^I$ as $s\rightarrow\infty$. Without loss of generality, at large $s$ we can parametrize the curve as:
\begin{equation}\label{pathbound}
X^I(s) = e^s P^I + \frac 12 e^{-s} V^I+\dots\,, \quad \text{ where }\quad  V_IP^I = -1\,,
\end{equation}
and the subleading terms go to zero faster than $e^{-s}$. We can then define the boundary dual $\Fk(P)$ of a bulk operator $\fk(X)$ by the equation
\begin{equation}\label{scboundlim}
\Fk(P) = \lim_{s\rightarrow\infty} \fcy N_{\Dk} e^{\Dk s}\fk\left(X(s)\right)\,,
\end{equation} 
where $\fcy N_\Dk = 2\pi^{3/2}\Gk(\Dk-1/2)$ is an overall normalization factor. With this definition, it is straightforward to check that
\begin{equation}
\lim_{s\rightarrow\infty} \fcy N_{\Dk}e^{\Dk s}\<\fk\left(X(s)\right)\fk(Y)\> = \lim_{s\rightarrow\infty} \fcy N_{\Dk}e^{\Dk s} G_\Dk\left(X(s);Y\right) = \frac{\Gk(\Dk)}{(-2P\cdot Y)^\Dk} = \<\Fk(P)\fk(Y)\>  \,,
\end{equation}
so that the boundary limit of the bulk-to-bulk scalar propagator gives us the bulk-boundary scalar propagator. By taking the bulk operator to the boundary in the bulk-boundary propagator, we can then derive the boundary-boundary two point function, \eqref{eq:scalarNorm}.

To extend our computation to the spinning case, let us first introduce the bispinor $S^A_a$ associated to $P^I$. We can then define
\begin{equation}\label{eq:pathspinors}\begin{split}
T^A_\ak(s) &= \lk^a_\ak(s)\left[e^{s/2} S^A_a - \frac {i e^{-s/2}}2\slashed V^A{}_BS^B_a+\dots\right]\,,\\
\bar T^A_\da(s) &=  \bar\lk^a_\da(s)\left[e^{s/2} S^A_a + \frac {i e^{-s/2}}2\slashed V^A{}_BS^B_a+\dots\right]\,,
\end{split}\end{equation}
where as usual $\slashed V^A{}_B = V^I(\Gk_I)^A{}_B$, and where the tensors $\lk^a_\ak$ and $\bar\lk^a_\da$ satisfy:
\begin{equation}
\epsilon_{ab}\lk^a_\ak\lk^b_\bk = \epsilon_{\ak\bk}\,,\qquad \epsilon_{ab}\bar\lk^a_\da\bar\lk^b_\db = \epsilon_{\da\db}\,.
\end{equation}
It is then straightforward to check that (up to subleading terms) \eqref{eq:pathspinors} satisfy the bispinor conditions and that
\begin{equation}
X^I(s) = -\frac14\<T_\ak(s)\Gk^I T^\ak(s)\> = -\frac14\<\bT_\da(s)\Gk^I\bT^\da(s)\>\,.
\end{equation}
Thus, when taking the boundary limit we can work with the variables $T^A_\ak(s)$ and $\bT^A_\ak(s)$ rather than using $X^I(s)$.

Taking boundary limits is particularly straightforward using the Poincar\'e patch coordinates \eqref{TTbarExplicit} and flat-space coordinates \eqref{eq:CFTSpinor} for the bulk and boundary bispinors respectively. In these coordinates, we find that
\begin{equation}\label{poinpath}\begin{split}
T^A_a(\vec x,z) &= \frac 1{\sqrt z}S^A_a(\vec x) - \frac{i\sqrt{z}}2\slashed{\tilde P}^A{}_B S^B_a(\vec x)\,, \\ 
\bT^A_a(\vec x,z) &= \frac 1{\sqrt z}S^A_a(\vec x) + \frac{i\sqrt{z}}2\slashed{\tilde P}^A{}_B S^B_a(\vec x) \,,
\end{split}\end{equation}
where $\tilde P = (0,0,0,-1,1)$ is the boundary point at infinity. We can recover \eqref{poinpath} from \eqref{eq:pathspinors} by taking
\begin{equation}
s = -\log(z)\,,\qquad \lk^\ak_a = \bar\lk^\da_a = \begin{pmatrix} 1&0\\0&1\end{pmatrix}\,,\qquad V^I = \tilde P^I\,,
\end{equation}
and setting the subleading terms to zero. Therefore, in Poincar\'e patch coordinates we can take the boundary limit by taking radial coordinate $z$ to zero.

Let us now consider a fermionic field $\yk_\ak(T,\bT)$ and its conjugate $\bar\yk_\da(T,\bT)$. Using the bulk-to-bulk propagators derived in Section \ref{BulkFermion}, we can compute
\begin{equation}\label{us532}\begin{split}
\lim_{s\rightarrow\infty} s^{\Dk_\Yk}\<\bar\yk^\da(T(s),\bT(s))\yk^\bk(T_2,\bT_2)\> &= \frac{i\Gk\left(\Dk_\Yk+\frac12\right)}{2\pi^{3/2}\Gk(\Dk_\Yk-1)}\frac{\bar\lk^\da_a\<S^a T_2^\bk\>}{(-2P\cdot X_2)^{\Dk_\Yk+\frac12}}\,,\\
\lim_{s\rightarrow\infty} s^{\Dk_\Yk}\<\yk^\ak(T(s),\bT(s))\yk^\bk(T_2,\bT_2)\> &= \frac{-i\Gk\left(\Dk_\Yk+\frac12\right)}{2\pi^{3/2}\Gk(\Dk_\Yk-1)}\frac{\lk^\ak_a\<S^a T_2^\bk\>}{(-2P\cdot X_2)^{\Dk_\Yk+\frac12}}\,.\\
\end{split}\end{equation}
By defining the boundary operator
\begin{equation}
\Yk_a(S) = \lim_{s\rightarrow\infty} i\fcy N_{\Yk} s^\Delta\Big[\lk^\ak_a\yk_\ak(T(s),\bT(s))-\bar\lk^\da_a\bar\yk_\da(T(s),\bT(s))\Big] \,,
\end{equation}
with $\fcy N_\Yk = \pi^{3/2}\Gk(\Dk-1)$, we can then easily use \eqref{us532} to show that the bulk-boundary propagators are given by \eqref{weylbubo}. Taking the bulk operator to the boundary in the bulk-boundary propagator, we can compute the boundary two-point function
\begin{equation}
\<\Yk(\sk_1,P_1)\Yk(\sk_2,P_2)\> = i2^{3-2\Dk_\Yk}\pi^2(2\Dk_\Yk-1)\Gk(2\Dk_\Yk-2) \frac{\<\sk_1\sk_2\>}{(-2P_1\cdot P_2)^{\Dk_\Yk+\frac12}}\,.
\end{equation}

\subsection{Photon propagator}
We obtain the photon bulk propagator using the method of \cite{DHoker:1999bve}, translated into our formalism. In the Poincar\'e patch, the propagator of \cite{DHoker:1999bve} has the form
\begin{equation}
\< A_i(x)A_j(x') \> = (\nb_{i}\nb_{j}'u)G_2(u) + \nb_{i}\nb_{j}' H(u)\,,\quad \text{ where } \quad G_2(u) = \frac 1 {u(u+2)}\,.
\end{equation}
The first term is the physical part, proportional to the AdS$_4$ propagator of a conformally coupled scalar with $m^2  = -2$ and $\Delta = 2$. The second term is an arbitrary pure gauge term which can be ignored. As in the previous section, we use $u = \frac 12 (X-X')^2$ to denote the chordal distance.

It is straightforward to translate this expression into the bispinor formalism:\footnote{We explain in detail how to relate the covariant derivative to derivatives with respect to arbitrary coordinates in Appendix~\ref{CONNECTION}.}
\begin{equation}\begin{split}\label{vectorProp}
\< A_{\ak\da}(T,\bT)A_{\bk\db}(T',\bT')\> &= (\Nb_{\ak\da}\Nb_{\bk\db}'u) G_2 (u) + \Nb_{\ak\da}\Nb_{\bk\db}' H(u)\\
&= \frac14\<T_\ak\Gk^I\bar T_\da\>\<T_\bk'\Gk_I\bT_\db'\>G_2(u)  + \Nb_{\ak\da}\Nb_{\bk\db}' H(u)\,.
\end{split}\end{equation}
With the assistance of the identity
\begin{equation}
\Nb_{\ak\da}\Nb_{\gk\dot\gk}f(u) = -\frac 14 \<T_\ak X'\bT_\da\>\<T_\gk X'\bT_{\dot\gk}\> f''(u) -2(u+1)\epsilon_{\ak\gk}\epsilon_{\da\dot\gk}f'(u)\,,
\end{equation}
one can then check that the Maxwell equation is satisfied:
\begin{equation}
\left\<\Nb^{\ak\da}\left(\Nb_{\ak\da}A_{\gk\dot\gk}(T,\bT) - \Nb_{\gk\dot\gk}A_{\ak\da}(T,\bT)\right)A_{\bk\db}(T',\bT')\right\> = \Nb_{\bk\db}'\Lk_{\gk\dot\gk}\,.
\end{equation}
The $\Lk_{\gk\dot\gk}(u)$ term on the right-hand side is a pure gauge term, required to avoid issues due to the non-invertability of the Maxwell operator.\footnote{The term delta function $\delta^4(X;X')$ is implicitly included because we will choose solutions with the required short distance behaviour as $u\rightarrow 0$.} 

We can take the boundary limit of our vector bulk-to-bulk propagator in order to compute the bulk-boundary propagator. The gauge field $A_{\ak\da}(T(s),\bT(s))$ at the boundary will split into $SO(2,1)$ vector and scalar operators. The scalar operator is an unwelcome gauge artefact, which we must eliminate. To achieve this goal, we expand the gauge terms in \eqref{vectorProp}:
\begin{equation}\label{vectorProp2}\begin{split}
\<A_{\ak\da}(T,\bT)A_{\bk\db}(T',\bT')\>
&= \frac14 \<T_\ak\Gk^I\bT_\da\>\<T_\bk'\Gk_I\bT_\db'\>\left[(G_2(u) + H'(u)+(u+2)H''(u)\right]\\
&-\<T_\ak T_\bk'\>\<\bT_\da\bT_\db'\>H''(u)\,.
\end{split}\end{equation}
Computing the limits
\begin{equation}\begin{split}
\lim_{s\rightarrow\infty}& e^{-s}\<T_\ak(s)\Gk^I\bT_\da(s)\> = -2\epsilon_{ab}\lk_\ak^a\bar\lk_\da^b P^I\,, \\
\lim_{s\rightarrow\infty}& e^{-s}\<T_\ak(s) T_\bk'\>\<\bT_\da(s)\bT_\db'\> = \lk_\ak^a\bar\lk_\da^b \<S_aT_\bk'\>\<S_b\bT_\db'\>\,,
\end{split}\end{equation}
we see that the first term in \eqref{vectorProp2} contributes to a scalar bulk-boundary propagator while the second term corresponds to a vector bulk-boundary propagator. Note however that for generic gauge choices the boundary scalar will dominate the boundary vector. To eliminate the boundary scalar,  we must fix the large $u$ behaviour of $H(u)$ to take the form:
\begin{equation}
H(u) = -\frac 1u + \frac{3}{2u^2} +\dots\,.
\end{equation}
For this specific choice, we find that
\begin{equation}
\lim_{s\rightarrow\infty} e^{2s}\<A_{\ak\da}(T(s),\bT(s))A_{\bk\db}(T',\bT')\> = \lk_\ak^a\bar\lk_\da^b \frac{2\<S_aT_\bk'\>\<S_b\bT_\db'\>}{(-2P\cdot X)^3}\,.
\end{equation}
By contracting both sides with $\lk^\ak_a\bar\lk^\da_b$ we can thus obtain the desired vector bulk-boundary propagator.

\subsection{Higher spin propagators}
Although there is interesting work on special cases in the literature \cite{Bena:1999py,Bena:1999be,Naqvi:1999va,Anguelova:2003kf,Leonhardt:2003qu,Basu:2006ti,Faizal:2011sa,Balitsky:2011tw,Costa:2014kfa}, a compact treatment of bulk propagators for general spin fields remains a challenge.  In this section we apply our bispinor machinery and obtain  simple results for bulk fields which describe the purely chiral and anti-chiral components of the principal Lagrangian field. These fields are obtained by differentiation of the principal fields for general spin.

For bosonic fields, i.e. integer $\ell$ we define
\begin{equation}\begin{split}
\fcy F_{\ak_1\dots\ak_{2\ell}}  &= \Nb_{(\ak_{\ell+1}}{}^{\da_1}\dots\Nb_{\ak_{2\ell}}{}^{\da_{\ell}} \varphi_{\ak_1\dots\ak_{\ell})\da_1\dots \da_{\ell}}\,, \\
\bar{\fcy F}_{\da_1\dots\da_{2\ell}}  &= \Nb^{\ak_1}{}_{(\da_{\ell+1}}\dots\Nb^{\ak_\ell}{}_{\da_{2\ell}} \varphi_{|\ak_1\dots\ak_{\ell}|\da_1\dots \da_{\ell})},
\end{split}\end{equation}
For fermionic fields with $\ell = n +\frac12$ we consider
\begin{equation}\begin{split}
\fcy F_{\ak_1\dots\ak_{2n+1}}  &= \Nb_{(\ak_{n+2}}{}^{\da_1}\dots\Nb_{\ak_{2n+1}}{}^{\da_{n}} \psi_{\ak_1\dots\ak_{n+1})\da_1\dots \da_{n}}\,, \\
\bar{\fcy F}_{\da_1\dots\da_{2n+1}}  &= \Nb^{\ak_1}{}_{(\da_{n+2}}\dots\Nb^{\ak_n}{}_{\da_{2n+1}} \bar\psi_{|\ak_1\dots\ak_n|\da_1\dots \da_{n+1})}\,, \\
\end{split}\end{equation}
Our treatment includes massless fields, $\Delta=\ell+1$,  whose chiral components describe (linearized) gauge invariant field strengths. So for $\ell = 1$ we have the usual electromagnetic field strength, for $\ell = \frac32$ we have the gravitino field strength and for $\ell = 2$ we have the Weyl curvatures $R^{\ak\bk\gk\dk}$ and $\bar R^{\da\db\dot\gk\dot\dk}$, see \eqref{eqRDef}.

Before computing the bulk-to-bulk propagator, we first consider the simpler case of the bulk-boundary propagator. The boundary dual of a fields $\varphi(T,\bT)$ or $\psi(T,\bT)$ is a real spin-$2\ell$ operator $\fcy O_{a_1\dots a_{2\ell}}(S)$. We normalize our bulk-boundary propagators so that
\begin{equation}\begin{split}\label{chiralbbp}
\<\fcy O(\sk,P)\fcy F(\tk,X)\>           &= (-i)^{\lfloor\ell\rfloor}\frac{\Gk(\Dk+\ell)\<\sk\tk\>^{2\ell}}{(-2P\cdot X)^{\Dk+\ell}}\,, \\ 
\<\fcy O(\sk,P)\bar{\fcy F}(\bar\tk,X)\> &= (-1)^{2\ell}i^{\lfloor\ell\rfloor}\frac{\Gk(\Dk+\ell)\<\sk\bar\tk\>^{2\ell}}{(-2P\cdot X)^{\Dk+\ell}}\,,
\end{split}\end{equation}
which in particular generalizes our previous conventions for scalars \eqref{sbbp} and Majorana fermions \eqref{weylbubo}. Note that because $\fcy F$ and $\bar{\fcy F}$ are Hermitian conjugates, the two bulk-boundary propagators in \eqref{chiralbbp} are related by Hermitian conjugation.

With the bulk-boundary propagators out of the way, let us now turn to the bulk-to-bulk propagator. In our formalism conformal symmetry fixes the two-point function of a chiral or antichiral field up to a single arbitrary function of $u = -1-X_1\cdot X_2$:
\begin{equation}\label{gpforms}
\begin{split}
\<\fcy F(\tk_1,X_1)\bar{\fcy F}(\bar\tk_2,X_2) \> &= \<\tk_1\bar\tk_2\>^{2\ell} F_1(u)\,,\\
\<\fcy F(\tk_1,X_1)\fcy F(\tk_2,X_2) \> &= \<\tk_1\tk_2\>^{2\ell} F_2(u)\,.
\end{split}\end{equation}
To determine $F_1(u)$ and $F_2(u)$, we use the equations of motion \eqref{eq:BBEoM}, which reduce to the second order differential equations 
\begin{equation}\label{photprop}\begin{split}
u(u+2)F_1''(u) +(4+4\ell+2(\ell+2)u) F_1'(u) - (\Dk+\ell)(\Dk-\ell-3)F_1(u) &= 0\,, \\
u(u+2)F_2''(u) +(4+2(\ell+2)u) F_2'(u) - (\Dk+\ell)(\Dk-\ell-3)F_2(u) &= 0\,. \\
\end{split}\end{equation}
At large $u$, these equations each have two linearly independent solutions, one which goes as $\propto u^{\Dk-\ell-3}$ and the other $\propto u^{-\Dk-\ell}$. In the boundary limit these correspond to boundary operators of dimension $3-\Dk$ and $\Dk$, assuming that the boundary dual has dimension $\Delta$,  only the latter solution is 
physical.\footnote{For fermions and scalars with sufficiently small values of $\Dk$, both solutions are allowed, and we must therefore specify which choice of boundary condition we have made in order to construct the bulk-bulk propagator.} 
One further degree of freedom can be eliminated by imposing 
\begin{equation}
\lim_{u\rightarrow0} u^{2\ell+1}F_1(u) = \frac{(-1)^\ell}{8\pi^2}\,,
\end{equation}
which normalizes the short distance behaviour of $F_1(u)$, and this completely fixes $F_1(u)$. To fix $F_2(u)$, we must impose a boundary condition relating $F_1(u)$ and $F_2(u)$. To derive this condition we note that 
\begin{equation}\label{boundFJ}
\fcy O_{a_1\dots a_{2\ell}}(S) = \lim_{s\rightarrow\infty} \fcy N_{\Dk,\ell}e^{\Dk s} \lk^{\ak_1}_{a_1}\dots\lk^{\ak_{2\ell}}_{a_{2\ell}}\fcy F_{\ak_1\dots\ak_{2\ell}}(T(s),\bT(s))
\end{equation}
for some constant $\fcy N_{\Dk,\ell}$. Taking the boundary limit of $\fcy F(\tk_1,X_1)$ for the two correlators \eqref{gpforms} and comparing to the bulk-boundary propagators \eqref{chiralbbp}, we find that
\begin{equation}
\lim_{u\rightarrow\infty}u^{\Delta+\ell}\left(F_1(u)-(-1)^{\lceil\ell\rceil} F_2(u)\right) = 0\,.
\end{equation}
With these conditions we can fully determine $F_1(u)$ and $F_2(u)$:
\begin{equation}\label{photprop2}\begin{split}
F_1(u) &= \frac{\Gk(\Dk+\ell)}{A_{\Dk,\ell}}(2u)^{-\Dk-\ell}{}_2F_1\left(\Dk-\ell-1,\Dk+\ell,2\Dk-2,-\frac2u\right)\\
F_2(u) &= \frac{(-1)^{\lceil\ell\rceil}\Gk(\Dk+\ell)}{A_{\Dk,\ell}}(2u)^{-\Dk-\ell}{}_2F_1\left(\Dk+\ell-1,\Dk+\ell,2\Dk-2,-\frac2u\right)\\
\text{ with } A_{\Dk,\ell} &= \frac{2\pi^{3/2}\Gk\left(\Dk-\frac12\right)\Gk(2\ell+1)}{(-1)^\ell\left(\Dk-1\right)_\ell}\,.
\end{split}\end{equation}
At the unitarity bound, $\Dk = \ell+1$,  these functions take a particularly simple form:
\begin{equation}\begin{split}
\<\fcy F(\tk_1,X_1)\bar{\fcy F}(\bar\tk_2,X_2) \> &= \frac{(-1)^\ell\<\tk_1\bar\tk_2\>^{2\ell}}{8\pi^2u^{2\ell+1}}\,,\\
\<\fcy F(\tk_1,X_1)\fcy F(\tk_2,X_2) \> &= \frac{(-1)^{\ell+\lceil \ell\rceil}\<\tk_1\tk_2\>^{2\ell}}{8\pi^2(u+2)^{2\ell+1}}\,.
\end{split}\end{equation}

Now that we know the bulk-to-bulk and bulk-boundary propagator, we can fix the normalization in \eqref{boundFJ} relating the bulk operator $\fcy F$ to the boundary operator $\fcy O$:
\begin{equation}
\fcy N_{\Dk,\ell} =  \frac{i^{2\ell+\lfloor\ell\lceil} 4^{\Dk-2}\Gk(2\Dk-2)\Gk(2\ell+1)}{\Gk(\Dk+\ell-1)}\,.
\end{equation}
By taking the boundary limit of the bulk operator in the bulk-boundary propagator, we can then compute the boundary two point function:
\begin{equation}
\<\fcy O(\sk_1,P_1)\fcy O(\sk_2,P_2)\> = i^\ell4^{2-\Dk}\pi^2(\Dk+\ell-1)\Gk(2\Dk-2)\Gk(2\ell+1)\frac{\<\sk_1\sk_2\>^{2\ell}}{(-2P_1\cdot P_2)^{\Dk+\ell}}\,.
\end{equation}

It has long been known \cite{PhysRevD.18.3565} that the equations governing bulk propagators enjoy the antipodal symmetry of AdS spacetime,  related to the reflection $X^I\rightarrow -X^I$ in embedding space.  Under this symmetry the chordal distance variable transforms as $u \rightarrow -(2+u)$ and $2+u\rightarrow -u$.  For instance, this symmetry is visible if we compute the Laplacian of a scalar quantity:
\begin{equation}
\Nb^2F(u) = u(u+2)F^{''}(u) + 4(u+1)F'(u)\,.
\end{equation}
Because of this symmetry, for any solution $F(u)$ which satisfies the equation of motion for a massive scalar field, $F(-u-2)$ is another solution. More generally the equations \eqref{photprop} also satisfy antipodal symmetry: if a pair of functions $F_1(u),~F_2(u)$ solve these equations, then so does $\tilde F_1(u)=F_2(-2-u),~ \tilde F_2(u)=F_1(-u-2)$. This means that once we have found a solution for $F_1(u)$ to the first equation, we immediately find that $F_1(-2-u)$ then solves the second equation.

\section{Discussion}
\label{DISCUSSION}

In this paper we have developed a new embedding space formalism for AdS$_4$/CFT$_3$ in which conformally covariant differential operators take particularly concise forms. This formalism has enabled us to greatly simplify the calculation of spinning Witten diagrams, as we illustrated with a number of examples.

An application of this formalism that we hope to pursue in the future is to the relation between CFT correlators of spinning operators and flat space scattering amplitudes of spinning particles.  This relation has been developed very explicitly for scalar correlators \cite{Penedones:2010ue,Fitzpatrick:2011ia,Fitzpatrick:2011hu,Fitzpatrick:2011dm}, building on the early ideas of \cite{Polchinski:1999ry,Susskind:1998vk,Giddings:1999jq}. In particular, the Mellin space representation of scalar CFT correlators  \cite{Mack:2009gy,Mack:2009mi}  gives a convenient representation that is easily related to the scattering amplitudes of the corresponding massless scalar particles in flat space, diagram by diagram.   (See also \cite{Goncalves:2014rfa} for some examples also including vector particles.)  We believe that because of its similarity to the spinor-helicity formalism for scattering amplitudes, our embedding-space formalism is particularly well-suited for exploring the generalization of the flat-space limit formula of \cite{Penedones:2010ue} to spinning correlators.  If one writes a spinning correlator in terms of a scalar correlator (whose flat space limit is understood) acted on by conformally-covariant differential operators, then the flat space limit of the spinning correlator would follow from identifying how these differential operators act on the scalar scattering amplitudes.

It would be particularly interesting to develop supersymmetric extensions of our formalism, since most top-down models of holography are supersymmetric. Spinor helicity variables can be easily extended to supersymmetric theories, at least in the massless case (for a textbook treatment see \cite{Elvang:2015rqa}), and so one may hope similar methods apply to AdS/CFT\@. For example, it is quite natural to extend our $Sp(4)\times GL(2,\mathbb R)$ spinor $S^A_a$ to an $OSp(\fcy N|4)\times GL(2,\mathbb R)$ spinor $S^{\dot A}_a \equiv (S^A_a,\theta^i_a)$, where $\theta^i_a$ is a Grassmannian field transforming as a vector under the $O(\fcy N)$ $R$-symmetry.  Such an extension could have important applications to the study of superconformal blocks, whose systematic understanding is one of the roadblocks for applying the conformal bootstrap program to supersymmetric theories more broadly than what has currently been done.

Finally, our AdS$_4$ and CFT$_3$ formalism utilizes special properties of spinors for the groups $SO(2,1)$, $SO(3,1)$, and $SO(3,2)$. It would be interesting to extend our methods to other spacetime dimensions. At the end of Appendix~\ref{COSET}, we have sketched how to extend the bulk formalism to other dimensions, but it is less clear how to extend the boundary formalism in general. Straightforward extensions do exist for AdS$_3$/CFT$_2$ and AdS$_2$/CFT$_1$, but whether a convenient formalism exists for AdS$_5$/CFT$_4$ or AdS$_7$/CFT$_6$ is less clear.

\section*{Acknowledgements}
We would like to thank Luca Iliesiu for useful discussions. The work of DJB and SSP was supported in part by the US NSF under Grant No. PHY-1820651 and by the Simons Foundation Grant No.~488653.    DJB was also supported in part by the General Sir John Monash Foundation.  The research of DZF is partially supported by US NSF grant PHY-1620045.  DZF and SSP would like to thank the organizers of ``Scattering amplitudes and the conformal bootstrap'' workshop and
the Aspen Center for Physics (ACP) for hospitality while this work was in progress. The ACP is supported by National Science Foundation Grant No.~PHY-1607611.

\appendix

\section{Conventions}
\label{sec:NORMSCONS}

\subsection{Group theory conventions}
\label{sec:GroupCon}
We work in mostly plus signature, with $SO(3,2)$ invariant
\begin{equation}
\eta_{IJ} = \text{diag}(-1,+1,+1,+1,-1)\,.
\end{equation}
Spinor indices can be raised and lowered with the $\epsilon$ tensor
\begin{equation}
\epsilon_{AB} = \begin{pmatrix}0&0&1&0\\0&0&0&1\\-1&0&0&0\\0&-1&0&0\end{pmatrix}\,,\qquad \epsilon^{AB} = - \epsilon_{AB} =  \begin{pmatrix}0&0&-1&0\\0&0&0&-1\\1&0&0&0\\0&1&0&0\end{pmatrix}\,.
\end{equation}
so that we raise and lower spinors using
\begin{equation}\label{eq:raiselower}
S_A = \epsilon_{AB}S^B\,,\qquad S^A = \epsilon^{AB}S_B\,,
\end{equation}
We work with explicitly real $SO(3,2)$ gamma matrices:
\begin{equation}\begin{split}
{(\Gk_0)^A}_B &= \begin{pmatrix}0&1&0&0\\-1&0&0&0\\0&0&0&-1\\0&0&1&0\end{pmatrix}\,,\quad {(\Gk_1)^A}_B = \begin{pmatrix}0&1&0&0\\1&0&0&0\\0&0&0&1\\0&0&1&0\end{pmatrix}\,,\quad {(\Gk_2)^A}_B = \begin{pmatrix}1&0&0&0\\0&-1&0&0\\0&0&1&0\\0&0&0&-1\end{pmatrix}\,, \\
{(\Gk_3)^A}_B &= \begin{pmatrix}0&0&0&1\\0&0&-1&0\\0&-1&0&0\\1&0&0&0\end{pmatrix}\,,\quad {(\Gk_4)^A}_B = \begin{pmatrix}0&0&0&1\\0&0&-1&0\\0&1&0&0\\-1&0&0&0\end{pmatrix}\,,\\
\end{split}\end{equation}
which satisfy the Clifford algebra
\begin{equation}
\{\Gk_I,\Gk_J\} = 2\eta_{IJ}\,.
\end{equation}
When both spinor indices are raised the gamma matrices are antisymmetric: $\Gk_I^{AB} = \Gk_I^{BA}$. We define the angle bracket
\begin{equation}
\langle ST\rangle = S^A T_A\,,\qquad \langle SP_1...P_nT\rangle = -S_{A_0}(\slashed P_1)^{A_0}_{\ A_1}\dots(\slashed P_n)^{A_{n-1}}_{\ A_n}S^{A_n}\,.
\end{equation}

Now let us consider differentiating with respect to some spinor variable $Q^A$. Using the identities
\begin{equation}
\frac{\nb Q^A}{\nb Q^B} = \delta^A_B\,,\qquad \frac{\nb Q_A}{\nb Q_B} = \delta_A^B\,,
\end{equation}
and our spinor raising and lowering conditions \eqref{eq:raiselower}, we find that 
\begin{equation}
\frac{\nb }{\nb Q^A} = -\epsilon_{AB}\frac{\nb}{\nb Q_B}\,.
\end{equation}
One must therefore be careful when raising and lowering indices in a differential operator. 

To raise and lower $SO(2,1)$ and $SO(3,1)$ spinors we use the tensors
\begin{equation}
\epsilon^{ab} = \epsilon^{\ak\bk} = \epsilon^{\dot\ak\dot\bk} = \begin{pmatrix} 0 & 1\\-1 & 0\end{pmatrix}\,,\quad \epsilon_{ab} = \epsilon_{\ak\bk} = \epsilon_{\dot\ak\dot\bk} = -\epsilon^{ab} = \begin{pmatrix} 0 & -1\\1 & 0\end{pmatrix}\,.
\end{equation}
Our raising and lower conventions are
\begin{equation}
s^a = \epsilon^{ab}s_b\,,\qquad s_a = \epsilon_{ab}s^b
\end{equation}
for $SO(2,1)$ spinors, and likewise for $SO(3,1)$ spinors.

It is more convenient for us to describe vector fields using a pair of spinor indices, as in $B_{\ak\da}$. We relate this to the more commonly used $B_\m$ through the equation
\begin{equation}
B_{\ak\dot\ak} \equiv \sigma_{\ak\dot\ak}^\mu B_\mu\,,
\end{equation}
where $\sigma^{\ak\dot\ak}_\mu$ are the sigma matrices\footnote{We have chosen our sigma matrices so that 
$$(\Gk_i)^A{}_B = \begin{pmatrix} (\sigma_i^T)^\da{}_\ak  & 0\\ 0&\sigma_i^\ak{}_\da\end{pmatrix} $$ for $i = 1, 2, 3$, which proves convenient when considering the explicit parametrizations \eqref{TTbarExplicit} and \eqref{eq:CFTSpinor} of $T^A_\ak$ and $S^A_a$ in terms of $x^i$ and $z$.
}
\begin{equation}
\sk^{\ak\dot\ak}_0 = \begin{pmatrix}-1&0\\0&-1\end{pmatrix}\,,\quad \sk^{\ak\dot\ak}_1 = \begin{pmatrix}-1&0\\0&1\end{pmatrix}\,,\quad \sk^{\ak\dot\ak}_2 = \begin{pmatrix}0&1\\1&0\end{pmatrix}\,,\quad \sk^{\ak\dot\ak}_3 = \begin{pmatrix}0&i\\-i&0\end{pmatrix}\,.
\end{equation}
which satisfy
\begin{equation}\label{eq:sigNorm}
\sigma^{\ak\dot\ak}_\mu\sigma^{\bk\dot\bk}_\nu\epsilon_{\ak\bk}\epsilon_{\dot\ak\dot\bk} = -2\eta_{\mu\nu}\,,\qquad \sigma^{\ak\dot\ak}_\mu\sigma^{\bk\dot\bk}_\nu\eta^{\mu\nu} = -2\epsilon^{\ak\bk}\epsilon^{\dot\ak\dot\bk}\,.
\end{equation}
With this convention, we find that $B^\mu B_\mu = -\frac 12B^{\ak\dot\ak}B_{\ak\dot\ak}$.

\subsection{Killing spinors}
\label{sec:KILLSPIN}
In this section we relate the bispinors $T^A_\ak$ and $\bar T^A_{\dot\ak}$ to Killing spinors. Recall that in AdS$_{d+1}$ a Killing spinor $\xi$ is defined to be a Dirac spinor satisfying the equation
\begin{equation}
\left(\Nb^\m - \frac 1{2L}\gamma^\m\right)\xi = 0\,,
\end{equation}
where $L = 1$ is the AdS radius. Working in AdS$_4$, we can write a Dirac spinor as a pair of left and right handed spinors $\xi = \begin{pmatrix} \chi_\ak \\ \varphi^{\da}\end{pmatrix}$, and can write the gamma matrices in terms of the $\sk^\mu_{\ak\dot\ak}$ matrices:  
\begin{equation}
\gamma^\mu = 
\begin{pmatrix} 0 & i\sigma^\mu_{\ak\da} \\ i(\bar\sigma^\mu)^{\da\ak} & 0 \end{pmatrix}\,,
\end{equation}
where $(\bar\sk^\mu)^{\da\ak} = (\sk^{T\mu})^{\ak\da}$ are the conjugate $\sk$ matrices, which are equal to the transposed $\sk$ matrices. These gamma matrices satisfy the Clifford algebra
\begin{equation}
\{\gk^\mu,\gk^\nu\} = 2\eta^{\mu\nu}\,.
\end{equation}

We then find that the spinors $\chi_\ak$ and $\varphi^{\da}$ satisfy the equations
\begin{equation}i\Nb^\m\ck_\ak = \frac 12 \sk^\m_{\ak\dot\ak}\varphi^{\dot\ak} \,,\qquad i\Nb^\m\varphi^{\dot\ak} = \frac 12 (\bar\sk^\m)^{\dot\ak\ak}\ck_\ak = \frac 12 (\sk^\m)^{\ak\da}\ck_\ak\,.
\end{equation}
Using the definition $\Nb_{\ak\dot\ak} =  \sk^\m_{\ak\dot\ak} \Nb_\m$ and then applying \eqref{eq:sigNorm}, we find that
\begin{equation}
\Nb_{\bk\dot\bk}\ck_\gk = i\epsilon_{\bk\gk}\varphi_{\dot\bk}\,,\qquad \Nb_{\bk\dot\bk}\varphi_{\dot\gk} = -i\epsilon_{\dot\bk\dot\gk}\ck_\bk\,.
\end{equation}
Comparing these equations to those satisfied by $T^A_\ak$ and $\bar T^A_{\dot\ak}$:
\begin{equation}
\Nb_{\bk\dot\bk} T^A_{\gk} = i\epsilon_{\bk\gk}\bar T^A_{\dot\bk}\,,\qquad \Nb_{\bk\dot\bk} \bar T^A_{\dot\gk} = -i\epsilon_{\dot\bk\dot\gk} T^A_\bk\,,
\end{equation}
we see that the Killing spinor equation has solutions
\begin{equation}
\chi_\ak = C_A T^A_\ak\,,\qquad \varphi_{\dot\ak} = C_A \bar T^A_{\dot\ak}\,,
\end{equation}
where $C^A$ is any arbitrary constant $Sp(4)$ spinor.

\section{Frame fields, metrics, and connections}

In this appendix we derive explicit, standard expressions for frame fields, metrics, and connections in intrinsic coordinate systems $x^\m$ from our bispinor formalism in embedding space.  We begin with the bulk and then extend to the more complicated case of the boundary.

\subsection{Bulk}
\label{CONNECTION}

Assume we have a parametrization of our bispinors $T^A_\alpha(x^i)$ and $\bar T^A_{\dot\alpha}(x^i)$ by four coordinates $x^0,\dots,x^3$, and that this mapping is one-to-one in some open subset of $\mathbb R^4$. For instance, we could consider the parametrization \eqref{TTbarExplicit},  taking $x^3 = z$. This corresponds to the Poincar\'e coordinates for $X^I$.

We begin by fixing the frame field $e^\mu_i(x)$ by demanding that covariant derivatives behave correctly on scalar fields:
\begin{equation}\label{eq:scalCoords}
\nabla_i\phi = -\frac12e^\mu_i\sigma^{\alpha\dot\alpha}_\mu\nabla_{\alpha\dot\alpha}\phi\,.
\end{equation}
Lifting the scalar field to a function $\phi(X)$ on $\mathbb R^{3,2}$, we can write the left-hand side as 
\begin{equation}
\nabla_i\phi(X) = \frac{\pa}{\pa x^i}\phi(X) = \frac{\pa X^I}{\pa x^i}\frac{\pa}{\pa X^I}\phi(X)\,,
\end{equation}
while on the right-hand side
\begin{equation}
-\frac12e^\mu_i\sigma^{\alpha\dot\alpha}_\mu\nabla_{\alpha\dot\alpha}\phi = -\frac i4e^\mu_i\sigma^{\alpha\dot\alpha}_\mu\<T_\alpha\Gamma^I\bar T_{\dot\alpha}\>\frac{\pa}{\pa X^I}\phi(X)\,.
\end{equation}
Equating these two expressions gives
\begin{equation}
\frac{\pa X^I}{\pa x^i} = -\frac i {4} e^\mu_i\sigma^{\alpha\dot\alpha}_\mu\<T_\alpha\Gamma^I\bar T_{\dot\alpha}\>\,.
\end{equation}
We can then contract both sides with $\<T_\beta\Gamma_I\bar T_{\dot\beta}\>$, and after using the identity
\begin{equation}
\<T_\alpha\Gamma^I\bar T_{\dot\alpha}\>\<T_\beta\Gamma_I\bar T_{\dot\beta}\> = 8\epsilon_{\alpha\beta}\epsilon_{\dot\alpha\dot\beta}
\end{equation}
we can finally isolate $e^\mu_i$:
\begin{equation}\label{eq:eimuDef}
e_i^{\mu} = -\frac12(\sigma^\mu)^{\alpha\dot\alpha}\left\< T_\alpha\frac{\pa \bar T_{\dot\alpha}}{\pa x^i}\right\> = -\frac12(\sigma^\mu)^{\alpha\dot\alpha}\left\< \bar T_{\dot\alpha}\frac{\pa T_{\alpha}}{\pa x^i}\right\> \,.
\end{equation}
The metric on AdS$_4$ is then defined by the equation:
\begin{equation}
g_{ij} \equiv \eta_{\mu\nu}e^\mu_i e^\nu_j = -\frac 12 \left\< T_\alpha\frac{\pa \bar T_{\dot\alpha}}{\pa x^i}\right\>\left\< \bar T^{\dot\alpha}\frac{\pa T^{\alpha}}{\pa x^i}\right\>\,.
\end{equation}
Substituting the Poincar\'e coordinates \eqref{TTbarExplicit} into these expressions, we can reproduce
\begin{equation}
e_i^\mu(\vec x,z) = \frac 1 z \delta_i^\mu\,,\qquad g_{ij}(\vec x,z) = \frac 1 {z^2}\eta_{ij} \,.
\end{equation}

Next we derive the spin connection. On spinor fields, the covariant derivative acts as
\begin{equation}\label{eq:spincon}
\Nb_i \yk_\alpha(x) = \nb_i\yk_\ak(x) + (\wk_i)_{\ak\bk}\yk^\bk(x)\,,\qquad \Nb_i \bar\yk_\da(x) = \nb_i\bar\yk_\da(x) + (\bar\wk_i)_{\da\db}\bar\yk^\db(x)
\end{equation}
where $(\wk_i)_{\ak\bk}$ is the spin connection and $(\bar\wk_i)_{\da\db}$ is its conjugate. The spin connection has symmetric spinorial indices, and so transforms in the adjoint of $\mathfrak{so}(3,1)$. The more conventional form with antisymmetric frame vector indices is obtained by:
\begin{equation}
(\wk_i)_{\mu\nu} \equiv (\wk_i)_{\ak\bk} (\sk_\mu)^{\ak\dot\gk}{(\sk_\nu)^\bk}_{\dot\gk} + (\bar\wk_i)_{\dot\ak\dot\bk} (\sk_\mu)^{\gk\dot\ak}{(\sk_\nu)_\gk}^\db\,,
\end{equation}
but the spinor indices are more convenient for us.

To compute the spin connection we require that the covariant derivative acts correctly on $T^A_\alpha$:
\begin{equation}
\nabla_i T^A_\beta = -\frac12e_i^\mu\sigma_\mu^{\alpha\dot\alpha}\nabla_{\alpha\dot\alpha}T^A_\beta\,.
\end{equation}
Using \eqref{eq:spincon} and \eqref{2kill} to simplify the left and right-hand sides of this equation, we find that
\begin{equation}
\pa_i T^A_\beta + (\omega_i)_{\beta\gamma} T^{A\gamma} = \frac i 2e^\mu_i{(\sigma_\mu)_\beta}^{\dot\alpha} \bar T_{\dot\alpha}
\end{equation}
and so 
\begin{equation}
(\omega_i)_{\beta\gamma} T^{A\gamma} = -\frac i 2e^\mu_i{(\sigma_\mu)_\beta}^{\dot\alpha} \bar T_{\dot\alpha}^A- \pa_i T^A_\beta \,.
\end{equation}
Contracting both sides with $T_A^\alpha$, we then find that
\begin{equation}\label{eq:wibg}
(\omega_i)_{\beta\gamma} = -\frac i 2 \left\<T_\beta\frac{\pa T_\gamma}{\pa x^i}\right\> \,.
\end{equation}
For the Poincar\'e coordinates \eqref{TTbarExplicit}, we find that
\begin{equation}\begin{split}
(\wk_0)_{\ak\bk} &= \frac{i}{2z} (\sigma_0)_{\ak\da}(\sigma_4)_\bk{}^\da =  \frac {i} {2z}\begin{pmatrix} 1&0\\0&1\end{pmatrix}\,,\quad  (\wk_1)_{\ak\bk}= \frac{i}{2z} (\sigma_1)_{\ak\da}(\sigma_4)_\bk{}^\da = \frac {i} {2z}\begin{pmatrix} -1&0\\0&1\end{pmatrix}\,,\\
(\wk_2)_{\ak\bk} &= \frac{i}{2z} (\sigma_2)_{\ak\da}(\sigma_4)_\bk{}^\da= \frac {i} {2z}\begin{pmatrix} 0&1\\1&0\end{pmatrix}\,, \quad (\wk_3)_{\ak\bk} = 0\,.
\end{split}\end{equation}

Finally, we check that the spin connection satisfies Cartan's structure equation:
\begin{equation}
de^\mu = {\wk^\mu}_\nu\wedge e^\nu\,.
\end{equation}
In index notation this reads
\begin{equation}\label{eq:cartanStructure}
\nb_{[i}e_{j]}^\mu = {(\wk_{[i})^\mu}_\nu e^\nu_{j]}\,.
\end{equation}
It is straightforward to check that this equation is satisfied when $e$ and $\omega$ are defined by \eqref{eq:eimuDef} and \eqref{eq:wibg}. Using these equations we obtain
\begin{equation}
\nb_{[i}e_{j]}^\mu = {(\wk_{[i})^\mu}_\nu e^\nu_{j]} = -\frac 12 \sigma^\mu_{\ak\da}\left\langle\frac{\nb T^\ak}{\nb x^{[i}}\frac{\nb \bar T^\da}{\nb x^{j]}}\right\rangle\,.
\end{equation}

\subsection{Boundary}
\label{sec:boundaryMet}

As in the previous section, we will use indices $i,j,\dots$ for coordinate indices and $\mu,\nu,\dots$ for $SO(2,1)$ gauge indices. However, since we now consider 3d manifolds these indices range from $0$ to $2$. We will take the 3d sigma matrices $\sk_\mu^{ab}$ to be 
\begin{equation}
\sk^{ab}_0 = \begin{pmatrix}-1&0\\0&-1\end{pmatrix}\,,\quad \sk^{ab}_1 = \begin{pmatrix}-1&0\\0&1\end{pmatrix}\,,\quad \sk^{ab}_2 = \begin{pmatrix}0&1\\1&0\end{pmatrix}\,.
\end{equation}

To study the boundary, let us consider some parametrization $S^A_a(x^i)$ of the bispinors, which is one-to-to in some open subset of $\mathbb R^3$. For instance, we could take the parametrization \eqref{eq:CFTSpinor}, which as we shall see corresponds to flat space.

By analogy to the bulk case, we will define the boundary frame field to be
\begin{equation}\label{fframebound}
e^\mu_i = -\frac 12 (\sk^\mu)^{ab}\left\<S_a\frac{\nb S_b}{\nb x^i}\right\>\,,
\end{equation}
and the metric is then given by
\begin{equation}
g_{ij} \equiv \eta_{\mu\nu}e_i^\mu e_j^\nu\,.
\end{equation}
While these equations may look similar to those for the bulk, we should emphasise a critical difference between the two. On the boundary, the little group $SO(2,1)\times\mathbb R_+$ includes Weyl rescalings. Under a Weyl transformation $\lambda(x)\in\mathbb R_+$, the bispinor transforms as $S^A_a(x) \rightarrow \sqrt{\lambda(x)}S^A_a(x)$. It is then straightforward to check that:
\begin{equation}
e^\mu_i(x) \rightarrow \lambda(x)e^\m_i(x)\,,\qquad g_{ij}(x) \rightarrow \lk(x)^2g_{ij}(x)\,,
\end{equation}
under this transformation. For this reason, the boundary of AdS$_4$ only possess a conformal structure, and dose not have a canonical choice of metric. It is the specific choice of a parametrization $S^A_a(x)$ of the bispinors that breaks the Weyl rescaling and allows us to fix a specific metric.

For the specific parametrization \eqref{eq:CFTSpinor} of $S^A_a$, it is straightforward to check that
\begin{equation}
e^\mu_i = \delta^\mu_i\,,\qquad g_{ij} = \eta_{ij}\,,
\end{equation}
so that we recover the flat metric on $\mathbb R^{2,1}$. By Weyl rescaling the bispinors we can study other conformally flat metrics on the AdS$_4$ boundary.

Our next task is to construct the boundary covariant derivative. To do this it is helpful to think about the embedding of our manifold, denoted by $\fcy M$, in $\mathbb R^{3,2}$. The tensor $\nb_i P_I(x)$ can be used to project arbitrary embedding vectors $Q^I(x)$ down onto the tangent space of $x\in\fcy M$. We say that $Q^I(x)$ is normal to $\fcy M$ if
\begin{equation}
Q_I\frac{\nb P^I(x)}{\nb x^i} = 0\,.
\end{equation}
Because we study a 3d manifold embedded in 5d, the space of normal vectors is two dimensional. It is easy to see that $P^I(x)$ is itself always normal to the manifold at $x$. Let us now define $V^I(x)$ be some vector field satisfying
\begin{equation}\label{eq:VIcond}
V^I(x)P_I(x) = -1\,,\qquad V^I(x)\frac{\nb P_I(x)}{\nb x^i} = 0\,.
\end{equation}
Together, these conditions imply that $V^I(x)$ is linearly independent of $P^I$ and normal to the manifold. Because the space of normal vectors is two-dimensional, they uniquely specify $V^I(x)$ up to ``gauge transformations'' ${V^I(x) \rightarrow V^I(x) + \mu(x)P^I(x)}$ for some arbitrary scalar function $\mu(x)$.

Next we construct the spin connection $(\wk_i)_{ab}$, which can equivalently be written with vector indices:
\begin{equation}
(\wk_i)_{\m\n} \equiv (\wk_i)_{ab}(\sk_{(\m})^{ac}(\sk_{\n)})^b{}_c\,.
\end{equation}
The spin connection is uniquely fixed by Cartan's structure equation \eqref{eq:cartanStructure}, which we used in the previous section to verify our computation of the AdS$_4$ spin connection. If we define
\begin{equation}\label{spinconbound}
(\wk_i)_{ab} \equiv -\frac 12\left\< S_aV\frac{\nb S_b}{\nb x^i} \right\>\,,
\end{equation}
then it is straightforward check that \eqref{eq:cartanStructure} is indeed satisfied:
\begin{equation}
\nb_{[i}e_{j]}^\mu = {(\wk_{[i})^\mu}_\nu e^\nu_{j]} = -\frac 12 (\sk^\m)^{ab}\left\<\frac{\nb S_a}{\nb x^i}\frac{\nb S_b}{\nb x^j}\right\> \,.
\end{equation}
Note that \eqref{spinconbound} is automatically symmetric in the $ab$ indices, and is invariant under shifts $V^I(x) \rightarrow V^I(x) + \mu(x)P^I(x)$. Now that we have found the spin connection, we can define the covariant derivative
\begin{equation}\label{cdbound}
\Nb_i \Yk_a(S) = \frac{\nb}{\nb x^i} \Yk_a(S) + (\wk_i)_{ab}\Yk^b(S)\,.
\end{equation}

For the coordinates \eqref{eq:CFTSpinor}, it is easy to check that \eqref{eq:VIcond} is satisfied by the constant vector:
\begin{equation}
V^I(x) \equiv P_{\infty} = (0,0,0,-1,1)\,,
\end{equation}
where $P_{\infty}$ is the point at infinity, and that the spin connection vanishes. In order to study more general conformally flat manifolds, we perform a Weyl transformation $S^A_a(x) \rightarrow \sqrt{\lk(x)}S^A_a(x)$. Under this transformation, $P^I(x) \rightarrow \lk(x)P^I(x)$ and the boundary metric becomes $g_{ij}(x) = \lk(x)^2\eta_{ij}$. We then find that
\begin{equation}
V^I(x) = \frac 1 {\lk(x)}\left(P_{\infty}^I + 2g^{ij}(x)\frac{\nb\lk(x)}{\nb x^i}\frac{\nb P^I(x)}{\nb x^j}\right)
\end{equation}
satisfies \eqref{eq:VIcond}, where $g^{ij}(x) = \lk(x)^{-2}\eta^{ij}$ is the inverse metric.

We now show that the covariant derivative \eqref{cdbound} takes a simple form when acting on functions $\fcy O(S)$ of the bispinor $S^A_a(x)$. To this end, let us define
\begin{equation}
\Nb_{ab} \equiv V_I(x)\nb^I_{ab}\,,
\end{equation}
where $\nb_{ab}^I$ is the differential operator introduced in Section \ref{sec:FurtherDOs}. Our claim is that for any $\fcy O(S)$, we can compute the covariant derivative $\Nb_i$ using $\Nb_{ab}$:
\begin{equation}\label{eq:bdiv}
\Nb_i\fcy O(S) = -\frac 12 e_i^\mu \sk_\mu^{bc}\Nb_{bc}\fcy O(S)\,.
\end{equation}
To prove this equation, we simply need to show that it is true for the bispinor $S^A_a(x)$, as the more general result then follows from the chain rule. Using the identities
\begin{equation}\label{eq:idb1}
V_I \nb^I_{bc} S^A_a = {\slashed V^A}_BS^B_{(b}\epsilon_{c)a}\,,\qquad e^\mu_i(\sk_\mu)_a{}^bS^A_b = {\slashed P^A}_B\frac{\nb S^B_a}{\nb x^i}\,,
\end{equation}
we find that
\begin{equation}
-\frac 12 e_i^\mu \sk_\mu^{bc}\Nb_{bc}S^A_a = -\frac 12 {\slashed V^A}_B{\slashed P^B}_C\frac{\nb S^C_a}{\nb x^i}\,.
\end{equation}
Using the Clifford algebra to swap the order of $\slashed V$ and $\slashed P$, we then find that
\begin{equation}
-\frac 12 e_i^\mu \sk_\mu^{bc}\Nb_{bc}S^A_a = \frac{\nb S^A_a}{\nb x^i}+\frac 12 {\slashed P^A}_B{\slashed V^B}_C\frac{\nb S^C_a}{\nb x^i}\,.
\end{equation}
Rewriting ${\slashed P^A}_B = S^A_bS^{Ab}$, we find 
\begin{equation}
-\frac 12 e_i^\mu \sk_\mu^{bc}\Nb_{bc}S^A_a = \frac{\nb S^A_a}{\nb x^i}+\frac 12 S^A_b \left\<S^bV\frac{\nb S_a}{\nb x^i}\right\> = \frac{\nb S^A_a}{\nb x^i}+(\wk_i)_{ab}S^{Ab}\,.
\end{equation}
Comparing this to the definition of the covariant derivative \eqref{cdbound}, we see that \eqref{eq:bdiv} is satisfied.

We can now use \eqref{eq:bdiv} to show that $S^A_a$ is a conformal Killing spinor. Recall that a conformal Killing spinor $\Sk_a$ satisfy the equation
\begin{equation}
e^i_\m\Nb_i\Sk_a = {(\sigma^\m)_a}^b \tilde \Sk_b 
\end{equation}
for some spinor $\tilde \Sk_b$. Using \eqref{eq:bdiv} and the first equation in \eqref{eq:idb1}, we find that
\begin{equation}
e^i_\m\Nb_iS^A_a = \sk_\m^{bc}\Nb_{bc}S^A_a = (\sk_\m)_a{}^b {\slashed V^A}_BS^B_b\,,
\end{equation}
and so $\Sk_a = C_AS^A_a$, $\tilde\Sk_a = \<CVS_a\>$ satisfies the conformal Killing equation for any constant $C_A$.

As a final task, we will derive the conservation condition \eqref{eq:conscon}. For simplicity we will consider a vector field $J^{ab}(S)$, as the generalization to higher spinning fields is straightforward. We can use the frame fields to rewrite the $J^{ab}(S)$ field as a tangent vector
\begin{equation}
J^i(x) = -\frac 12 e^i_\mu \sk^\mu_{ab}J^{ab}(S)\,.
\end{equation}
A conserved vector field satisfies the condition $\Nb_iJ^i(x) = 0$. Using \eqref{eq:bdiv} and the identity $\Nb_i e^j_\m = 0$, we find that
\begin{equation}
\Nb_iJ^i(x) = -\frac 12 e^i_\mu\sk^\mu_{ab}\Nb_i J^{ab}(S) = -\frac 12\Nb_{ab}J^{ab}(S) = -\frac 12 V_I\nb^I_{ab}J^{ab}(S)\,.
\end{equation}
Hence, if $\nb^I_{ab}J^{ab}(S) = 0$ we see that $J^i(S)$ is conserved regardless of the boundary metric. Conversely, if $V_I\nb^I_{ab}J^{ab}(S) = 0$ is satisfied by any given vector $V^I$ satisfying $P_I V^I = -1$, then using conformal invariance one can see that it is true for all such $V^I$. This, combined with the fact $P_I\nb^I_{ab}$ is always zero, implies that $\nb^I_{ab}J^{ab}(S) = 0$.

\section{Coset construction of AdS$_4$}
\label{COSET}

In this appendix we present a coset construction of AdS$_4$ which directly produces our bispinor formalism. Recall that in any spacetime dimension, AdS$_{d}$ can be constructed as the coset space $SO(d-1,2)/SO(d-1,1)$. When $d = 4$ the double cover of $SO(3,2)$ is $Sp(4)$, while the double cover of $SO(3,1)$ is $SL(2,\mathbb C)$, and so we can write AdS$_4\approx Sp(4)/SL(2,\mathbb C)$. 

The group $Sp(4)$ has a natural action of $Sp(4)\times Sp(4)$ given by left and right multiplication. We can therefore write any element of $Sp(4)$ as a matrix $M^{A\hat A}$, where the unhatted and hatted index transforms under the left and right $Sp(4)$ respectively, which satisfy the conditions
\begin{equation}\label{eq:MCond}
\epsilon_{AB}M^{A\hat A}M^{B\hat B} = \epsilon^{\hat A\hat B}\,, \qquad \epsilon_{\hat A\hat B}M^{A\hat A}M^{B\hat B} = \epsilon^{AB}\,.
\end{equation}

To quotient $Sp(4)$ by $SL(2;\mathbb C)$, let us define an embedding of $SL(2;\mathbb C)$ in the right $Sp(4)$:
\begin{equation}
M^{A\hat A} = \frac 1 {\sqrt 2}\begin{pmatrix} T^{A\ak} \\ \bar T^{A\da} \end{pmatrix}\,,\qquad \epsilon^{\hat A\hat B} = \begin{pmatrix} i\epsilon^{\ak\bk} & 0 \\ 0 & -i\epsilon^{\da\db} \end{pmatrix}\,,\qquad \epsilon_{\hat A\hat B} = \begin{pmatrix} -i\epsilon_{\ak\bk} & 0 \\ 0 & i\epsilon_{\da\db} \end{pmatrix}\,.
\end{equation}
Here, as in the main text, we use $\ak$ for left-handed $SO(3,1)$ spinors (which are equivalent to $SL(2;\mathbb C)$ fundamentals) and $\da$ for the conjugate representation. In this language the first condition in \eqref{eq:MCond} becomes
\begin{equation}\label{eq:bicond}
\epsilon_{AB}T^{A\ak}T^{B\bk} = 2i\epsilon^{\ak\bk}\,,\qquad \epsilon_{AB}\bar T^{A\da}\bar T^{B\db} = -2i\epsilon^{\da\db}\,,\qquad  \epsilon_{AB}T^{A\ak}\bar T^{B\db} = 0 \,.
\end{equation}
We now recognize the first two equations as \eqref{Proj3} and the last equation as \eqref{eq:TbTcond2}, which are the constraints which we imposed on the bispinors $T^A_\ak$ and $\bT^A_\da$ in the main text. The second condition in \eqref{eq:MCond} becomes
\begin{equation}\label{eq:bicond2}
-i\epsilon_{\ak\bk} T^{A\ak}T^{B\bk} + i\epsilon_{\da\db}\bar T^{A\da}\bar T^{B\db} = 2\epsilon^{AB}\,,
\end{equation}
which one can derive from \eqref{GotX}. We therefore conclude that an element $M^{A\hat A}$ of $Sp(4)$ is equivalent to the bispinors $T^A_\ak$ and $\bT^A_\da$.

Equipped with this more abstract understanding of the bispinor variables, let us now consider bulk derivative operators. The generators of the left and right $Sp(4)$ symmetry can be written as
\begin{equation}
(G_L)^{AB} = -iM^{(A|\hat A}\frac{\nb}{\nb {M_{B)}}^{\hat A}}\,,\qquad (G_R)^{\hat A\hat B} = M^{A(\hat A}\frac{\nb}{\nb {M^A}_{\hat B)}}\,,
\end{equation}
which manifestly preserve the conditions \eqref{eq:MCond}. If we rewrite these in terms of $T^A_\ak$ and $\bT^A_\da$, we find that
\begin{equation}
(G_L)^{AB} = D^{AB}\,,\qquad (G_R)^{\hat A\hat B} = \begin{pmatrix} -L^{\ak\bk} & \frac12\Nb^{\ak\db} \\ \frac12\Nb^{\da\bk} & \bar L^{\da\db} \end{pmatrix}\,.
\end{equation}
We therefore see that the right $Sp(4)$ generator splits into the three differential operators $L_{\ak\bk}$, $\bar L_{\da\db}$ and $\Nb_{\ak\da}$, and that the commutator relations \eqref{DcomV} and \eqref{Nbcom} follow from those of ${\mathfrak{sp}(4)\times\mathfrak{sp}(4)}$. In particular, the covariant derivative corresponds to the generators of the right $Sp(4)$ which are orthogonal to the quotient $SL(2;\mathbb C)$.

Now that we understand how our bispinors arise from the coset construction of AdS$_4$, it is straightforward to generalize to any AdS$_d$. We write the group elements of $SO(d-1,2)$ as matrices $M^{A\hat A}$ satisfying certain quadratic conditions. By quotienting on the right with an $SO(d-1,1)$ subgroup, we can hence describe AdS$_4$ by $SO(d-1,2)\times SO(d-1,1)$ bispinors satisfying quadratic constraints. The leftover generators of the right $SO(d-1,2)$ become the covariant derivative on AdS$_d$. 

Our discussion has been schematic because the exact details depend on the properties of the spinor representations of $SO(d-1,2)$ and $SO(d-1,1)$. Computations are easy for AdS$_3$ and AdS$_2$, but for larger spacetime dimensions the constraints on $M^{A\hat A}$ become increasingly difficult to work with.

\section{Conformally-covariant differential operators}
\label{sec:diffOpAp}

In this section we provide a more abstract way to understand the boundary differential operators in section \ref{DIFFOPS}. This will allow us to prove the two results of section \ref{sec:FurtherDOs}.

Although we have been thinking of $S^A_a$ as an $Sp(4)\times GL(2;\mathbb R)$ bispinor, it will prove useful to rewrite it as an 8 dimensional vector
\begin{equation}
Z^i = R^{i a}_A S^A_a\,,\qquad S^A_a = (R^{-1})_{i a}^A Z^i\,,
\end{equation}
where $R^{i a}_A$ implements the change of basis. Note that unlike in previous sections, in this appendix we will use indices $i,j,\dots$ taking values $1,2,\dots,8$ for these 8 dimensional vectors. The condition $\langle S_a S^a\rangle$ corresponds to imposing
\begin{equation}\label{eq:SO44Cond}
\delta_{ij}Z^i Z^j = 0\,,
\end{equation}
where $\delta_{ij} = \epsilon^{AB}\epsilon_{ab}(R^{-1})_{i a}^A(R^{-1})_{j b}^B$ is a metric with signature $(4,4)$. The group of linear transformations acting on $Z^i$ which preserve \eqref{eq:SO44Cond} is $SO(4,4)\times \mathbb R$, and their generators are the differential operators
\begin{equation}
M_{ij} = 2Z_{[i}\frac{\nb}{\nb Z^{j]}}\,,\qquad U = -\frac 12 Z^i\frac{\nb}{\nb Z^i}\,.
\end{equation}
These operators satisfy the commutator relations
\begin{equation}\begin{aligned}\label{eq:so44coms}
&[U,Z^i] = -\frac12 Z^i\,,&\qquad [M^{ij},Z^k] &= 2\delta^{k[i} Z^{j]} \\
&[Z^i,Z^j] = [U,M_{ij}] = 0\,,&\qquad
[M_{ij},M_{kl}] &= -(\delta_{ik}M_{jl}-\delta_{il}M_{jk}-\delta_{jk}M_{il}+\delta_{jl}M_{ik})\,.
\end{aligned}\end{equation}
Rewriting $Z^i$ in terms of $S^A_a$, we see that $U$ corresponds to the operator $\fcy D$, while $M_{ij}$ splits into the three operators $\fcy L_{ab}$, $\fcy D_{AB}$, and $\nb^I_{ab}$. As a simple check, we note that
$${1+28 = 1+3+10+3\times 5}\,,$$
so that the number of differential operators match in both languages. From this, we conclude that operators $\fcy L_{ab}$, $\fcy D_{AB}$, and $\nb^I_{ab}$ together generate the Lie algebra $\mathfrak{so}(4,4)$.

We can likewise rewrite $\fcy E^A_a$ as an $SO(4,4)\times U(1)$ vector
\begin{equation}
Y^i = R^{i a}_AS^A_a = \frac 14 Z^i\frac{\nb}{\nb Z^j}\frac{\nb}{\nb Z_j} + \left(U - \frac32\right)\frac{\nb}{\nb Z_i}\,.
\end{equation} 
From this, we can compute the commutation relations:
\begin{equation}\label{eq:so55Coms}\begin{aligned}
&[U, Y^i] = \frac12 Y^i\,,&\qquad [M^{ij},Y^k] &= 2\delta^{k[i} Y^{j]}\,,\\
&[Y^i,Y^j]  = 0\,,&\qquad [Z^i,Y^j] &= \delta^{ij}\left(\frac 32 - U\right)+\frac12M^{ij}\,.
\end{aligned}\end{equation}
Rewriting \eqref{eq:so44coms} and \eqref{eq:so55Coms} in terms of $S^A_a$, we can rederive the commutation relations \eqref{opComs0} and \eqref{opComs} given in the main text.  

Examining \eqref{eq:so44coms} and \eqref{eq:so55Coms} more closely, we see that they actually form the commutators relations of $\mathfrak{so}(5,5)$. To see this, let us define the operator $N^{pq}$, for $p,q = 1,\dots 10$, as
\begin{equation}\begin{aligned}
N^{ij} &= M^{ij}\,,\qquad & N^{9\,j} & = - (Y^i+Z^i)\,,\\
N^{10j} &= -(Y^j-Z^j)\,,\qquad & N^{9\,,10} & = - 2(U-3/2)\,.
\end{aligned}\end{equation}
We can then check that $N^{pq}$ satisfies the $\mathfrak{so}(5,5)$ commutation relations
\begin{equation}
[N^{pq},N^{rs}] = -(\eta^{pr}N^{qs} - \eta^{qr}N^{ps} - \eta^{ps}N^{qr} + \eta^{qs}N^{pr})\,,
\end{equation}
where $\eta^{ij}$ is the diagonal tensor:
\begin{equation}
\eta^{ij} = \delta^{ij}\,, \qquad \eta^{9\,9} = -1\,,\qquad \eta^{10\,10} = 1\,.
\end{equation}
Since all we have done is rewritten $S^A_a$, $\fcy E^A_a$, $\fcy D^{AB}$, $\fcy D-\frac32$, $\fcy L_{ab}$, and $\nb^I_{ab}$ in a manifestly $\mathfrak{so}(4,4)$ language, we immediately deduce that these operators also together generate the Lie algebra $\mathfrak{so}(5,5)$.

We will now study the properties of more general conformally-covariant differential operators, or CCDOs for short. Recall that a CCDO $\fcy G$ is a differential operator satisfying \eqref{eq:diffidCond}, which in the $\mathfrak{so}(4,4)$ language becomes
\begin{equation}\label{CCDOcond}
\left.\fcy G\big(Z^iZ_if(Z_i)\big)\right|_{Z^iZ_i = 0} = 0\,.
\end{equation}
Let us expand out $\fcy G$ formally as a series in $\frac{\nb}{\nb Z^{i}}$:
\begin{equation}
\fcy G = \sum_{k = 0}^\infty g^{i_1\dots i_k}(Z) \frac{\nb}{\nb Z^{i_1}}\dots\frac{\nb}{\nb Z^{i_k}}
\end{equation}
where $g^{i_1\dots i_m}(Z)$ are arbitrary functions of the $Z_i$. Our task for the rest of this section is to show that $\fcy G$ can be rewritten purely in terms of $Y_i$ and $Z_i$. Because $\fcy G$ is an arbitrary operator and $Y_i$ is the operator $\fcy E^A_a$ in the $Sp(4)\times SO(2,1)$ language, this proves that every CCDO can be expressed in terms of $S^A_a$ and $\fcy E^A_a$.

Let us begin by introducing the differential operator
\begin{equation}
\hat Y^i = \left(U-\frac32\right)^{-1}Y^i = \frac1{2U-3}Y^i\frac{\nb}{\nb Y^j}\frac{\nb}{\nb Y_j}  + \frac{\nb}{\nb Y_i}\,,
\end{equation}
which is just the $\mathfrak{so}(4,4)$ version of the operator $\hat{\fcy E}^A_a$ defined in \eqref{hatE}. We use this to define
\begin{equation}
\tilde{\fcy G} = \sum_{k = 0}^\infty g^{i_1\dots i_k}(Z) \hat Y_{i_1}\dots \hat Y_{i_k}\,,
\end{equation}
and our goal is to argue that $\fcy G = \tilde{\fcy G}$.

To this end, introduce an arbitrary polarization vector $z_i$ and define $\zk = Z^iz_i$. The action of $\fcy G$ on the tensor $Z_{i_1}\dots Z_{i_m}$ can be computed as:
\begin{equation}\label{lmz1}
\fcy G\left[Z_{i_1}\dots Z_{i_m}\right] = \frac 1 {m!}\frac{\nb }{\nb z^{i_1}}\dots\frac{\nb}{\nb z^{i_m}} \fcy G[\zk^m]\,,
\end{equation}
and \eqref{CCDOcond} then implies that
\begin{equation}\label{lmz2}
\fcy G[\zk^m] = \left.\fcy G[\zk^m]\right|_{z^iz_i = 0}\,.
\end{equation}
These equations of course also hold for $\tilde{\fcy G}$.

Let us now compute:
\begin{equation}\begin{split}
\frac{\nb}{\nb Z^{i_1}}\dots\frac{\nb}{\nb Z^{i_k}} \zk^m\Bigg|_{z^iz_i = 0} &= \frac{m!}{(m-k)!}z_{i_1}\dots z_{i_k}\zk^{m-k}\Bigg|_{z^iz_i = 0} \\
\hat Y_{i_1}\dots \hat Y_{i_k} \zk^m\Bigg|_{z^iz_i = 0} &= \frac{m!}{(m-k)!}z_{i_1}\dots z_{i_k}\zk^{m-k}\Bigg|_{z^iz_i = 0} \,.\\ 
\end{split}\end{equation}
Using these equations, it is then easy to verify that for all $m\in\mathbb Z^+$
\begin{equation}
\fcy G[\zeta^m]\Big|_{z^iz_i = 0} = \tilde {\fcy G}[\zeta^m]\Big|_{z^iz_i = 0} \,.
\end{equation}
We finally apply both \eqref{lmz1} and \eqref{lmz2} to conclude that for any $m$,
\begin{equation}
\fcy G\left[Z_{i_1}\dots Z_{i_m}\right] = \tilde{\fcy G}\left[Z_{i_1}\dots Z_{i_m}\right] \,,
\end{equation}
and hence conclude that $\fcy G = \tilde{\fcy G}$.

We should note a potentially puzzling aspect of previous argument. Take the case where $\fcy G$ is a polynomial in $\frac{\nb}{\nb Z_i}$. It not obvious that $\tilde{\fcy G}$ is also a polynomial in $\frac{\nb}{\nb Z_i}$, because the definition $\hat Y^i = (U-\frac32)^{-1}Y^i$ requires us to invert a differential operator. Nevertheless, it must be the case that if $\fcy G$ is a CCDO, then $\fcy G = \tilde{\fcy G}$ and so it must be the case that the factors of $(U-\frac32)^{-1}$ cancel in $\tilde{\fcy G}$. This in particular implies that $\tilde{\fcy G}$ is a polynomial in $Y_i$.

\section{Ward identities}
\label{sec:wards}

In this appendix we discuss how the Ward identities for conserved currents are realized in Witten diagrams. We have already shown in \eqref{eq:conscon} that the operator $\partial^I_{ab}$ defined in \eqref{divIab} implements current conservation on boundary currents $J^{ab\ldots}(S)$ in embedding space. At the end of Appendix \ref{sec:boundaryMet} we derived the conventional conservation law in an intrinsic coordinate chart from \eqref{eq:conscon}. Here we first discuss how $\nb_{ab}^I$ acts on general massless bulk-boundary propagators, and then use our results to derive the Ward identities from Witten diagrams in scalar QED.

\subsection{Bulk-boundary propagators}
Let us begin by considering a bulk gauge field $A_{\ak\da}(T,\bar T)$ that is dual to a conserved current $J^{ab}(S)$ on the boundary. \footnote{DZF thanks Elliot Hijano for a useful suggestion.} Let us first consider the Poincar\'e patch. In (50) of \cite{Freedman:1998tz} it was established that the bulk-boundary propagator of a photon $G_{\m i}(z,\vec x)$ has the property
\begin{equation}\label{wardpoinpatch}
\frac{\pa}{\pa x^i} G_{\m i}(z,\vec x) = \frac{\pa}{\pa z^\m} H(z,\vec x),
\end{equation}
where $H(z,\vec x)$ involves a hypergeometric function. This describes the important feature that \emph{the boundary divergence of $G_{\m i}$ is compensated by a bulk gauge transformation}.  The bulk derivative $\pa_\m$ is then integrated by parts, and the Ward identity follows after use of Green's theorem and further analysis.

We will show that in embedding space, \eqref{wardpoinpatch} generalizes to the equation
\begin{equation}\label{ward1}
\pa^I_{ab}\<J^{ab}(S) A_{\ak\da}(T,\bT)\> = \nb^I_{ab}\frac{2\langle S^aT^\ak\rangle\langle S^b\bar T^\da\rangle}{(-2P\cdot X)^3} = -2P^I\Nb_{\ak\da}\left(\frac1{(-2P\cdot X)^3}\right)\,.
\end{equation}
The factor $P^I$ is a necessary ``spectator'' in this relation. Using the Leibniz rule to compute \eqref{ward1}, we need to evaluate
\begin{equation}\label{w2}\begin{split}
\nb_{ab}^I\left(S^a_AS^b_B\right) = \frac32[\Gk^I,\slashed P]_{AB}\,, \qquad 
\nb_{ab}^I\frac{1}{(-2P\cdot X)^3} = -3 \frac{\<S_a\Gk^I\slashed X S_b\>}{(-2P\cdot X)^4}\,.
\end{split}\end{equation}
Using \eqref{bdycons} and the $\Gk$-matrix algebra, we can contract the second expression with $S^a_AS^b_B$ to compute
\begin{equation}\label{w3}
S^a_AS^b_B\nb_{ab}^I\frac{1}{(-2P\cdot X)^3} = -\frac{3[\Gk^I,\slashed P]_{AB}}{2(-2P\cdot X)^3}-\frac{3P^I [\slashed X,\slashed P]_{AB}}{(-2P\cdot X)^4} \,.
\end{equation}
Combining \eqref{w2} and \eqref{w3}, we see that the first term in both cancels, and hence that
\begin{equation}\label{ward3}
\pa^I_{ab}\<J^{ab}(S) A_{\ak\da}(T,\bT)\> = -6P^I\frac{\<T_\ak[\slashed X,\slashed P]\bT_\da\>}{(-2P\cdot X)^4}= -6iP^I\frac{\<T_\ak\slashed P\bT_\da\>}{(-2P\cdot X)^4}.
\end{equation}
To finish the job, we recognise that\footnote{As discussed in Section \ref{sec:bulkbound}, \eqref{eq:derivScalar} is fixed by conformal invariance up to an overall coefficient. We can simply substitute $\ell = 0$ and $m = n = \frac 12 $ into equation \eqref{eq:bulkBound}.}
\begin{equation}\label{eq:derivScalar}
\Nb_{\ak\da}\frac 1{(-2P\cdot X)^3} = -3i\frac{\langle T_\ak\slashed P\bar T_\da\rangle}{(-2P\cdot X)^4}
\end{equation}
and so we find the desired bulk gauge transformation
\begin{equation}\label{ward5}
\pa^I_{ab}\<J^{ab}(S) A_{\ak\da}(T,\bT)\> = -2\nabla_{\ak\da}\left(\frac{P^I}{(-2P\cdot X)^3}\right) = -2P^I\Nb_{\ak\da}\left(\frac1{(-2P\cdot X)^3}\right)\,.
\end{equation}

The proof of the Ward identity for general $\ell$ follows the argument above, but the details are more complicated. For this reason we will only provide an outline of the procedure. It is helpful to use the polarized bulk-boundary propagator
\begin{equation}\label{eq:propell}
\<\fcy J(\sk,P)\varphi(\tk,\bar\tk,X) \> = 2\Gamma(\ell+1)\frac{\<\sk\tk\>^\ell\<\sk\bar\tk\>^\ell}{(-2 P\cdot X)^{2\ell+1}},
\end{equation}
and to strip polarizations only for one pair of indices, which we contract with $\nb^I_{ab}$. Defining 
\begin{equation}
\fcy J^{ab}(\sk,P) = \frac 1 {\ell\Gamma(\ell+1)}\frac{\nb^2}{\nb s_a\nb s_b}\fcy J(\sk,P)\,,
\end{equation}
(we include the factor $(\ell\Gamma(\ell+1))^{-1}$ merely for convenience), we can compute
\begin{equation}\begin{split}\label{eq:genJab}
\langle \fcy J^{ab}&(\sk,P)\varphi(\tk,\bar\tk,X)\rangle = \frac{2n^{ab}}{(-2P\cdot X)^{2\ell+1}} \\
n^{ab} &= \langle\sk\tk\rangle^{\ell-2}\langle\sk\bar\tk\rangle^{\ell-2}\Big(\ell\langle S_{(a}\tk\rangle\langle S_{b)}\bar\tk\rangle\langle\sk\tk\rangle\langle\sk\bar\tk\rangle + (\ell-1)\left[\langle S_a\tk\rangle\langle S_b\tk\rangle \langle\sk\bar\tau\rangle^2+ \text{H.c.}\right]\Big)\,.
\end{split}\end{equation}
Contracting both sides of \eqref{eq:genJab} with $\nb_{ab}^I$ and applying the Leibniz rule, we find that
\begin{equation}\label{eq:gwardL}
\nb_{ab}^I\langle \fcy J^{ab}(\sk,P)\varphi(\tk,\bar\tk,X)\rangle = n^{ab}\nb^I_{ab}\frac {2} {(-2P\cdot X)^{2\ell+1}} + \frac{2} {(-2P\cdot X)^{2\ell+1}}\nb^I_{ab}n^{ab}\,.
\end{equation}
Generalizing \eqref{w3}, we find that the first term in \eqref{eq:gwardL} can be computed using
\begin{equation}\label{eq:w4}
S_A^aS_B^b\nb_{ab}^I \frac 1{(-2P\cdot X)^{2\ell+1}} = \frac{2\ell+1}2\frac{[\slashed P,\Gk^I]^{AB}}{(-2P\cdot X)^{2\ell+1}} - (2\ell+1)\frac{P^I[\slashed X,\slashed P]^{AB}}{(-2P\cdot X)^{2\ell+2}}\,.
\end{equation}
A detailed calculation of $\nb_{ab}^In^{ab}$ is more difficult, revealing terms which are proportional to $[\slashed P,\Gk^I]$. These exactly cancel with those terms in \eqref{eq:w4}, and so we find that
\begin{equation}\label{eq:reduct}
\nb_{ab}^I\langle \fcy J^{ab}(\sk,P)\varphi(\tk,\bar\tk,X)\rangle  = 2i\ell(2\ell+1)P^I\frac{\langle\sk\tk\rangle^{\ell-1}\langle\sk\tk\rangle^{\ell-1}\langle\tk P\bar\tk\rangle}{(-2P\cdot X)^{2\ell+2}}\,.
\end{equation}
We are almost done; now all we must show is that \eqref{eq:reduct} is a total bulk divergence. For this we can first observe that:
\begin{equation}
t^\ak\bar t^\da\Nb_{\ak\da} \frac 1 {(-2P\cdot X)^{2\ell+1}} = -i(2\ell+1)\frac{\langle\tk P\bar\tk\rangle}{(-2P\cdot X)^{2\ell+2}}\,.
\end{equation}
We then note that, because $t^\ak t_\ak = \bar t^\da \bar t_\da = 0$, it follows that 
\begin{equation}
t^\ak \bar t^\da \Nb_{\ak\da}\langle \sk\tk\rangle = t^\ak \bar t^\da \Nb_{\ak\da}\langle \sk\bar\tk\rangle = 0\,,
\end{equation}
and so we conclude that
\begin{equation}\label{eq:finWardBB}
\nb_{ab}^I\langle \fcy J^{ab}(\sk,P)\varphi(\tk,\bar\tk,X)\rangle  = -2\ell P^I t^\ak \bar t^\da\Nb_{\ak\da}\left(\frac{\langle\sk\tk\rangle^{\ell-1}\langle\sk\bar\tk\rangle^{\ell-1}}{(-2P\cdot X)^{2\ell+1}}\right)\,.
\end{equation}

As a final task, let us be more concrete and consider the stress tensor $T_{abcd}(S)$, whose bulk dual is the metric perturbation field $h_{\ak\bk\da\db}(T,\bT)$. In this case our Ward identity \eqref{eq:finWardBB} becomes
\begin{equation}
\nb_{ab}^I\langle \fcy T^{ab}(\sk,P)\varphi(\tk,\bar\tk,X)\rangle = -4 P^I t^\ak \bar t^\da\Nb_{\ak\da}\left(\frac{\langle\sk\tk\rangle\langle\sk\bar\tk\rangle}{(-2P\cdot X)^5}\right)\,.
\end{equation}
Extracting the last two remaining polarizations, we can rewrite that as
\begin{equation}\begin{split}
\nb_{ab}^I\langle \fcy T^{ab}(\sk,P)\varphi(\tk,\bar\tk,X)\rangle &=  -4 P^I t^\ak t^\bk\bar t^\da\bar t^\db\Nb_{\ak\da}\left(\frac{\langle\sk T_\bk\rangle\langle\sk\bT_\db\rangle}{(-2P\cdot X)^5}\right) \\ 
&= -P^I t^\ak t^\bk\bar t^\da\bar t^\db\left(\Nb_{\ak\da}v_{\bk\db} + \Nb_{\bk\db}v_{\ak\da}+\Nb_{\ak\db}v_{\bk\da} + \Nb_{\bk\da}v_{\ak\db}\right)\,.
\end{split}\end{equation}
The quantity 
\begin{equation}
v_{\ak\da} = \frac{\langle \sk T_\ak\rangle\<\sk \bT_\da\>}{(-2P\cdot X)^5}
\end{equation}
is effectively a (linearized) diffeomorphism, under which the metric perturbation transforms as ${\delta h_{\mu\nu} = \Nb_{\mu}v_{\n} + \Nb_{\n}v_{\m}}$.

\subsection{Ward identities in scalar QED}

Now that we have established the Ward identity for bulk-boundary propagators, we can apply them to Witten diagrams. For simplicity, we shall restrict our discussion to scalar QED, as the generalization to other examples is straightforward.

Consider first the diagram
\begin{equation}\label{warddi1}
\Sk^{ab}(S_1,Y_2,Y_3) = {\scriptsize\feynmandiagram [inline=(b.base), horizontal = a to c] {
d [particle=$J(S_1)$] -- [photon] b ,
a [particle=$\fk(Y_2)$] -- [fermion] b -- [fermion] c [particle = $\fk^*(Y_3)$]
};}\,,
\end{equation}
in scalar QED, and compute
\begin{equation}\begin{split}
\nb_{ab}^I&\Sk^{ab}(S_1,Y_2,Y_3) = -\frac12 e P_1^I \\
&\times\int dX\left(\Nb_{\ak\da}\frac 1 {(-2P_1\cdot X)^3}\left[G_\Dk(X;Y_2)\Nb^{\ak\da}G_{\Dk}(X;Y_3)-G_\Dk(X;Y_3)\Nb^{\ak\da}G_{\Dk}(X;Y_2)\right]\right)\\
&= -eP^I_1\int\frac{dX}{(-2P_1\cdot X)^3}\left(G_\Dk(X;Y_2)\Nb^2G_\Dk(X;Y_3)-G_\Dk(X;Y_3)\Nb^2G_\Dk(X;Y_2)\right)\,,
\end{split}\end{equation}
where $\Delta$ is the conformal dimension of the boundary dual $\Fk(P)$ of $\fk(X)$. After integration by parts, Green's theorem produces the equations of motion for the bulk-bulk-propagator, we find that
\begin{equation}\label{witdiev}
\nb_{ab}^I\Sk^{ab}(S_1,Y_2,Y_3) = eP_1^I\left(\frac1{(-2P_1\cdot Y_3)^3}-\frac 1 {(-2P_1\cdot Y_2)^3}\right)G_\Delta(Y_2;Y_3)\,.
\end{equation}

Now consider the Witten diagram
\begin{equation}\label{warddi2}
\Xi^{ab}(S_1,P_2,Y_3) = {\scriptsize\feynmandiagram [inline=(b.base), horizontal = a to c] {
d [particle=$J(S_1)$] -- [photon] b ,
a [particle=$\Fk(P_2)$] -- [fermion] b -- [fermion] c [particle = $\fk^*(Y_3)$]
};}\,.
\end{equation}
This diagram is identical to \eqref{warddi1}, except that we have taken the boundary limit of the bulk point
\begin{equation}
Y_2(s) = e^{s} P_2 + \frac 12e^{-s} V_2^I\,, \qquad \text{ with }V_2\cdot P_2 = -1\,.
\end{equation}
We can therefore use \eqref{scboundlim} to compute:
\begin{equation}\begin{split}
\nb_{ab}^I\Xi^{ab}(S_1,P_2,Y_3) &= \lim_{s\rightarrow\infty} \fcy N_\Dk e^{s\Dk}\nb_{ab}^I\Sk^{ab}\left(S_1,Y_2(s),Y_3\right)\,,
\end{split}\end{equation}
Taking the limit as $s\rightarrow0$ of the first term in \eqref{witdiev} is straightforward, so let us focus on the second term:
\begin{equation}
\lim_{s\rightarrow\infty} \frac 1{\left(-2P_1\cdot Y_2(s)\right)^3} = \lim_{s\rightarrow\infty} \frac {1}{(-2e^{s}P_1\cdot P_2 - e^{-s} P_1\cdot V_2)^3}\,.
\end{equation}
As we take $s\rightarrow \infty$ we find that this vanishes unless $P_1 \cdot P_2 = 0$, but that for $P_1\cdot P_2 = 0$ the limit is singular. Using flat-space coordinates \eqref{eq:PForm} to parametrize $P_1$ and $P_2$, and setting $V^I$ to be the point at infinity, $P_\infty^I =(\vec{0},-1,1)$,  we find that
\begin{equation}
\lim_{s\rightarrow\infty} \frac 1{\left(-2P_1(\vec x_1)\cdot Y_2(s)\right)^3} = \lim_{s\rightarrow\infty}\frac{1}{\left(e^{s}(\vec x_1-\vec x_2)^2+e^{-s}\right)^3} = \frac{\pi^2}4 \delta^{(3)}(\vec x_1-\vec x_2)\,.
\end{equation}
To compute the limit on arbitrary conformally flat manifolds we use a Weyl transformation on this result, obtaining
\begin{equation}
\lim_{s\rightarrow\infty} \frac 1{\left(-2P_1(\vec x_1)\cdot Y_2(s)\right)^3} = \frac{\pi^2}4\frac{\delta^{(3)}(\vec x_1-\vec x_2)}{\sqrt{-g}}
\equiv \frac{\pi^2}4\delta^{(3)}(P_1;P_2)\,.\end{equation}
We therefore find that
\begin{equation}\label{witdiev2}
\nb_{ab}^I\Xi^{ab}(S_1,P_2,Y_3) = \frac{e P^I_1\Gk(\Dk)}{(-2P_2\cdot Y_3)^\Dk}\left(\frac{1}{(-2P_1\cdot Y_3)^3}-\frac{\pi^2}4\delta^{(3)}(P_1;P_2)\right)\,.
\end{equation}

Using \eqref{witdiev} and \eqref{witdiev2}, it is straightforward to check that the Ward identities hold in order by order in $e$. As is well known that gauge invariance requires that we consider the sum of diagrams in which the external photon is inserted in all possible places along a charged line. In such a sum we find that all terms in \eqref{witdiev} and \eqref{witdiev2} will cancel out, except for the delta functions in \eqref{witdiev2} which always come from the end-points of a charged line. As a simple illustration of this point, consider the sum of the two diagrams
\begin{equation}
\Pi^I(S_0,S_1,S_2,\dots) = {\scriptsize\feynmandiagram [inline=(b2.base), horizontal = a to c] {
d [particle=$\nb_{ab}^IJ^{ab}(S)$] -- [photon] b1,
d -- [photon,opacity = 0] b3,
a [particle=$\Fk(P_1)$] -- [fermion] b1 -- [fermion] b2 -- [fermion] b3 -- [fermion] c [particle = $\Fk^*(P_2)$],
b2 -- h [blob] -- e,
f -- h -- g,
b1 -- [opacity = 0] f,
b3 -- [opacity = 0] g
};} + {\scriptsize\feynmandiagram [inline=(b2.base), horizontal = a to c] {
d [particle=$\nb_{ab}^IJ^{ab}(S)$] -- [photon,opacity = 0] b1,
d -- [photon] b3,
a [particle=$\Fk(P_1)$] -- [fermion] b1 -- [fermion] b2 -- [fermion] b3 -- [fermion] c [particle = $\Fk^*(P_2)$],
b2 -- h [blob] -- e,
f -- h -- g,
b1 -- [opacity = 0] f,
b3 -- [opacity = 0] g
};}\,.
\end{equation}
By using \eqref{witdiev2}, we can deduce that
\begin{equation}
\Pi^I(S_0,S_1,S_2,\dots)  = \frac{e\pi^2}4P^I\left(\delta(P_0;P_1)-\delta(P_0;P_2)\right)\Pi_1(S_1,S_2,\dots) \,,
\end{equation}
where $\Pi_1(S_1,S_2,\dots)$ is the diagram with no photon
\begin{equation}
\Pi_1(S_1,S_2,\dots) = {\scriptsize\feynmandiagram [inline=(b2.base), horizontal = a to c] {
d -- [photon,opacity = 0] b1,
d -- [photon,opacity = 0] b3,
a [particle=$\Fk(P_1)$] -- b1 -- [fermion] b2 -- [fermion] b3 -- c [particle = $\Fk^*(P_2)$],
b2 -- h [blob] -- e,
f -- h -- g,
b1 -- [opacity = 0] f,
b3 -- [opacity = 0] g
};} \,.
\end{equation}

\section{Scalar Witten diagrams}
\label{sec:SCLWITDIA}

In this appendix we collect some basic results about scalar field theories in AdS$_4$.

Consider a free scalar field $\fk(X)$ dual to a boundary field $\Fk(P)$ of dimension $\Dk$. The free scalar propagator is given by
\begin{equation}
\<\fk(X_1)\fk(X_2)\> = G_\Dk(u)
\end{equation} 
where $u = -1 - X_1\cdot X_2$ and $G_\Dk(u)$ is the function \cite{BURGESS1985137}:
\begin{equation}
G_\Dk(u) = \frac{\Gk(\Dk)}{2\pi^{3/2}\Gk(\Dk-1/2)}(2u)^{-\Dk}{}_2F_1\left(\Dk,\Dk-1,2\Dk-2,-\frac2u\right)\,.
\end{equation}
This propagator satisfies the free scalar equation of motion
\begin{equation}
\left[(\Nb_1)^2-m^2\right]\<\fk(X)\fk(Y)\> = -\dk(X,Y)\,,
\end{equation}
where, as we showed in \eqref{sec:bulkbound}, $m^2 = \Dk(\Dk-3)$.

We define the $D$-function \cite{DHoker:1999kzh} by the integral
\begin{equation}
D_{\Dk_1\dots\Dk_n}(P_1,\dots,P_n) \equiv \int_{AdS}dX \prod_{i = 1}^n \frac 1 {(-2P_i\cdot X)^{\Dk_i}}\,.
\end{equation}
In Mellin space this corresponds to a constant Mellin amplitude \cite{Symanzik:1972wj,Penedones:2010ue}:
\begin{equation}
D_{\Dk_1\dots\Dk_n}(P_1,\dots,P_n) = \frac{\pi^{3/2}}2\Gk\left(\frac{\sum_i\Dk_i-3}2\right)\left(\prod_{i = 1}^n\Gk(\Dk_i)\right)\int [d\gk]\prod_{i<j} \frac{\Gk(\gk_{ij})}{(-2P_i\cdot P_j)^{\gk_{ij}}}\,,
\end{equation}
where $[d\gk]$ is the usual Mellin space measure
\begin{equation}
[d\gk] = \prod_{i = 1}^n \frac{d\gk_i}{2\pi i}\, \dk\left(\Dk_i-\sum_j\gk_{ij}\right)\,.
\end{equation}
Scalar contact Witten diagrams can be evaluated using $D$-functions. When considering four point functions, it is useful to define the ``Mandelstam invariants''
\begin{equation}
s = \Dk_1+\Dk_2-2\gk_{12}\,,\qquad t = \Dk_1+\Dk_3-2\gk_{13}\,,\qquad u = \Dk_1+\Dk_4-2\gk_{14}\,,
\end{equation}
which satisfy $s+t+u = \sum \Delta_i$. 

Let us next turn to the tree-level scalar scattering:
\begin{equation}
\Pi_{s\text{-exch}}(P_i) = \feynmandiagram [inline=(b.base), horizontal = b to e] {
a [particle=$\Fk_1$] --  b --  c [particle=$\Fk_2$],
d [particle=$\Fk_3$] --   e --  f [particle=$\Fk_4$],
b -- [edge label=$\fk$] e
};\,.
\end{equation}
We will denote the conformal dimension of $\Fk_i$ by $\Dk_i$, and the conformal dimension of the exchanged field $\fk$ by $\Dk$. These diagrams were first evaluated in \cite{DHoker:1999mqo} for conformal dimensions satisfying ${{(\Delta_1+\Delta_2-\Delta)/2}\in\mathbb N}$, as a finite sum of $D$-functions, while the more general case was evaluated in \cite{Penedones:2010ue,Fitzpatrick:2011ia,Paulos:2011ie}, and was found to be:
\begin{equation}
	\Pi_{s\text{-exch}}(P_i) = \frac{\pi^{3/2}\Gk\left(\frac{\Dk_1+\Dk_2+\Dk-3}2\right)\Gk\left(\frac{\Dk_3+\Dk_4+\Dk-3}2\right)}{2\Gk\left(\Dk-\frac12\right)}\int [d\gk]\fcy M_{s\text{-exch}}(\gk_{ij})\prod_{i<j}\frac{\Gk(\gk_{ij})}{(-2P_i\cdot P_j)^{\gk_{ij}}}
\end{equation}
where the Mellin amplitude is given by:
\begin{equation}\begin{split}
\fcy M_{s\text{-exch}} &= \sum_{m = 0}^\infty \frac {a_m}{s-\Delta-2m}\, \\
\text{with } a_m &= \frac{\left(\frac{2+\Dk-\Dk_1-\Dk_2}2\right)_m\left(\frac{2+\Dk-\Dk_3-\Dk_4}2\right)_m}{m!\left(\Dk-\frac12\right)_m}\,.
\end{split}\end{equation}
The method used to evaluate scalar exchange diagrams comes down to the following observation. When we act with $(D_1+D_2)^2$ on $\Pi_{s\text{-exch}}(P_i)$, we can use conformal invariance to trade the boundary conformal transformations for a bulk transformation, $\fcy D_X^2$, acting on the bulk field $\fk(X)$. The bulk-to-bulk propagator is Green's function of this differential operator:
\begin{equation}
(\Nb_X^2-m^2)\langle\fk(X)\fk(Y)\rangle = (\fcy D^2_X-\Delta(\Delta-d))\langle \fk(X)\fk(Y)\rangle = -\delta(X,Y)\,,
\end{equation}
and so this reduces the exchange diagram into a contact diagram. We thus find that
\begin{equation}\begin{split}
\Big[-(D_1+D_2)^2&+\Delta(\Delta-d)\Big]\Pi_{s\text{-exch}}(P_i) = \feynmandiagram [inline=(b.base), horizontal = f to b] {
a [particle=$\Phi_1$] --  b --  c [particle=$\Phi_2$],
d [particle=$\Phi_3$] --  b -- [opacity=0] e ,
f -- [opacity=0] b --  g [particle=$\Phi_4$],
h   -- [opacity=0]  b -- [opacity=0] i
}; \\
&= \frac{\pi^{3/2}}2\Gk\left(\frac{\sum_i\Dk_i-3}2\right)\left(\prod_{i = 1}^n\Gk(\Dk_i)\right) D_{\Dk_1\Dk_2\Dk_3\Dk_4}(P_i)\,.
\end{split}\end{equation}
When this identity is converted into Mellin space, it becomes a finite difference equation, which one can solve in order to compute $\fcy M_{s\text{-exch}}$.

\bibliographystyle{ssg}
\bibliography{Bispinor}

\end{document}